\documentclass[12pt,a4paper]{article}

\title{Branes wrapped on quadrilaterals}
\author{Federico Faedo, Alessio Fontanarossa, Dario Martelli}

\usepackage[hmargin=2.5cm,vmargin=3cm]{geometry}
\usepackage[english]{babel}
\usepackage[utf8]{inputenc}
\usepackage{amsmath,amssymb,amsfonts,amsthm}
\usepackage{mathrsfs}
\usepackage{caption}
\usepackage{subcaption}

\usepackage[bbgreekl]{mathbbol}
\DeclareSymbolFontAlphabet{\mathbb}{AMSb}
\DeclareSymbolFontAlphabet{\mathbbl}{bbold}

\numberwithin{equation}{section}
\usepackage[bottom]{footmisc}
\usepackage{cite}

\usepackage{xcolor}
\usepackage[
  bookmarksnumbered,
  linktocpage,
  colorlinks,
  citecolor={green!50!black},
	pdfusetitle
]{hyperref}

\theoremstyle{plain}
\newtheorem{theorem}{Theorem}
\newtheorem{lemma}{Lemma}
\newtheorem{proposition}{Proposition}

\theoremstyle{remark}
\newtheorem*{remark}{Remark}

\usepackage{tikz}
\tikzset{
	dot/.style={circle, minimum size=3pt, inner sep=0, fill=black},
	every label/.append style={font=\footnotesize, label distance=-5pt}
}
\usetikzlibrary{calc}

\usepackage{array}
\newcolumntype{C}[1]{>{\centering\let\newline\\\arraybackslash\hspace{0pt}}m{#1}}

\usepackage{etoolbox}
\makeatletter
\patchcmd\end@float{\@cons\@currlist\@currbox}
   {\@cons\@currlist\@currbox
    \global\holdinginserts\@ne}
    {}{}

\apptocmd\@specialoutput{\global\holdinginserts\z@}
\makeatother

\newcommand{\ie}{\textit{i.e.}}
\newcommand{\eg}{\textit{e.g.}}
\newcommand{\cf}{\textit{cf.}}

\DeclareMathOperator{\sign}{sign}

\newcommand{\NN}{\mathbb{N}}
\newcommand{\ZZ}{\mathbb{Z}}
\newcommand{\QQ}{\mathbb{Q}}
\newcommand{\RR}{\mathbb{R}}
\newcommand{\CC}{\mathbb{C}}
\newcommand{\spindle}{\mathbbl{\Sigma}}

\newcommand{\hemi}{{\mathbbl{S}^4}}
\newcommand{\Morb}{\mathbb{M}}
\newcommand{\loci}{\mathcal{L}}

\newcommand{\vol}[1]{\mathrm{vol}(#1)}
\newcommand{\AdS}{\mathrm{AdS}}

\newcommand{\dd}{\mathrm{d}}
\newcommand{\ee}{\mathrm{e}}
\newcommand{\ii}{\mathrm{i}}

\newcommand{\Nx}{N_x}
\newcommand{\Ny}{N_y}

\newcommand{\XA}{x_-}
\newcommand{\XB}{x_+}
\newcommand{\XAB}{x_\mp}
\newcommand{\XBA}{x_\pm}
\newcommand{\YA}{y_-}
\newcommand{\YB}{y_+}
\newcommand{\YAB}{y_\mp}
\newcommand{\YBA}{y_\pm}

\newcommand{\killJ}{\mathtt{j}}
\newcommand{\killK}{\mathtt{k}}

\newcommand{\sij}{\varkappa_J}
\newcommand{\sik}{\varkappa_K}
\newcommand{\sid}{\varkappa_D}
\newcommand{\kk}{\kappa}

\newcommand{\x}{\mathtt{x}}

\newcommand{\flp}{\mathfrak{p}}
\renewcommand{\flq}{\mathfrak{q}}

\newcommand{\Ispindle}{F}

\newcommand{\fan}{n}
\newcommand{\fpN}{n}
\newcommand{\vv}{v}
\newcommand{\facet}{\mathcal{F}}

\begin{document}

\begin{titlepage}

\begin{center}

\vspace*{2cm}
{\Large \bf Branes wrapped on quadrilaterals}

\vspace{1.5cm}

{Federico Faedo${}^a$, Alessio Fontanarossa${}^{b,c}$ and Dario Martelli${}^{b,c}$}

\vspace{0.7cm}

${}^a$\textit{International Centre for Theoretical Physics Asia--Pacific,\\
University of Chinese Academy of Sciences, 100190 Beijing, China}

\vspace{0.2cm}

${}^b$\textit{Dipartimento di Matematica ``Giuseppe Peano'', Universit\`a di Torino,\\
Via Carlo Alberto 10, 10123 Torino, Italy}

\vspace{0.2cm}

${}^c$\textit{INFN, Sezione di Torino,\\
Via Pietro Giuria 1, 10125 Torino, Italy}

\end{center}

\vspace{2.5cm}

\begin{abstract}

\noindent
We construct new families of supersymmetric $\AdS_2\times\Morb_4$ solutions of $D=6$ gauged supergravity and $\AdS_3\times\Morb_4$ solutions of $D=7$ gauged supergravity, where $\Morb_4$ are
four-dimensional toric orbifolds with four fixed points. These are presented in a unified fashion, that highlights their common underlying geometry.
The $D=6$ solutions uplift to massive type IIA and describe the near-horizon limit of D4-branes wrapped on $\Morb_4$, while the $D=7$ solutions uplift to $D=11$ supergravity and describe the near-horizon limit of M5-branes wrapped on $\Morb_4$.
We reproduce the entropy and gravitational central charge of the two families by extremizing a function constructed gluing the orbifold gravitational blocks proposed in \cite{Faedo:2022rqx}.

\end{abstract}

\end{titlepage}

\tableofcontents

\section{Introduction}

In the history of the AdS/CFT correspondence very often the construction of explicit supergravity solutions with an anti-de Sitter (AdS) factor has prompted
the discovery of new classes of superconformal field theories (SCFTs) and has led to tremendous progress in understanding general aspects of the holographic dictionary.
An important entry in this dictionary is the relationship between extremization problems in SCFTs and their geometric realization. An early example of such  relationships
is that between $a$-maximization in four-dimensional SCFTs  \cite{Intriligator:2003jj} and volume extremization of Sasaki--Einstein manifolds \cite{Martelli:2005tp,Martelli:2006yb}, followed by many other instances involving SCFTs in different dimensions on one side and geometries of different kinds on the other side.

The study of this feature of holography  has been the subject of a conspicuous body of work over the last twenty years and it received a boost recently, following
the uncovering of supergravity solutions comprising geometries with orbifold singularities.
While in the initial examples of these constructions \cite{Ferrero:2020laf,Ferrero:2020twa,Ferrero:2021wvk,Faedo:2021nub} the main actor was the spindle, a number of supergravity
solutions  based on geometries comprising four-dimensional orbifolds $\Morb_4$ have appeared subsequently. These include solutions describing either D4-branes
or  M5-branes wrapped on $\Morb_4$  \cite{Giri:2021xta, Cheung:2022ilc,Suh:2022olh,Couzens:2022lvg,Faedo:2022rqx}.

In this paper we will extend these  constructions by presenting more general classes of supersymmetric  $\AdS_2\times\Morb_4$ solutions of $D=6$ gauged supergravity and  $\AdS_3\times\Morb_4$ solutions of $D=7$ gauged supergravity, where $\Morb_4$ are four-dimensional toric orbifolds with four fixed points for a torus action, that we refer to as \emph{quadrilaterals}, because they are
characterized by  certain labelled
two-dimensional polytopes\footnote{We will be more precise later about how this labelled polytope is constructed.} with four edges \cite{legendre2009,Apostolov:2013}.
We will present the two families of solutions in a unified fashion, exploiting the fact that the four-dimensional factor of the geometry involves in both cases the same data, namely a Hermitian orbifold metric on $\Morb_4$, together with two gauge fields corresponding
to local connections on line orbibundles, as well as two real scalar fields, governed by very similar equations of motion and supersymmetry conditions. In particular, we will give a unified description of the underlying toric orbifolds  in terms of their associated labelled polytopes~\cite{Lerman:1995aaa}.

An important role in the stream of ideas surrounding the geometrization of the extremization principles characterizing SCFTs has been played by the \emph{gravitational blocks} \cite{Hosseini:2019iad}. These are elementary building blocks in terms of which one can assemble various functions of the gravitational  data
that are interpreted as dual to SCFT observables and whose extremization corresponds to necessary (and, sometimes, also sufficient) conditions for the existence of the supergravity solutions. This generally yields a valuable abstract  understanding of the geometry, and allows one  to extract relevant properties of supergravity solutions without their explicit knowledge. Orbifold gravitational blocks for various branes wrapped on the spindle were conjectured in   \cite{Faedo:2021nub} and, in some cases,  subsequently proved using GK geometry \cite{Boido:2022mbe} (for M2 and D3-branes) or equivariant localization \cite{BenettiGenolini:2023kxp} (for M5-branes).

In our previous paper \cite{Faedo:2022rqx} we conjectured the form of the orbifold gravitational blocks for D4-branes as well as M5-branes wrapped on
toric orbifolds $\Morb_4$ and successfully tested it against a number of explicit supergravity solutions.
Subsequently, the validity of this proposal has been confirmed using  the idea  of equivariant localization.
In particular, in \cite{Martelli:2023oqk} it was shown that equivariant integration of the M5-branes anomaly polynomial on four-dimensional toric orbifolds yields an off-shell trial central charge of the corresponding  two-dimensional SCFT that matches precisely the gravitational  block decomposition conjectured in \cite{Faedo:2022rqx}.
Moreover, in \cite{BenettiGenolini:2023kxp} and \cite{Martelli:2023oqk}  has been proposed  that equivariant localization plays a key role for
characterizing   the geometry of supersymmetric solutions. As an application, it has later been shown that  the gravitational blocks associated with M5 and D4-branes wrapped on four-dimensional orbifolds can be
extracted either from the equivariant volume of an associated geometry  \cite{Colombo:2023fhu} or by fixed-point evaluation\footnote{A computation of the gravitational block expression for D4-branes wrapped on orbifolds using the approach of \cite{BenettiGenolini:2023kxp,BenettiGenolini:2023ndb} is  not yet available.} of equivariant integrals of certain
equivariantly closed forms  constructed with Killing spinor bilinears \cite{BenettiGenolini:2023ndb}.

The results of the present paper allow us to test the validity of the
orbifold gravitational blocks and the corresponding extremization principles
in the context of some sophisticated explicit supergravity solutions. In particular, one should bare in mind that, at present, the only extremization principle applied to supergravity
that has been rigorously  proved to guarantee the existence
of solutions is that of volume extremization of Sasaki--Einstein manifolds \cite{Martelli:2006yb}.%
\footnote{Analogous extremization principles have been formulated in geometry, but so far they have not been shown to be directly relevant to supergravity or holography. See, \eg, \cite{Futaki:2006cc,vanCoevering:2012dg,Li:2016qwl,Boyer:2016yae,futaki2017volume}.}
Thus, the construction of explicit solutions in all other cases is a non-trivial result and serves as further evidence of the validity
of the respective extremization principles. In general, it would be wonderful to obtain a proof that the solution of one of these  extremal problems is necessary and sufficient for the existence of supergravity
solutions.

The rest of the paper is organized as follows. In section~\ref{sec:toric} we discuss some preliminaries about the geometry of toric orbifolds.
In section~\ref{sec:sugra-sol} we present the local form of our new solutions  as a single $D$-dependent solution (with $D=6,7$). In order to gain complete analytic control over the complicated algebraic expressions, we will restrict to the subfamilies with equal charges.
In section~\ref{sec:global_structure} various global aspects are discussed, allowing us to uncover the underlying toric orbifolds.
We will show how an artful combination of analytic and topological gadgets will allow us to extract from the solutions the data necessary for implementing the extremization procedure. This is done in section~\ref{sec:extremization}, where the off-shell free energies are computed and their extremization is shown to reproduce the  entropy and the gravitational central charge of the respective explicit solutions.
Section~\ref{sec:discussion} discusses our findings and gives some suggestions for future work.
Five technical appendices complement the paper.
In appendix~\ref{app:spindle-limit} we show how previous solutions can be recovered in particular limits of the present ones.
Appendix~\ref{app:uplifts}  discusses the uplift of  the solutions to $D=10$ and $D=11$ supergravities.
In appendix~\ref{app:6d_KSE} we demonstrate the supersymmetry  of the $D=6$ solutions by constructing  explicit Killing spinors solving the BPS conditions. In appendix~\ref{app:diofa} we show that
the set of solutions to some Diophantine equations that need to be satisfied to construct the supergravity solutions is non-empty.
Finally, in appendix~\ref{app:adjunction} we report a proof of the adjunction formula used in the paper, valid for toric orbifolds in arbitrary dimension, that we have not found in the literature.

\section{Four-dimensional toric orbifolds}
\label{sec:toric}

\subsection{Toric geometry ingredients}
\label{subsec:toric-intro}

Before presenting the compact four-dimensional orbifolds $\Morb_4$ that will be the main focus of this paper, we will recall some results of toric geometry that will be used in the analysis,
highlighting some aspects pertaining to orbifolds. We  start by noting  that in the solutions that we will discuss the metric on $\Morb_4$ solving the supersymmetry conditions will be compatible with an integrable complex structure (thus will be Hermitian), however we will not find a compatible K\"ahler structure. As emphasized in \cite{Martelli:2023oqk,Colombo:2023fhu} this  fact does not prevent one to employ the formalism of symplectic toric geometry as a tool for computing
topological quantities,  on which the corresponding extremal problems, and ultimately the dual field theories, are based.
It should also be possible to reformulate our analysis entirely in terms of complex geometry for orbifolds, and in particular to use the language of toric stacks~\cite{Borisov:2005dm,Fantechi:2010dm,Hochenegger:2012,Sakai:2013,Geraschenko:2015ts1}
to describe the underlying toric geometry in rigorous terms. However,  as we shall see, this will  not be necessary for the purposes of constructing the solutions.

We now give a brief review of the geometry of symplectic toric orbifolds, referring to~\cite{Martelli:2023oqk} and references therein for a more detailed discussion.
Let us consider a $2m$-dimensional symplectic toric orbifold $(\Morb_{2m},\omega)$, equipped with an effective Hamiltonian action of the real torus $\mathbb{T}^m = \RR^m/2\pi\ZZ^m$. A convenient set of coordinates to parameterize $\Morb_{2m}$ are the symplectic coordinates $(y_I,\phi_I)$, with $I=1,\ldots,m$, where $y_I$ are the moment maps associated with $\mathbb{T}^m$ and $\phi_I$ are angular coordinates on the torus, with canonical $2\pi$ periodicity.
According to the construction of~\cite{Lerman:1995aaa},  compact symplectic toric orbifolds are completely characterized by the associated \emph{labelled polytope $\mathcal{P}$}, and the total space $\Morb_{2m}$ arises as a torus fibration over~$\mathcal{P}$. A labelled polytope is a rational simple convex polytope in~$\mathbb{R}^m$, obtained as the image of $\Morb_{2m}$ under the moment maps $y_I$, equipped with a label attached to each of its facets.
This object can be described concisely as the subset
\begin{equation} \label{polytope}
	\mathcal{P} = \bigcap_{a=1}^\fan \bigl\{ y \in \RR^m \; : \; l_a(y) \equiv y_I v_a^I - \lambda_a \ge 0 \bigr\} \,,
\end{equation}
with $\vec{\vv}_a \in \ZZ^m$ and $\lambda_a$ real constants.
The linear equations $l_a(y) = 0$ define the facets~$\mathcal{F}_a$ of the polytope and we denote with $\fan$ their total number. In our conventions, $\vec{\vv}_a$ are the inward-pointing normal vectors to the facets $\mathcal{F}_a$. The labels $m_a$ are positive integers such that the structure group of every point in the preimage of the interior of $\mathcal{F}_a$ under the moment maps is $\ZZ_{m_a}$.
In particular, the labels $m_a$  enter in the above construction as a common factor in the entries of the vectors $\vec{\vv}_a$, which can be written as
$\vec{\vv}_a = m_a \vec{\hat{\vv}}_a$, where  $\vec{\hat{\vv}}_a$ are primitive vectors and therefore define an ordinary fan.%
\footnote{Notice that we are using the conventions of \cite{Martelli:2023oqk}, which are opposite to the ones in~\cite{Faedo:2022rqx}!}
In the context of algebraic geometry the fan  $\{ \vec{\hat{\vv}}_a\}$ defines a toric variety, which by construction can have
orbifold singularities only  in complex codimension higher than one.
In order for $\mathcal{P}$ to be simple, exactly $m$ facets meet at each vertex $p$ and the $m$ vectors associated with these facets form a basis for $\RR^m$.
These vertices correspond to fixed points of the $\mathbb{T}^m$ action. Assuming that a given fixed point is obtained as the intersection of the facets $\mathcal{F}_{a_I}$, with $I=1,\ldots,m$, the order of the orbifold singularity at this point is given by $d_A = \left|\det(\vec{\vv}_{a_1}, \ldots, \vec{\vv}_{a_m})\right|$. The toric orbifold $\Morb_{2m}$ viewed as a torus fibration over the polytope $\mathcal{P}$ is non-degenerate in the interior of $\mathcal{P}$.
At each facet~$\mathcal{F}_a$ a particular circle in $\mathbb{T}^m$ collapses, thus determining a symplectic subspace of $\Morb_{2m}$ of (real) codimension two, which is the preimage of $\mathcal{F}_a$ under the moment maps.
Note that in algebraic geometry, the fan defined by $\vec{\vv}_a$
(equivalently, by $(\vec{\hat{\vv}}_a, m_a)$) does not give rise to a toric variety,
but to a generalization known as a \emph{toric stack} \cite{Borisov:2005dm,Hochenegger:2012,Geraschenko:2015ts1}.
The data  $(\vec{\hat{\vv}}_a, m_a)$ are then referred to as a ``stacky fan'' or an associated  ``stacky polytope'' \cite{Sakai:2013}.

As we shall see, the  data  that we will not be able to extract directly from the metric can be obtained exploiting the fact that there exists a compatible complex structure, thus the underlying spaces are actually (compact) complex orbifolds.
In particular, given a $k$-dimensional complex orbifold~$X$ (that can be either $\Morb_4$ or one of its divisors) from the holomorphic $(k,0)$-form $\Omega_X$ one can compute the connection~$P_X$ on the anti-canonical line bundle through the relation
\begin{equation}
 	\dd\Omega_X = \ii P_X \wedge \Omega_X \,.
\end{equation}
The Ricci form follows as $\rho_X = \dd P_X$ and, from this point onward, we shall refer to $P_X$ as the (real) ``Ricci potential''  corresponding to its curvature $\rho_X$. The first Chern class of the tangent bundle of~$X$ is then obtained taking the  explicit representative $c_1(TX) = [\rho_X]/2\pi$.
From the complex point of view, the preimage of the $\mathcal{F}_a$ described above are \emph{toric divisors}. More precisely, these are the divisors associated with the vectors~$\vec{\vv}_a$ and have to be identified with the ramification divisors, which we will denote as $D_a$. In contrast, the divisors $\hat{D}_a$, associated with the vectors~$\vec{\hat\vv}_a$, are to be identified with the branch divisors (see \eg~\cite{Boyer:2003pe}).
Formally, the two sets of divisors are related as $\hat{D}_a = m_a D_a$, which slightly more accurately may be written as $D_a = \hat D_a \times \mathrm{pt}/\ZZ_{m_a}$, indicating that at each point in $D_a$ the structure group~$\ZZ_{m_a}$ is acting on the complex direction $\CC$ normal to $D_a$.
Equivalently, this relation may be also understood from the expression of the Poincaré dual two-forms, given by the first Chern classes of a set of line bundles $L_a$.
Their explicit form is~\cite{Guillemin:1994kae,Abreu:2001to}
\begin{equation} \label{c1(La)}
	c_1(L_a) = -\frac{\ii}{2\pi} \bigl[ \partial\bar\partial \log l_a \bigr] \,,
\end{equation}
where $[\alpha]$ denotes the cohomology class of a given differential form $\alpha$ and $\partial$, $\bar\partial$ are the (anti-)holomorphic differentials.
Using an explicit expression in symplectic coordinates one can show
that indeed $c_1(\hat{L}_a)= m_a c_1(L_a)$, where $\hat{L}_a$ are the line bundles associated with the divisors $\hat D_a$.\footnote{We note in passing that
in the framework of toric stacks, the divisors $D_a$ should be understood in terms of a ``root construction'', namely as the $m_a$-th root of the $\hat  D_a$ \cite{Fantechi:2010dm}. However, we shall refrain from making use of this language in our subsequent discussions.}
One can also see that the following $m$ relations among the Chern classes hold~\cite{Martelli:2023oqk}
\begin{equation} \label{cohomo-rel}
	\sum_a \vec{\vv}_a c_1(L_a) = 0 \,,
\end{equation}
thus only $\fan-m$ of the line bundles are independent.
These  are equivalent to the familiar $m$ linear relations among the divisors
\begin{equation} \label{homo-rel}
	\sum_a \vec{\vv}_a D_a = 0 \,,
\end{equation}
which imply that there are only $\fan-m$ independent ($2m-2$)-cycles in homology.

It will also be useful to note that given a toric ramification divisor $D_a$ of the total space~$\Morb_{2m}$, the first Chern class of the tangent bundle $T\Morb_{2m}$ can be decomposed by means of the \emph{adjunction formula} as
\begin{equation} \label{adjunction}
	c_1(T\Morb_{2m}) \bigr|_{D_a} = c_1(TD_a) + c_1(L_a) \bigr|_{D_a} \,.
\end{equation}
In this formula the first Chern class of the line bundles $L_a$ is given by~\eqref{c1(La)}, while $c_1(T\Morb_{2m})$ follows from
\begin{equation} \label{c1(TM)}
	c_1(T\Morb_{2m}) = \sum_a c_1(L_a) \,.
\end{equation}
In the case of four-dimensional toric orbifolds, which is relevant for this paper, a proof can be found in~\cite{Chen:2004abc}.
In general dimension we are not aware of a standard reference for an adjunction formula valid for orbifolds.
However, in the toric setting, endowing each divisor with its natural complex structure, from the corresponding Ricci form $\rho_{D_a}$ we can compute  $c_1(TD_a)=[\rho_{D_a}]/2\pi$, and then the  adjunction formula~\eqref{adjunction} can be verified explicitly. For completeness, the details of the computation are reported in appendix~\ref{app:adjunction}.

\subsection{Quadrilaterals}
\label{quadrilat_section}

We now continue requiring  that $\Morb_4$ has four real dimensions, namely setting $m=2$. With this assumption, the polytope becomes planar and, therefore, enjoys the special feature of having the same number of facets (edges) and fixed points (vertices), that we denote by~$\fpN$. When working with four-dimensional toric orbifolds, the intersection matrix of two toric divisors is \cite{Martelli:2023oqk}
\begin{equation} \label{inter-num}
  D_{ab} \equiv D_a \cdot D_b = \int_{\Morb_4} c_1(L_a) \wedge c_1(L_b) =
	\left\{
	\begin{aligned}
		& \frac{1}{d_{a-1,a}}    \qquad  && \text{if} \ b = a - 1 \,, \\
		& \frac{1}{d_{a,a+1}}    \qquad  && \text{if} \ b = a + 1 \,, \\
		& -\frac{d_{a-1,a+1}}{d_{a-1,a} \, d_{a,a+1}}    \qquad  && \text{if} \ b = a \,, \\
		& 0    \qquad  && \text{otherwise} \,,
	\end{aligned}
	\right.
\end{equation}
where we defined $d_{a,b} \equiv \det(\vec{\vv}_a, \vec{\vv}_b)$.
One can verify explicitly that
\begin{equation} \label{matrix-vectors}
	\sum_b D_{ab} \, \vec{\vv}_b = 0 \,,
\end{equation}
which is a consequence of the cohomological relation~\eqref{cohomo-rel}.

It is straightforward to show that $d_{a,b}$ is invariant under $SL(2,\ZZ)$ transformations applied to all the vectors $\vec{\vv}_a$ of the fan. This property will be particularly useful for us,
because it implies that also the intersection matrix $D_{ab}$ is not affected by such transformations. The fan corresponding to a given toric orbifold is thus determined only up to $SL(2,\ZZ)$ transformations, and, in a similar way, also the related polytope.
In fact, this group of invariance can be formally extended to $SL(2,\RR)$. In this case, the new fan is not well-defined, since the vectors are not integer-valued, but the
$d_{a,b}$ are unchanged and, in particular, we still have $d_{a,b}\in \ZZ$. This property will turn out to be crucial, as it will allow
 us to work with ``fake'' vectors $\vec{V}_a$, that are not $\ZZ^2$-valued,%
\footnote{This implies that they are expressed in a ``wrong'' basis for the $\mathbb{T}^2$ action, in which the orbits in general do not close.}
in intermediate stages of the computations.

Using  the adjunction formula~\eqref{adjunction}, we can relate the toric data,
encoded in the intersection matrix $D_{ab}$, and the complex formalism, embodied by $c_1(T\Morb_4)$ and $c_1(TD_a)$, by means of the following two relations%
\footnote{The second equation can be obtained from the first using~\eqref{c1(TM)} together with $\sum_b D_{ab} = \int_{D_a} c_1(T\Morb_4)$.}
\begin{align}
	\label{chern-D-1}
	& D_{aa} = \int_{D_a} c_1 (L_a) = \int_{D_a} \bigl[ c_1(T\Morb_4) - c_1(TD_a) \bigr] \,, \\
	\label{chern-D-2}
	& D_{a\,a-1} + D_{a\,a+1} = \sum_{b \neq a} \int_{D_a} c_1(L_b) = \int_{D_a} c_1(TD_a) = \chi_{(a)} \,,
\end{align}
where, by definition, $\chi_{(a)}$ is the orbifold Euler characteristic of the divisor $D_a$.

We now specialize to the case of four fixed points, that we refer to as quadrilaterals. Indeed, these configurations are completely characterized by a generic polytope depicted in figure~\ref{fig:polytope}, which is a convex polygon with four vertices and four edges.
\begin{figure}[ht]
	\centering

	\begin{tikzpicture}
		\draw[-latex] (0,0) -- (2.5,0)     node[below left]  {$\vec{\vv}_1$};
		\draw[-latex] (0,0) -- (0.5,1.5)   node[below right] {$\vec{\vv}_2$};
		\draw[-latex] (0,0) -- (-2.5,-1.5) node[above left]  {$\vec{\vv}_3$};
		\draw[-latex] (0,0) -- (1.5,-1)    node[below left]  {$\vec{\vv}_4$};

		\node[dot] (p12) at (1,0.5) {};
		\node[dot] (p23) at (-2,1.5) {};
		\node[dot] (p34) at (-0.010,-1.816) {};
		\node[dot] (p41) at (1,-0.3)   {};

		\draw[-] (p12) -- (p23) -- (p34) -- (p41) -- (p12);
	\end{tikzpicture}

	\caption{Outward-pointing fan and polytope of a generic quadrilateral. The vectors $\vec{v}_a$ are \emph{not} primitive.}
	\label{fig:polytope}
\end{figure}

Their basis-independent information is encoded in the following $4 \times 4$ intersection matrix~\eqref{inter-num}
\begin{equation}
	D_{ab} =  \begin{pmatrix}
		\frac{d_{2,4}}{d_{4,1} \, d_{1,2}}  &  \frac{1}{d_{1,2}}  &  0  &  \frac{1}{d_{4,1}} \\
		\frac{1}{d_{1,2}}  &  -\frac{d_{1,3}}{d_{1,2} \, d_{2,3}}  &  \frac{1}{d_{2,3}}  &  0 \\
		0  &  \frac{1}{d_{2,3}}  &  -\frac{d_{2,4}}{d_{2,3} \, d_{3,4}}  &  \frac{1}{d_{3,4}} \\
		\frac{1}{d_{4,1}}  &  0  &  \frac{1}{d_{3,4}}  &  \frac{d_{1,3}}{d_{3,4} \, d_{4,1}} \\
	\end{pmatrix} \, ,
\end{equation}
where a vector identity implies that the $d_{a,b}\in \ZZ$ are actually not independent, but satisfy the relation
\begin{equation} \label{drelation}
	d_{1,2} \, d_{3,4} - d_{2,3} \, d_{4,1} = d_{1,3} \, d_{2,4} \,.
\end{equation}

It is convenient to parameterize the integers $d_{a,b}$ as
\begin{equation} \label{dvslabels}
	d_{a,b} = m_a m_b \, \hat{d}_{a,b} \,,
\end{equation}
where $\hat d_{a,b} \equiv \det (\vec{\hat \vv}_a, \vec{\hat \vv}_b)\in \ZZ$ and satisfy
\begin{equation} \label{dhatrelation}
	\hat{d}_{1,2} \, \hat{d}_{3,4}  - \hat{d}_{2,3} \, \hat{d}_{4,1} = \hat{d}_{1,3} \, \hat{d}_{2,4} \,.
\end{equation}
Thus, a generic quadrilateral is characterized by five independent integer parameters $\hat{d}_{a,b}$ (subject to~\eqref{dhatrelation}) and four integer labels $m_a$, which we refer to as the toric data of the orbifold.
Later we will show that the local solutions that we are about to introduce can be extended to well-defined \emph{global} solutions on a quadrilateral whose toric data are
 subject to some additional ``Diophantine constraints'' that we shall discuss. Moreover, we will show that the magnetic fluxes of the supergravity solutions will be written in terms of these toric data and some
 twisting parameters, precisely as conjectured in \cite{Faedo:2022rqx}.

\section{Supergravity solutions}
\label{sec:sugra-sol}

In this section we discuss solutions of specific $D=6,7$ matter-coupled gauged supergravity theories with gauge group $U(1)^2$. Both of them comprise two gauge fields $A_1$, $A_2$, a $(D-4)$-form $B$ and two real scalar fields $\vec{\varphi}=(\varphi_1,\varphi_2)$.

The six-dimensional model can also be obtained as a sub-sector of an extension of Romans $F(4)$ gauged supergravity~\cite{Romans:1985tw}, coupled to three vector multiplets~\cite{DAuria:2000afl,Andrianopoli:2001rs}.
The bosonic part of the action reads\footnote{Here and in what follows we define, for any $p$-form $\omega$, $|\omega|^2 = \frac{1}{p!} \, \omega_{\mu_1\ldots\mu_p} \omega^{\mu_1\ldots\mu_p}$.}
\begin{equation} \label{6d_action}
	\begin{split}
		S_\text{6D} &= \frac{1}{16\pi G_{(6)}} \int \dd^6x \, \sqrt{-g} \biggl[ R - V_6 - \frac12 |\dd\vec{\varphi}|^2 - \frac12 \sum_{i=1}^2 X_i^{-2} |F_i|^2 - \frac18 (X_1 X_2)^2 |H|^2 \\
		& - \frac{m^2}{4} (X_1 X_2)^{-1} |B|^2 - \frac{1}{16} \frac{\varepsilon^{\mu\nu\rho\sigma\tau\lambda}}{\sqrt{-g}} B_{\mu\nu} \Bigl( F_{1\,\rho\sigma} F_{2\,\tau\lambda} + \frac{m^2}{12} B_{\rho\sigma} B_{\tau\lambda} \Bigr) \biggr] \,,
	\end{split}
\end{equation}
where $F_i=\dd A_i$, $H=\dd B$ and the scalar fields $\vec{\varphi}$ are parameterized as
\begin{equation}
	X_i = \ee^{-\vec{a}_i\cdot\vec{\varphi}}  \qquad  \text{with}  \qquad  \vec{a}_1 = \bigl(2^{-1/2}, 2^{-3/2}\bigr) \,,  \qquad  \vec{a}_2 = \bigl(-2^{-1/2}, 2^{-3/2}\bigr) \,.
\end{equation}
The scalar potential is
\begin{equation}
	V_6 = m^2 X_0^2 - 4g_c^2 X_1 X_2 - 4mg_c \, X_0 (X_1 + X_2) \,,
\end{equation}
with $g_c$ the gauge coupling and $m$ the mass parameter, and where, for later convenience, we defined $X_0=(X_1 X_2)^{-3/2}$. In what follows we shall set $m=2g_c/3$.%
\footnote{As pointed out in~\cite{Faedo:2021nub} and~\cite{Faedo:2022rqx}, this can always be done without loss of generality when $g_c,m>0$.}

The seven-dimensional theory can be constructed as a consistent truncation~\cite{Liu:1999ai} of the $D=7$, maximal $SO(5)$ gauged supergravity~\cite{Pernici:1984xx}. The bosonic part of the action is~\cite{Chong:2004dy}
\begin{equation} \label{7d_action}
	\begin{split}
		S_\text{7D} &= \frac{1}{16\pi G_{(7)}} \int \dd^7x \, \sqrt{-g} \biggl[ R - V_7 - \frac12 |\dd\vec{\varphi}|^2 - \frac12 \sum_{i=1}^2 X_i^{-2} |F_i|^2 - \frac12 (X_1 X_2)^2 |H|^2 \\
		& - \frac{1}{24} \frac{\varepsilon^{\mu\nu\rho\sigma\tau\lambda\eta}}{\sqrt{-g}} B_{\mu\nu\rho} \Bigl( F_{1\,\sigma\tau} F_{2\,\lambda\eta} - \frac{g_c}{12} H_{\sigma\tau\lambda\eta} \Bigr) \biggr] \,.
	\end{split}
\end{equation}
Again, the field strengths are given by $F_i=\dd A_i$ and $H=\dd B$, while now the scalar fields~$\vec{\varphi}$ take the parameterization
\begin{equation}
	X_i = \ee^{-\vec{a}_i\cdot\vec{\varphi}}  \qquad  \text{with}  \qquad  \vec{a}_1 = \bigl(2^{-1/2}, 10^{-1/2}\bigr) \,,  \qquad  \vec{a}_2 = \bigl(-2^{-1/2}, 10^{-1/2}\bigr) \,.
\end{equation}
The scalar potential is
\begin{equation}
	V_7 = \frac{g_c^2}{2} \bigl[ X_0^2 - 8X_1 X_2 - 4X_0 (X_1 + X_2) \bigr] \,,
\end{equation}
with $g_c$ the gauge coupling and, in this context, $X_0=(X_1 X_2)^{-2}$.
Additionally, the following self-duality condition must hold
\begin{equation} \label{7d_duality}
	(X_1 X_2)^2 \star\!H = -g_c B + \frac12 A_1 \wedge F_2 + \frac12 A_2 \wedge F_1 + \dd\lambda \,,
\end{equation}
for some two-form $\lambda$.

\subsection{Solutions in $D$ dimensions}
\label{subsec:sugra-solutions}

We now present a family of solutions to both the $D=6$ and the $D=7$ theories in which the spacetime has the structure of a warped product $\AdS_{D-4}\times\Morb_4$. Their local form can be obtained as an
analytic continuation of the asymptotically AdS$_D$, non-extremal, charged and rotating\footnote{It is crucial that they have a single non-zero angular momentum, so that the solutions contain
a round $(D-4)$-sphere, which results in an $\AdS_{D-4}$ factor after analytic continuation.} black holes constructed in~\cite{Chow:2011fh}, with the addition of a NUT parameter.
These backgrounds can be written in a unified fashion, with the line element reading
\begin{equation} \label{nut_metric}
	\dd s^2 = (H_1 H_2)^{1/(D-2)} \, \biggl( \frac{x^2 y^2}{a^2} \, \dd s_{\AdS_{D-4}}^2 + \dd s_{\Morb_4}^2 \biggr) \,,
\end{equation}
where $\dd s_{\AdS_{D-4}}^2$ denotes the unit radius metric on~$\AdS_{D-4}$ and we defined the four-dimensional metric
\begin{equation} \label{nut_metric-M4}
	\begin{split}
		\dd s_{\Morb_4}^2 &= \frac{(H_1 H_2)^{-1}}{\Xi^2 (x^2 - y^2)} \biggl[ (V_y^2 \Delta_x - V_x^2 \Delta_y) \, \dd\psi^2 + (\widetilde{V}_y^2 \Delta_x - \widetilde{V}_x^2 \Delta_y) \, \dd\phi^2 \\
		& - \frac{4c_1 c_2 \tilde{c}_1 \tilde{c}_2}{a^2} \biggl( \frac{\Ny \Delta_x}{y^{D-5}} - \frac{\Nx \Delta_y}{x^{D-5}} \biggr) \, \dd\psi \, \dd\phi \biggr] + \frac{y^2 - x^2}{\Delta_y} \, \dd y^2 + \frac{x^2 - y^2}{\Delta_x} \, \dd x^2 \,.
	\end{split}
\end{equation}
The scalar fields are
\begin{equation} \label{nut_scalars}
	X_i = (H_1 H_2)^{(D-3)/2(D-2)} H_i^{-1} \,,
\end{equation}
the gauge potentials are given by
\begin{equation} \label{nut_gauge}
	\begin{split}
		A_i &= \frac{2\Ny s_i c_i \tilde{c}_i}{\Xi (x^2 - y^2) H_i y^{D-5}} \biggl( \frac{\tilde{c}_1 \tilde{c}_2}{\tilde{c}_i^2} \, V_x^{(i)} \, \dd\psi - \frac{c_1 c_2}{c_i^2} \, \widetilde{V}_x^{(i)} \, \dd\phi \biggr) \\
		& - \frac{2\Nx s_i c_i \tilde{c}_i}{\Xi (x^2 - y^2) H_i x^{D-5}} \biggl( \frac{\tilde{c}_1 \tilde{c}_2}{\tilde{c}_i^2} \, V_y^{(i)} \, \dd\psi - \frac{c_1 c_2}{c_i^2} \, \widetilde{V}_y^{(i)} \, \dd\phi \biggr) \,,
	\end{split}
\end{equation}
and the $(D-4)$-form is
\begin{equation}
	\begin{aligned}
		B &= -\frac{4s_1 s_2 (\Ny x^3 - \Nx y^3)}{a^2 (x^2 - y^2)} \, \vol{\AdS_2}   \qquad  && (D=6) \,, \\
		B &= -\frac{2s_1 s_2 (\Ny x^4 - \Nx y^4)}{a^3 (x^2 - y^2)} \, \vol{\AdS_3} - \frac{2\Ny g s_1 s_2 x}{\Xi a^2 y^2} \, \dd\psi \wedge \dd\phi \wedge \dd x \\
		& - \frac{2\Nx g s_1 s_2 y}{\Xi a^2 x^2} \, \dd\psi \wedge \dd\phi \wedge \dd y  \qquad  && (D=7) \,.
	\end{aligned}
\end{equation}
The functions are
{\allowdisplaybreaks
\begin{align}
	\Delta_y &= -y^2 + a^2 - \frac{2\Ny}{y^{D-5}} + g^2 \biggl( y^2 - \frac{2\Ny s_1^2}{y^{D-5}} \biggr) \biggl( y^2 - \frac{2\Ny s_2^2}{y^{D-5}} \biggr) - a^2 g^2 y^2 - \frac{2\Ny a^2 g^2 s_1^2 s_2^2}{y^{D-5}} \,, \nonumber \\
	\Delta_x &= -x^2 + a^2 - \frac{2\Nx}{x^{D-5}} + g^2 \biggl( x^2 - \frac{2\Nx s_1^2}{x^{D-5}} \biggr) \biggl( x^2 - \frac{2\Nx s_2^2}{x^{D-5}} \biggr) - a^2 g^2 x^2 - \frac{2\Nx a^2 g^2 s_1^2 s_2^2}{x^{D-5}} \,, \nonumber \\[0.3em]
	V_y^2 &= V_y^{(1)} V_y^{(2)} \,,  \qquad  \widetilde{V}_y^2 = \widetilde{V}_y^{(1)} \widetilde{V}_y^{(2)} \,,  \qquad  V_x^2 = V_x^{(1)} V_x^{(2)} \,,  \qquad  \widetilde{V}_x^2 = \widetilde{V}_x^{(1)} \widetilde{V}_x^{(2)} \,, \nonumber \\[0.5em]
	V_y^{(i)} &= 1 - g^2 \biggl(y^2 - \frac{2\Ny s_i^2}{y^{D-5}}\biggr) \,,  \qquad  \widetilde{V}_y^{(i)} = 1 - \frac{1}{a^2} \biggl(y^2 - \frac{2\Ny s_i^2}{y^{D-5}}\biggr) \,, \\
	V_x^{(i)} &= 1 - g^2 \biggl(x^2 - \frac{2\Nx s_i^2}{x^{D-5}}\biggr) \,,  \qquad  \widetilde{V}_x^{(i)} = 1 - \frac{1}{a^2} \biggl(x^2 - \frac{2\Nx s_i^2}{x^{D-5}}\biggr) \,, \nonumber \\
	H_i &= 1 + \frac{2s_i^2}{x^2 - y^2} \biggl(\frac{\Ny}{y^{D-5}} - \frac{\Nx}{x^{D-5}} \biggr) \,, \nonumber \\
	\Xi &= 1 - a^2 g^2 \,,  \qquad  s_i = \sinh\delta_i \,,  \quad  c_i = \cosh\delta_i \,,  \quad  \tilde{c}_i = \sqrt{1 + a^2 g^2 s_i^2} \nonumber \,.
\end{align}
}
The constant $g$ is related to the gauge coupling as $g=2g_c/3$ in $D=6$ and as $g=g_c/2$ in $D=7$. In our discussion we shall assume $g_c>0$  in both theories, therefore $g>0$.
The background is completely determined by a real parameter $a$, two charges $\delta_{1,2}$ and two parameters $\Ny$ and $\Nx$ (corresponding to the NUT and the  mass of the Lorentzian solution).
When $x=0$ or $y=0$ a curvature singularity is encountered, therefore the range of definition of these two coordinates will have to avoid the origin of the real axis.
Moreover, in the rest of the paper we shall consider these two ranges to be disjoint, in order to prevent the occurrence of a singularity when $x=y$. We refer to section~\ref{sec:discussion} for a very brief discussion about the latter case.

The equations of motion that descend from the actions~\eqref{6d_action} and~\eqref{7d_action}, as well as the self-duality condition~\eqref{7d_duality}, have been checked explicitly, with the spacetime orientation induced by the coordinate ordering $(\AdS,y,\phi,x,\psi)$.

The $\AdS_{D-4}\times\Morb_4$ background is invariant under the exchange of the coordinates $x$ and $y$, accompanied by $\Nx \leftrightarrow \Ny$.
Additionally, the system is also invariant under the discrete inversion symmetry
\begin{equation} \label{inv_sym}
	\begin{gathered}
		y \mapsto \gamma\,y \,,  \qquad  x \mapsto \gamma\,x \,,  \qquad  \psi \mapsto \gamma\,\phi \,,  \qquad  \phi \mapsto \gamma\,\psi \,, \\
		a \mapsto \gamma^2 a \,,  \qquad  \Ny \mapsto \gamma^{D-1} \Ny \,,  \qquad  \Nx \mapsto \gamma^{D-1} \Nx \,,  \qquad  s_i \mapsto \gamma^{-1} s_i \,,
	\end{gathered}
\end{equation}
with $\gamma = \pm1/(ag)$, which connects solutions with positive and negative values of~$\Xi$.

To show how our backgrounds are related to the solutions in \cite{Chow:2011fh}, we must perform an analytic continuation of $\AdS_{D-4}$ into the sphere $S^{D-4}$ and define
the time and radial coordinates $t$ and $r$ as well as the  mass parameter $M$ as
\begin{equation} \label{Wick}
	t = \psi \,,  \qquad  r = \ii \, x \,,  \qquad  M = \ii^{D-5} \Nx \,.
\end{equation}
The result is the family of black holes of~\cite{Chow:2011fh} for $D=6,7$, with the addition of a NUT parameter, here denoted as~$\Ny$.
In the same way it is possible to retrieve the seven-dimensional solutions presented in~\cite{Wu:2011gp}, equivalent to the ones of~\cite{Chow:2011fh}, but written in generalized Boyer--Lindquist coordinates.

Moreover, the solutions presented in this paper and the Kerr--NUT--AdS black holes constructed in~\cite{Chen:2006xh} match in an overlapping region of  parameters. Indeed, the uncharged subfamily of the former coincides with the $r\to0$ limit of the latter, when the mass parameter is set to zero and only one angular momentum is retained.
Lastly, in both six and seven dimensions, the $\AdS$ vacuum solution is recovered setting $\Ny=\Nx=0$, and its radius is $L_{\AdS} = 1/g$.

In~\cite{Chow:2011fh}, the supersymmetry conditions for the non-extremal black holes constructed therein were derived by studying the vanishing of the eigenvalues of the Bogomolny matrix. The BPS condition obtained is $a g s_1 s_2 = \pm1$, with $\delta_1$ and $\delta_2$ of the same sign, either positive or negative.%
\footnote{These combinations can be obtained considering the whole set of eigenvalues of the Bogomolny matrix, \ie\ all the possible combinations $E \pm g J \pm (Q_1 + Q_2)$, not only the one studied in~\cite{Chow:2011fh} (see~\cite{Cvetic:2005zi}, setting to zero all the angular momenta, but one).}
Our solution is the analytic continuation of the black holes of~\cite{Chow:2011fh}, with the addition of NUT. Since the mass parameter of the latter does not appear explicitly in the supersymmetry condition, we expect that the introduction of a NUT parameter does not affect this constraint. Furthermore, since the BPS condition is not altered by~\eqref{Wick}, we conjecture that the solution we presented is supersymmetric when $a g s_1 s_2 = \pm1$, with $\delta_1$ and $\delta_2$ of the same sign.
This assumption is confirmed by the explicit computation we performed in $D=6$ when the two charges are equal ($\delta_1=\delta_2\equiv\delta$), which shows that a non-trivial Killing spinor can be constructed if and only if $ags^2 = \pm1$, with $s=\sinh\delta$ (see appendix~\ref{app:6d_KSE} for  details).
Additionally, as shown in appendix~\ref{app:spindle-limit}, the supersymmetric $\AdS \times \spindle_1 \ltimes \spindle_2$ solutions of~\cite{Faedo:2022rqx,Couzens:2022lvg} and~\cite{Cheung:2022ilc} can be retrieved if and only if the condition $a g s_1 s_2 = \pm1$ is imposed, thus giving evidence for the validity of our hypothesis.

\subsection{Solutions with equal charges}
\label{subsec:equal-charges}

The study of the general system presented in the previous subsection is computationally involved due to the complicated  form of the $(\psi,\phi)$ part of the metric~\eqref{nut_metric-M4}. For this reason, from now on we shall focus on the more tractable subfamily of solutions with equal charges $\delta_1=\delta_2\equiv\delta$.  Despite being simpler, this system still captures all the important features of the four-dimensional toric orbifold~$\Morb_4$, allowing for a detailed analysis of its structure and a reliable test of the gravitational blocks conjecture of~\cite{Faedo:2022rqx}.
We are confident that with some extra effort (and a good motivation) it should be possible to analyse the more general solutions with unequal charges.

When the two charges are set equal, the metric on the toric orbifold~\eqref{nut_metric-M4} acquires the following diagonal structure
\begin{equation} \label{equal_metric-M4}
	\begin{split}
		\dd s_{\Morb_4}^2 &= \frac{1}{\Xi^2 H^2} \biggl[ \frac{\Delta_y}{y^2 - x^2} \bigl( V_x \, \dd\psi - \widetilde{V}_x \, \dd\phi \bigr)^2 + \frac{\Delta_x}{x^2 - y^2} \bigl( V_y \, \dd\psi - \widetilde{V}_y \, \dd\phi \bigr)^2 \biggr] \\
		& + \frac{y^2 - x^2}{\Delta_y} \, \dd y^2 + \frac{x^2 - y^2}{\Delta_x} \, \dd x^2 \,,
	\end{split}
\end{equation}
where we defined the functions $H \equiv H_1 = H_2$, $V_\bullet \equiv V_\bullet^{(1)} = V_\bullet^{(2)}$ and $\widetilde{V}_\bullet \equiv \widetilde{V}_\bullet^{(1)} = \widetilde{V}_\bullet^{(2)}$.
This class of solutions is characterized by having equal scalar fields and equal gauge fields, where the latter now read
\begin{equation} \label{equal_gauge}
	A_1 = A_2 = \frac{2s c \tilde{c}}{\Xi (x^2 - y^2) H} \biggl[ \frac{\Ny}{y^{D-5}} \bigl( V_x \, \dd\psi - \widetilde{V}_x \, \dd\phi \bigr) - \frac{\Nx}{x^{D-5}} \bigl( V_y \, \dd\psi - \widetilde{V}_y \, \dd\phi \bigr) \biggr] \,,
\end{equation}
with $s = \sinh\delta$, $c = \cosh\delta$ and $\tilde{c} = \sqrt{1 + a^2 g^2 s^2}$.
For later convenience, we define the orthonormal frame
\begin{equation} \label{M4-frame}
	\begin{aligned}
		\hat{e}^1 &= \sqrt{\frac{y^2 - x^2}{\Delta_y}} \, \dd y \,,  \qquad  &
		\hat{e}^2 &= \frac{1}{\Xi \, H} \sqrt{\frac{\Delta_y}{y^2 - x^2}} \, \bigl( V_x \, \dd\psi - \widetilde{V}_x \, \dd\phi \bigr) \,, \\
		\hat{e}^3 &= \sqrt{\frac{x^2 - y^2}{\Delta_x}} \, \dd x \,,  \qquad  &
		\hat{e}^4 &= \frac{1}{\Xi \, H} \sqrt{\frac{\Delta_x}{x^2 - y^2}} \, \bigl( V_y \, \dd\psi - \widetilde{V}_y \, \dd\phi \bigr) \,.
	\end{aligned}
\end{equation}

When the system is supersymmetric, additional properties arise. Assuming $ags^2 = -\kk$, with $\kk=\pm1$, we can decompose $\Delta_x$ and~$\Delta_y$ as
\begin{equation} \label{Delta_deco}
	\Delta_x = \Delta_x^+ \, \Delta_x^- \,,  \qquad  \Delta_y = \Delta_y^+ \Delta_y^- \,,
\end{equation}
where
\begin{equation} \label{Delta+-}
	\begin{split}
		\Delta_x^\pm &= g \biggl( x^2 - \frac{2\Nx s^2}{x^{D-5}} \biggr) - \kk\,a \pm (1 - \kk\,ag) x \,, \\
		\Delta_y^\pm &= g \biggl( y^2 - \frac{2\Ny s^2}{y^{D-5}} \biggr) - \kk\,a \pm (1 - \kk\,ag) y \,.
	\end{split}
\end{equation}
Considering, \eg, $\Delta_x$, every root of this function is necessarily a root of either $\Delta_x^+$ or $\Delta_x^-$, therefore, given a generic root $x_*$ of $\Delta_x$, we have
\begin{equation}
	x_*^2 - \frac{2\Nx s^2}{x_*^{D-5}} = \frac{\kk\,a}{g} - \tau \, \frac{1 - \kk\,ag}{g} \, x_* \,,
\end{equation}
with $\tau=\pm$ according to the fact that $x_*$ is a root of $\Delta_x^\pm$, respectively.
Thanks to this relation and the corresponding one for~$y_*$, we can obtain simplified expressions for many functions, when evaluated at one of the roots $\XBA$ or $\YBA$. In particular, the function~$H$ reads
\begin{equation} \label{root_H}
	(x_*^2 - y_*^2) H(x_*,y_*) = -\frac{1 - \kk\,ag}{g} \, (\tau^{(x)} x_* - \tau^{(y)} y_*) \,,
\end{equation}
while the derivatives of $\Delta_x$ and~$\Delta_y$ become, collectively,
\begin{equation} \label{root_Delta'}
	\Delta_z'(z_*) = -2\tau (1 - \kk\,ag) \bigl[ (D-3) g z_*^2 + \tau (D-4) (1 - \kk\,ag) z_* - \kk (D-5) a \bigr] \,,
\end{equation}
where $z$ can be either $x$ or $y$.

For the sake of simplicity, from now on we shall restrict the supersymmetry condition to $ags^2 = +1$, discarding the case $-1$ and with no assumption on the sign of~$\delta$. The discussion could be carried out in the other case with minor modifications.

\subsection{Conditions of existence}
\label{subsec:existence}

We will now begin  investigating  global aspects of the solutions, starting from some basic necessary conditions of existence. In particular, in order for the solutions to  well-defined we require that all the quantities are  real-valued, the signature of the metric is Lorentzian and the scalar fields are positive.
These requirements are met if and only if
\begin{equation} \label{exist-cond}
	\frac{\Delta_x}{x^2 - y^2} > 0 \,,  \qquad  \frac{\Delta_y}{y^2 - x^2} > 0 \,,  \qquad  H > 0 \,,
\end{equation}
in two closed intervals $[\XA,\XB]$ and $[\YA,\YB]$ not intersecting and not containing the curvature singularities in $x=0$ and $y=0$. Without loss of generality we restrict to positive values of $x$ and $y$. This can always be done in $D=7$, while in $D=6$ a change in the sign of $\Nx$ or $\Ny$ may be needed.%
\footnote{In the latter case, the two-form $B$ changes its sign, but the original form can always be restored changing the sign of the $\AdS_2$ time coordinate.}
Taking advantage of the exchange symmetry between $x$ and $y$ we furthermore assume $x>y$, and the first two conditions in~\eqref{exist-cond} boil down to $\Delta_x>0$ and $\Delta_y<0$.
Having restricted to $x,y>0$, we can analyse the loci of the roots and the positivity of $\Delta_x$ and $\Delta_y$ also studying the polynomials $P_x \equiv x^{2(D-5)}\Delta_x$ and $P_y \equiv y^{2(D-5)}\Delta_y$. The positivity of $H$ can be examined considering the function $P_H \equiv (x^2-y^2) H$ because $x>y$.

For the rest of this subsection, we shall assume the background is supersymmetric, \ie\ $ags^2=1$. This relation implies that $a>0$, being $g$ positive.
By means of the decompositions~\eqref{Delta_deco}, we can write
\begin{equation}
	P_x = P_x^+(x) \, P_x^-(x) \,,  \qquad  P_y = P_y^+(y) \,P_y^-(y) \,,
\end{equation}
where we defined the polynomials
\begin{equation} \label{P+-}
	P_z^\pm(z) = \left\{
	\begin{aligned}
		& g z^3 \pm(1 + ag) z^2 + a z - \frac{2N_z}{a}    \qquad  & (D&=6) \,, \\
		& g z^4 \pm(1 + ag) z^3 + a z^2 - \frac{2N_z}{a}  \qquad  & (D&=7) \,,
	\end{aligned}
	\right.
\end{equation}
in which $z$ can be either $x$ or $y$ and $N_z$ is $\Nx$ or $\Ny$ accordingly.
Notice that, \eg, $P_x^\pm(x) = x^{D-5} \Delta_x^\pm$.

As it can be shown, $P_x>0$ in $x=0$ and $P_x\to +\infty$ when $x$ goes to infinity. Since we need $P_x$ to be positive in a closed interval in the positive $x$-axis, this polynomial must have at least four positive roots. By Descartes' rule of signs, a necessary condition for this to happen is that $\Nx>0$.
In general, when $N_z>0$, $P_z^+(z)$ has one positive root, while $P_z^-(z)$ can have one or three of them. In the desired configuration $P_x^-(x)$ must have three positive roots, so to bring the total amount to four.

Applying Descartes' rule, we see that, in $D=6$, $P_x^-(x)$ has no negative root, hence all the existing ones are positive. In six dimensions  $P_x^-(x)$ is a cubic polynomial, which admits three distinct real roots if and only if its discriminant is positive.
Restricting to $\Nx>0$, this condition is met when
\begin{equation} \label{Nx6}
	0 < \Nx < a^3 \, \frac{-2q^3 + 3q^2 + 3q - 2 + 2(q^2 - q + 1)^{3/2}}{54q^2} \,,
\end{equation}
where we defined $q = ag$.

In seven dimensions $P^-_x(x)$ has one negative root, therefore the existence of four roots would ensure that three of them are positive. In this case $P^-_x(x)$ is a quartic polynomial and its discriminant is positive if and only if
\begin{equation} \label{Nx7}
	0 < \Nx < \frac{a^3}{g} \, \frac{-27q^4 + 36q^3 - 2q^2 +36q -27q + (1 + q) (9q^2 - 14q + 9q)^{3/2}}{1024q^2} \,,
\end{equation}
where, again, we restricted to $\Nx>0$ and introduced the constant $q = ag$.
In the case of a quartic polynomial this alone does not guarantee the existence of four distinct real roots. We need the additional conditions
\begin{equation}
	-3q^2 + 2q - 3q < 0 \,,  \qquad  -3q^4 + 4q^3 - 2(1 + 64\Nx) q^2 + 4q - 3 < 0 \,.
\end{equation}
The first one is always satisfied for every real $q$, while the second one gives
\begin{equation} \label{Nx7-add}
	\Nx > -\frac{(1 - q)^2 (3 + 2q + 3q^2)}{128q^2} \,.
\end{equation}
Since the right-hand side of the above inequality is never positive, the condition \eqref{Nx7-add} is automatically implied by the constraint $\Nx>0$.

Considering $a$ and $\Nx$ such that $P_x$ has four distinct real roots, the structure of $P_x^\pm(x)$ is depicted in figure~\ref{fig:P+-}. Specifically, $\XA=x_2$, the smallest root of $P_x^-(x)$, and $\XB=x_3$, the middle root of $P_x^-(x)$, since in this interval $P_x^\pm(x)$ are both positive and so is their product $P_x$. The location of the two functions follows from the fact that $P_z^+(z) > P_z^-(z)$ where $z>0$ and that they are equal in $z=0$.
Because $\XBA$ are both roots of $P_x^-(x)$, we have $\tau^{(\XBA)}=-1$.
\begin{figure}[ht]
	\centering

	\begin{tikzpicture}
		\draw [-stealth] (-1,0) -- (7,0) node[below] {$x$};
		\draw [-stealth] (0,-2) -- (0,3);

		\draw[xscale=1, yscale=0.5, domain=-0.4:6.8, smooth, variable=\x] plot ({\x}, {0.1*\x*\x*\x - 1.1*\x*\x + 3.4*\x - 2.4});
		\draw[xscale=1, yscale=0.5, domain=-0.5:1.2, smooth, variable=\x] plot ({\x}, {0.1*\x*\x*\x + 1.1*\x*\x + 3.4*\x - 2.4});

		\node [dot, label=135  : {$x_1$}] at (0.589,0) {};
		\node [dot, label=-45  : {$x_2$}] at (1,0) {};
		\node [dot, label=-135 : {$x_3$}] at (4,0) {};
		\node [dot, label=-45  : {$x_4$}] at (6,0) {};

		\node [dot, label=135 : {\scriptsize $-\dfrac{2\Nx}{a}$}] at (0,-1.2) {};

		\node [] at (1.6,2.2) {\footnotesize $P_x^+$};
		\node [] at (6.7,1.1) {\footnotesize $P_x^-$};
	\end{tikzpicture}

	\caption{Example of graphs of $P_x^\pm(x)$ in the configuration in which $P_x$ admits four positive roots.}
	\label{fig:P+-}
\end{figure}

We now move to the analysis of~$P_y$, which has to be negative in a closed positive interval $[\YA,\YB]$ such that $\YB<\XA$.
First of all, we notice from their definition in~\eqref{P+-} that $P_x^+(x)$ and $P_y^+(y)$ have the same shape, but are shifted vertically; similarly happens for $P_x^-(x)$ and $P_y^-(y)$.
The strategy is to start from the configuration $\Ny=\Nx$, in which $P_y^\pm(y)$ and $P_x^\pm(x)$ are coincident, and to deform it, studying how the system modifies. In the initial case, $P_y$ is negative in the intervals $[y_1,y_2]$ and $[y_3,y_4]$, with $y_i=x_i$. Because $[y_1,y_2]$ lies to the left of $[\XA,\XB]$, it will be our candidate.
As $\Ny$ increases, $P_y^-(y)$ moves below $P_x^-(x)$ and the root $y_2$ is shifted to the right of~$x_2$, yielding $\YB=y_2>\XA$. When, moreover, $\Ny$ exceeds a specific value, $y_2$ and $y_3$ disappear, giving the identification $\YB=y_4>\XB$. Both cases are not allowed.
On the contrary, when $\Ny$ decreases, both $P_y^\pm(y)$ move up and the roots $y_1$ and $y_2$ are shifted to the left of $x_1$ and $x_2$, respectively, resulting in the desired closed interval with $\YA=y_1$ and $\YB=y_2$.
However, when $\Ny$ becomes negative, the roots $y_1$ and $y_2$ disappear, while $y_3$ and $y_4$, if still existing, are brought closer  to one another. In this case, the only possible interval has $\YA=y_3>\XB$ and is, thus, not acceptable.
Ultimately, the roots of $P_y$ meet all the requirements for the metric to have a correct signature if and only if%
\footnote{In the special configuration $\Ny = \Nx$ we have $y_2 = x_2$, hence $\YB = \XA$.}
\begin{equation}
	0 < \Ny < \Nx \,,
\end{equation}
with $\Nx$ satisfying~\eqref{Nx6}, in $D=6$, or~\eqref{Nx7}, in $D=7$. In this case $\YA=y_1$, the unique root of $P_y^+(y)$, and $\YB=y_2$, the smallest root of $P_y^-(y)$. These results yield $\tau^{(\YA)}=+1$ and $\tau^{(\YB)}=-1$.

The last condition we examine is the positivity of~$H$ or, equivalently, of~$P_H$. It can be shown that, in the range of definition of the coordinates, $\partial_x P_H > 0$ and $\partial_y P_H < 0$, therefore the minimum value is reached in $x=\XA$ and $y=\YB$.
When supersymmetry is imposed, we can make use of equation~\eqref{root_H}, with $\kk=-1$ and $\tau^{(\XA)}=-1$, because $\XA$ is a root of~$P_x^-(x)$, hence of $\Delta_x^-$. Since $x>y$, it follows that $P_H(\XA,\YB)>0$, hence $P_H$ is positive in the region of interest.

\section{Global structure of the solutions}
\label{sec:global_structure}

We will now analyse global aspects of the solutions, uncovering the geometry of the underlying
four-dimensional toric orbifolds $\Morb_4$. This will be done combining the information obtained from the Killing vectors of the metric, the compatible complex structure, as well as the flux quantization conditions, ensuring that the gauge fields are well-defined
connections on orbifold line bundles.
Although this is a standard practice, we will find that the analysis for our new solutions is unusually sophisticated, due to the lack of a fibre bundle structure (present in the solutions of~\cite{Cheung:2022ilc,Couzens:2022lvg,Faedo:2022rqx}).~Below we will concentrate on the global regularity of the solutions in $D=6,7$ dimensions. Although in principle one should perform this analysis in ten or eleven dimensions, the regularity of the metrics in string/M-theory will follow from the regularity of the lower-dimensional metrics together with the regularity of the scalars and the correct quantization conditions for the gauge fields. The explicit uplifted solutions are discussed in appendix~\ref{app:uplifts}, where we also address the quantization conditions of all the fluxes and compute the relevant gravitational observables.

\subsection{Warm-up: revisiting the solutions of \cite{Faedo:2022rqx}}

In order to illustrate our strategy for analysing the solutions presented here, we will first apply the same logic to the solutions in~\cite{Faedo:2022rqx}.
The geometry of the toric orbifolds associated with those solutions has some additional structure that makes the analysis considerably simpler.
In order to be consistent with the conventions of~\cite{Faedo:2022rqx}, all the vectors will be outward-pointing in this section.

The four-dimensional orbifold $\Morb_4$ of~\cite{Faedo:2022rqx}, which describes the fibration of a spindle over another spindle  $\spindle_1 (x,\psi)\ltimes \spindle_2(y,z)$, is locally described by the metric
\begin{equation}\label{2spin_metric}
	\begin{split}
		\dd s_{\Morb_4}^2 = \frac{x^2}{q} \, \dd x^2 + \frac{q}{4x^2} \, \dd\psi^2 + \frac{y^2}{F} \, \dd y^2 + \frac{F}{h_1 h_2} \Bigl( \dd z - \frac{1}{2m} \Bigl(1 - \frac{\mathtt{a}}{x}\Bigr) \, \dd\psi \Bigr)^2 \,.
	\end{split}
\end{equation}
The coordinates $x$ and $y$ range in $[\XA,\XB]$ and $[\YA,\YB]$, respectively, where $q(\XBA)=0$ and $F(\YBA)=0$. The singularities of these spindles are encoded by the following quantization conditions
\begin{equation}
	\begin{aligned}
		\frac{q'(\XA)}{4\XA^2} \, \Delta\psi &= \frac{2\pi}{m_-} \,,  \qquad  & -\frac{q'(\XB)}{4\XB^2} \, \Delta\psi &= \frac{2\pi}{m_+} \,, \\
		\frac{g F'(\YA)}{3\YA^3} \, \Delta z &= \frac{2\pi}{n_-} \,,  \qquad  & -\frac{g F'(\YB)}{3\YB^3} \, \Delta z &= \frac{2\pi}{n_+} \,,
	\end{aligned}
\end{equation}
where the minus signs are due to $\sign(q'(\XB))=\sign(F'(\YB))=-1$. The fibration is required to be well-defined in the orbifold sense, that is
\begin{equation}
	\frac{1}{2\pi} \int_{\spindle_1} \dd\eta = \frac{t}{m_+ m_-} \,, \qquad  \eta = \frac{2\pi}{\Delta z} \Bigl( \dd z - \frac{1}{2m} \Bigl(1 - \frac{\mathtt{a}}{x}\Bigr) \, \dd\psi \Bigr) \,,  \qquad  t \in \ZZ \,,
\end{equation}
and we assume, for the moment, that $t$ is coprime to $m_\pm$. In this way $\eta$ is a connection on $\mathcal{O}(t)$, and at a fixed value of $y$ the metric~\eqref{2spin_metric} describes a lens space\footnote{Notice that here $t$ is negative.} $L(-t,1)=S^3/\ZZ_{-t}$ with base $\spindle_1$. This metric possesses four Killing vectors $\xi_a$ which degenerate on the loci
\begin{equation} \label{2spin_loci}
	\loci_1 = \{y = \YA\} \,,  \qquad  \loci_2 = \{x = \XA\} \,,  \qquad  \loci_3 = \{y = \YB\} \,,  \qquad  \loci_4 = \{x = \XB\} \,.
\end{equation}
To focus on these $\loci_a$, we define adapted coordinates
\begin{equation} \label{2spin_adapted-coords}
	\tilde{\psi} = m_+ \phi_+ - m_- \phi_- \,,  \quad
	\tilde{z} = -m_+ \frac{\mathtt{a} - \XB}{2m \XB} \frac{\Delta\psi}{\Delta z} \, \phi_+ + m_- \frac{\mathtt{a} - \XA}{2m \XA} \frac{\Delta\psi}{\Delta z} \, \phi_- \,,\quad \frac{(2\pi)^2}{-t}= \Delta \phi_+ \Delta\phi_-\,,
\end{equation}
where $\tilde{\psi}=\frac{2\pi}{\Delta \psi}\psi$  and $\tilde{z}=\frac{2\pi}{\Delta z}z$ are $2\pi$-periodic and the total metric, restricted to $\loci_a$, reads
\begin{equation} \label{2spin_hat-met}
	\begin{split}
		\dd s_{(1)}^2 &= \dd s_{(3)}^2 = \frac{x^2}{q} \, \dd x^2 + \frac{q}{4x^2} \, \dd\psi^2 \,, \\
		\dd s_{(2)}^2 &= \frac{y^2}{F} \, \dd y^2 + \frac{t^2}{m_-^2} \frac{F}{h_1 h_2} \biggl(\frac{\Delta z}{2\pi}\biggr)^2 \dd\phi_+^2 \,, \\
		\dd s_{(4)}^2 &= \frac{y^2}{F} \, \dd y^2 + \frac{t^2}{m_+^2} \frac{F}{h_1 h_2} \biggl(\frac{\Delta z}{2\pi}\biggr)^2 \dd\phi_-^2 \,.
	\end{split}
\end{equation}
Looking, \eg, at  $\dd s_{(2)}^2$, we see that it describes $\spindle_2/\ZZ_{m_-}$ if we require $(\Delta\phi_+)^{(2)}=-2\pi/t$, where $(\Delta\phi_+)^{(2)}$ is the periodicity of $\phi_+$ when referred to $\loci_2$. From~\eqref{2spin_adapted-coords}, we have then $(\Delta\phi_-)^{(2)}=2\pi$.
Similarly, the line element $\dd s_{(4)}^2$ describes $\spindle_2/\ZZ_{m_+}$ when $(\Delta\phi_-)^{(4)}=-2\pi/t$ and $(\Delta\phi_+)^{(4)}=2\pi$.~The reason for which the periodicities seem not to be globally defined and ``jump'' to different values on the loci $\loci_2$ and $\loci_4$ is that the quotients must be done locally in patches, and then suitably glued. Thus, we should have defined local coordinates $\phi_{\pm}^{(2)}$ and $\phi_{\pm}^{(4)}$ on a patch that contains $\loci_{2}$ or $\loci_{4}$, for which $(\Delta\phi_+)^{(2)}$ is more precisely $\Delta(\phi_+^{(2)})$. Eventually, we will write $\Delta\phi_{+}^{(2)}$ when it is clear what we mean. This will happen again in the following, when we will use our procedure to extract the labels associated with the new solutions.

From the degenerating Killing vectors on $\loci_a$ we extract  the $\ZZ^2$-valued outward-pointing vectors  \cite{Faedo:2022rqx}
\begin{equation} \label{2spin_vec-n-stack}
	\vec{w}_1 = (n_-, 0) \,,  \qquad  \vec{w}_2 = (t \,a_+, m_-) \,,  \qquad  \vec{w}_3 = (-n_+, 0) \,,  \qquad  \vec{w}_4 = (t \,a_-, -m_+) \,,
\end{equation}
and an associated labelled polytope (or labelled fan), depicted in figure~\ref{fig:2spin_polytope_coprime}.
\begin{figure}
	\centering
	\begin{subfigure}[b]{0.48\textwidth} \centering

		\begin{tikzpicture}[scale=1.75]
			\node [label=below : {
				$p_3$
			}] (x3) at (3.5,7/6)   {};
			\node [label=below : {
				$p_4$
			}] (x4) at (6,6/3)     {};
			\node [label=above : {
				$p_1$
			}] (x1) at (6,6/2)     {};
			\node [label=above : {
				$p_2$
			}] (x2) at (3.5,21/12) {};

			\draw (x2.center) -- (x3.center) node[midway,anchor=center] (d3) {};
			\draw (x3.center) -- (x4.center) node[midway,anchor=center] (d4) {};
			\draw (x4.center) -- (x1.center) node[midway,anchor=center] (d1) {};
			\draw (x1.center) -- (x2.center) node[midway,anchor=center] (d2) {};

			\node (c1) at (3.7,1.49)   {$n_+$};
			\node (c2) at (5.8,2.45)   {$n_-$};

			\draw [-latex] (d3.center) -- node[above left]{$\vec{\hat{w}}_3$}  ($(d3)+(-1/2,0)$);
			\draw [-latex] (d4.center) -- node[below left]{$\vec{\hat{w}}_4$}  ($(d4)+(1/6,-1/2)$);
			\draw [-latex] (d1.center) -- node[below right]{$\vec{\hat{w}}_1$} ($(d1)+(1/2,0)$);
			\draw [-latex] (d2.center) -- node[above right]{$\vec{\hat{w}}_2$} ($(d2)+(-1/4,1/2)$);

		\end{tikzpicture}

		\subcaption{$\gcd(t,m_\pm)=1$.}
		\label{fig:2spin_polytope_coprime}
	\end{subfigure}
	\begin{subfigure}[b]{0.48\textwidth} \centering

		\begin{tikzpicture}[scale=0.59pt]
			\node [label=right : {$p_1$}] (x1) at (4,8)   {};
			\node [label=left  : {$p_2$}] (x2) at (1.5,3) {};
			\node [label=left  : {$p_3$}] (x3) at (1.5,0) {};
			\node [label=right : {$p_4$}] (x4) at (4,0)   {};

			\draw (x4.center) -- node[left]{$n_-$}  (x1.center) node[midway,anchor=center] (d1) {};
			\draw (x1.center) --    node[right]{$m_-$}                 (x2.center) node[midway,anchor=center] (d2) {};
			\draw (x2.center) -- node[right]{$n_+$} (x3.center) node[midway,anchor=center] (d3) {};
			\draw (x3.center) --    node[above]{$m_+$}                 (x4.center) node[midway,anchor=center] (d4) {};

			\draw [-latex] (d1.center) -- node[above] {$\vec{\hat{n}}_1$} ($(d1)+(1,0)$);
			\draw [-latex] (d2.center) -- node[above] {$\vec{\hat{n}}_2$} ($(d2)+(-2,1)$);
			\draw [-latex] (d3.center) -- node[above] {$\vec{\hat{n}}_3$} ($(d3)+(-1,0)$);
			\draw [-latex] (d4.center) -- node[right] {$\vec{\hat{n}}_4$} ($(d4)+(0,-1)$);
		\end{tikzpicture}

		\subcaption{$t = m_+m_-\overline{t}$.}
		\label{fig:2spin_polytope_non_coprime}
	\end{subfigure}

	\caption{Polytope of the  $\spindle_1\ltimes\spindle_2$ orbifold corresponding to the vectors~\eqref{2spin_vec-n-stack} and~\eqref{2spin_vec-n-stack-new}, respectively.}
	\label{fig:stack_geometry}
\end{figure}
Here $a_\pm\in\ZZ$ are chosen such that $ a_+\,m_+ + a_-\,m_-=1$, which always exist by Bézout's lemma for coprime $m_{\pm}$.  The set of labels derived by~\eqref{2spin_vec-n-stack} is $m_a=(n_-,1,n_+,1)$. To understand  better why this is the case and the reason for which $m_{\pm}$ are not labels, we look directly at the metric~\eqref{2spin_metric}. The four loci~\eqref{2spin_loci}, with metrics~\eqref{2spin_hat-met}, correspond to the divisors $\hat{D}_a$ and topologically are
\begin{equation}
	\hat{D}_1 = \spindle_1\,, \qquad \hat{D}_2 = \frac{\spindle_2}{\ZZ_{m_-}}\,, \qquad \hat{D}_3 = \spindle_1\,, \qquad \hat{D}_4 = \frac{\spindle_2}{\ZZ_{m_+}}\,,
\end{equation}
where $\hat{D}_{1,3}$ are copies of the base $\spindle_1$ at the north and south poles of the fibre $\spindle_2$, respectively. On the other hand, at the poles of $\spindle_1$, the four-dimensional orbifold is locally modelled by $(\CC\times\spindle_2)/\ZZ_{m_\mp}$, and as a consequence $\hat{D}_{2,4}$ are  \textit{global} quotients of the fibre~$\spindle_2$. Notice that, even if $\hat{D}_{2,4}=\spindle_2/\ZZ_{m_\mp}$, there are no orbifold singularities at generic points (\ie\ different from the north and south poles) on $\hat{D}_{2,4}$ and thus the labels are $m_{2,4}=1$.
These are then divisors associated with an ordinary fan, with primitive vectors $\vec{\hat{w}}_{2,4} = \vec{w}_{2,4}$.
The situation is different on $\loci_{1}$ and $\loci_{3}$. Zooming in for example near $y=\YA$, the metric at leading order in $R^2=\frac{4\YA^2}{F'(\YA)}(y-\YA)$ reads\footnote{See section 4.4 of \cite{Abreu:2001to} for an analogous computation.}
\begin{equation} \label{2spin_stack-metric}
	\dd s_{\Morb_4}^2 \underset{\YA}{\simeq} \frac{x^2}{q} \, \dd x^2 + \frac{q}{4x^2} \, \dd\psi^2 +\, \dd R^2 + \frac{R^2}{n_- ^2} \Bigl( \dd \tilde{z} - \frac{1}{2m} \Bigl(1 - \frac{\mathtt{a}}{x}\Bigr)\frac{1}{\Delta z} \, \dd\psi \Bigr)^2 \,.
\end{equation}
When we set $y=\YA$, the metric reduces to $\dd s^2_{(1)}$ in~\eqref{2spin_hat-met}, which is $\hat{D}_1=\spindle_1$. However, at each point  $\XA \leq x \leq \XB$, from
\eqref{2spin_stack-metric} we see that there is a normal conical singularity giving, \textit{locally}, a copy of  $\spindle_1 \ltimes (\CC/\ZZ_{n_-})$.
As a consequence, $n_-$ is a label with associated divisor  $D_1 =\hat{D}_1 \times \mathrm{pt}/\ZZ_{n_-}$.~The metric~\eqref{2spin_stack-metric} can be compared with its counterpart when approaching the locus $\loci_2$. In this case, we zoom in near $x=\XA$ defining the new coordinate $R$ such that $x = \XA + \frac{q'(\XA)}{4\XA^2}R^2$. In the adapted coordinates~\eqref{2spin_adapted-coords}, the metric on~$\Morb_4$ at leading order in $R^2$ reads
\begin{equation}\label{2spin_facet2}
	\dd s_{\Morb_4}^2 \underset{\XA}{\simeq} \dd R^2 + R^2 \bigl(\dd\phi_-^{(2)} + c(y) \, \dd\phi_+^{(2)}\bigr)^2 + \frac{y^2}{F} \, \dd y^2 + \frac{t^2}{m_-^2} \frac{F}{h_1 h_2} \biggl(\frac{\Delta z}{2\pi}\biggr)^2 (\dd\phi_+^{(2)})^2 \,,
\end{equation}
with $c(y)$ a regular function on $[\YA,\YB]$ which at the endpoints of the interval takes the values $c(y_\pm)=-m_+/m_-$. Assuming $(\Delta\phi_-)^{(2)}=2\pi$ as before, the first contribution to the total metric is the smooth complex plane in polar coordinates, therefore no singularity along the divisor~$D_2$ is present.
Setting $R=0$, which accounts in moving exactly on the divisor, the line element reduces to the second line of~\eqref{2spin_hat-met}, which, as already discussed, corresponds to the metric on $\spindle_2/\ZZ_{m_-}$, once the correct periodicity for $\phi_+^{(2)}$ is considered. An analogous reasoning can be applied to $\loci_3$ and $\loci_4$, showing that $D_3=\hat{D}_3\times \text{pt}/\ZZ_{n_{+}}$, but no singularity along $D_4$.

We now consider  the case when $t$ and $m_\pm$ have common factors, plugging a small gap in the discussion  \cite{Faedo:2022rqx}.  On general grounds, it is known that in this case the fibration does not smooth the orbifold points on the base~\cite{Ferrero:2020twa}. Then, the four-dimensional orbifold will be characterized by four different labels with $m_2=\gcd(t,m_-)$ and $m_4=\gcd(t,m_+)$, as we shall understand in a moment. To see explicitly the difference with the previous case, we can zoom in near $y=\YBA$ in~\eqref{2spin_facet2}, obtaining
\begin{equation}
	\begin{aligned}
		\dd s_{\Morb_4}^2 &\underset{\XA,\YA}{\simeq} \dd R^2 + R^2 \Bigl(\dd\phi_-^{(2)} -\frac{m_+}{m_-} \ \dd\phi_+^{(2)}\Bigr)^2 + \dd P^2+\Bigl(\frac{t}{m_- n_-}\Bigr)^2P^2 (\dd\phi_+^{(2)})^2 \,,
		\\
		\dd s_{\Morb_4}^2 &\underset{\XA,\YB}{\simeq}  \dd R^2 + R^2 \Bigl(\dd\phi_-^{(2)} -\frac{m_+}{m_-}  \ \dd\phi_+^{(2)}\Bigr)^2 + \dd P^2+\Bigl(\frac{t}{m_- n_+}\Bigr)^2P^2 (\dd\phi_+^{(2)})^2 \,.
	\end{aligned}
\end{equation}
For coprime $m_\pm$, we can now take $t = k_+\overline{t}_+ = k_-\overline{t}_-$ and $m_\pm = k_\pm\overline{m}_\pm$ (with $\gcd(k_+,k_-)=1$), which implies $t = k_+k_-\overline{t}$, such that $\gcd(t,m_\pm)=k_\pm$. The requirement $\Delta{\phi}_{+}^{(2)}=2\pi/(-k_+\overline{t})$ translates into
\begin{align}
	\frac{2\pi}{-k_+\overline{t}}\,\Delta\phi_- ^{(2)}= \Delta{\phi}_{+}^{(2)}\Delta\phi_{-}^{(2)}=\frac{(2\pi)^2}{-t} \quad \implies \quad \Delta\phi_{-}^{(2)}=\frac{2\pi}{k_-}\, .
\end{align}
Then the space is $(\spindle_2/\ZZ_{\overline{m}_{-}})\ltimes(\CC/\ZZ_{k_-})$ and $k_-$ acts as a label on the transverse space to the divisor.~Subsequently, there exists an associated divisor $D_2$ with $\hat{D}_2=\spindle_2/\ZZ_{\overline{m}_{-}}$. This can also be seen from the vectors~\eqref{2spin_vec-n-stack}, which become
\begin{equation} \label{2spin_vec-n-stack_noncoprime}
	\vec{w}_1 = n_- (1, 0) \,,  \quad  \vec{w}_2 = k_- (a_+ k_+ \overline{t}, \overline{m}_-) \,,  \quad
	\vec{w}_3 = n_+ (-1, 0) \,,  \quad  \vec{w}_4 = k_+ (a_- k_- \overline{t}, -\overline{m}_+) \,,
\end{equation}
and the labels are $m_a=(n_-,k_-,n_+,k_+)$, as stated before. The case $\overline{m}_\pm=1$ (that is $t=m_+m_-\overline{t}$) is particularly familiar, in that the connection is on $\mathcal{O}(m_+ m_-\overline{t})$ and at a fixed value of $y$ the metric~\eqref{2spin_metric} describes a branched lens space $L_{m_{\pm}}(-\overline{t},1)$~\cite{Inglese:2023tyc}. Using an $SL(2,\ZZ)$ transformation, the vectors~\eqref{2spin_vec-n-stack_noncoprime} can then be rotated to
\begin{equation} \label{2spin_vec-n-stack-new}
	\vec{n}_1 = n_-(1, 0) \,,  \qquad  \vec{n}_2 = m_-(\overline{t}, 1) \,,  \qquad  \vec{n}_3 = n_+(-1, 0) \,,  \qquad  \vec{n}_4 =m_+ (0, -1) \,,
\end{equation}
which are in $\ZZ^2$, and the labels are simply $m_a=(n_-,m_-,n_+,m_+)$. The associated labelled polytope is sketched in figure~\ref{fig:2spin_polytope_non_coprime}, and is clearly
a labelled (or, stacky) version of the polytope associated with the Hirzebruch surfaces $\mathbb{F}_{-\overline{t}}$.

To summarize, in these simple examples we can see that the presence of a fibration can result in different types of polytopes, associated with different $\Morb_4$, according to the values of $\gcd(t,m_\pm)$. When $t$ is relatively coprime with $m_\pm$, the integers $m_{\pm}$ characterizing the base spindle are not labels, but result in $\mathbb{Z}_ {m_\pm}$ quotients on the fixed points; \textit{vice versa}, if $t$ has common factors with one or both $m_\pm$, orbifold points are present also on the base, with more than two labels associated with it. In our more general solutions~\eqref{equal_metric-M4}, all the hatted divisors will be global quotients of some spindle, but also with  labels attached.

\subsection{Degenerating Killing vectors}

The four-dimensional orbifold~$\Morb_4$ possesses a $U(1)^2$ isometry associated with the two Killing vectors~$\partial_\psi$ and~$\partial_\phi$ of the metric~\eqref{equal_metric-M4}, therefore it is natural to study this system within the framework of toric geometry. For the solutions in~\cite{Faedo:2022rqx} we showed  that the toric data  characterizing a given orbifold, as well as the fluxes, could be derived from the analysis of degenerate Killing vectors of the metric. For our new solutions we will need to complement the study of the degenerating Killing vectors with some additional information, in order to extract all the data required to implement the extremization procedure.

The metric~\eqref{equal_metric-M4} has in total four degenerating Killing vectors
\begin{align}
	\label{killing-j}
	\killJ_\pm &= \frac{2a^2}{\Delta_x'(\XBA)} \bigl[ \widetilde{V}_x(\XBA) \, \partial_\psi + V_x(\XBA) \, \partial_\phi \bigr] \equiv J^{(\psi)}_\pm \partial_\psi + J^{(\phi)}_\pm \partial_\phi \,, \\
	\label{killing-k}
	\killK_\pm &= \frac{2a^2}{\Delta_y'(\YBA)} \bigl[ \widetilde{V}_y(\YBA) \, \partial_\psi + V_y(\YBA) \, \partial_\phi \bigr] \equiv K^{(\psi)}_\pm \partial_\psi + K^{(\phi)}_\pm \partial_\phi \,,
\end{align}
whose norm vanishes at $x=\XBA$ and $y=\YBA$, respectively. All of them are normalized so to have unitary surface gravity.
In order to present the results in a uniform way  we introduce the four loci
\begin{equation} \label{divisors}
	\hat{D}_1 = \{x = \XA\} \,,  \qquad  \hat{D}_2 = \{y = \YA\} \,,  \qquad
	\hat{D}_3 = \{x = \XB\} \,,  \qquad  \hat{D}_4 = \{y = \YB\} \,,
\end{equation}
and define the four vectors
\begin{equation}
	\xi_{1} = \killJ_- \,,  \qquad  \xi_{2} = \killK_- \,,  \qquad  \xi_{3} = \killJ_+ \,,  \qquad  \xi_{4} = \killK_+ \,,
\end{equation}
so that $\xi_{a}$ is the Killing vector that degenerates at $\hat{D}_a$.

For later convenience, we define the two alternative sets of coordinates $\psi_\pm$ and $\chi_\pm$ such that $\killJ_\pm = \sij\,\partial_{\psi_\pm}$ or $\killK_\pm = \sik\,\partial_{\chi_\pm}$, with $\sij=\pm$ and $\sik=\pm$. These coordinates are adapted to the direction generated by the corresponding Killing vector $\xi_a$ and, for this reason, they will play a fundamental role when zooming in on the different loci~$\hat{D}_a$.
They are  defined  by
\begin{alignat}{2} \label{psi-chi1}
	\psi &= \sij \sum_{\sigma=\pm} J_\sigma^{(\psi)} \psi_\sigma \,,  \qquad  & \phi &= \sij \sum_{\sigma=\pm} J_\sigma^{(\phi)} \psi_\sigma\,,
\end{alignat}
or, alternatively, by
\begin{alignat}{2} \label{psi-chi2}
	\psi &= \sik \sum_{\sigma=\pm} K_\sigma^{(\psi)} \chi_\sigma \,,  \qquad  & \phi &= \sik \sum_{\sigma=\pm} K_\sigma^{(\phi)} \chi_\sigma \,.
\end{alignat}
Here $\psi_\sigma$ should be intended as well-defined coordinates on a patch which contains $\hat{D}_1$ or~$\hat{D}_3$, and subsequently we will denote them as $\psi_\sigma^{(1)}$ or $\psi_{\sigma}^{(3)}$, with periodicities $\Delta\psi_\sigma^{(1)}$ and $\Delta\psi_\sigma^{(3)}$, as explained in the previous section. The same applies also to $\chi_\sigma$, which will be written as $\chi_\sigma^{(2)}$ and $\chi_\sigma^{(4)}$ on $\hat{D}_{2}$ and $\hat{D}_{4}$.
The reason for the introduction of $\sij$ and $\sik$ and the prescription for their choice will be explained shortly.
The Jacobian matrices of these two transformations have determinant
\begin{equation}
	\det(\mathbf{J}) = J_+^{(\psi)} J_-^{(\phi)} - J_-^{(\psi)} J_+^{(\phi)} ,  \quad\;\;\;
	\det(\mathbf{K}) = K_+^{(\psi)} K_-^{(\phi)} - K_-^{(\psi)} K_+^{(\phi)} ,
\end{equation}
and the periodicities of the old and new coordinates are connected by
\begin{equation} \label{periods}
	\begin{aligned}
	\Delta\psi \, \Delta\phi &= \bigl|\det(\mathbf{J})\bigr| \, \Delta\psi_+^{(1)} \, \Delta\psi_-^{(1)}= \bigl|\det(\mathbf{J})\bigr| \, \Delta\psi_+^{(3)} \, \Delta\psi_-^{(3)} \,,
	\\
	\Delta\psi \, \Delta\phi &= \bigl|\det(\mathbf{K})\bigr| \, \Delta\chi_+^{(2)} \, \Delta\chi_-^{(2)}= \bigl|\det(\mathbf{K})\bigr| \, \Delta\chi_+^{(4)} \, \Delta\chi_-^{(4)} \,.
	\end{aligned}
\end{equation}
The signs~$\sij$ and~$\sik$ are given by
\begin{equation} \label{cond_signs}
	\begin{split}
		\sij &= \sign\bigl[ \Xi \, (x^2 - y^2) \det(\mathbf{J}) \bigr] = \sign(x^2 - y^2) \,, \\
		\sik &= -\sign\bigl[ \Xi \, (x^2 - y^2) \det(\mathbf{K}) \bigr] = -\sign(x^2 - y^2) \,,
	\end{split}
\end{equation}
where, in the last steps, we restricted to $x,y>0$ and imposed the signature conditions $\Nx,\Ny>0$; in this case, both $\det(\mathbf{J})$ and $\det(\mathbf{K})$ have the same sign of~$\Xi$.
By means of~\eqref{cond_signs}, all the coefficients of the orthonormal frame~\eqref{M4-frame} are positive when restricted to $\hat{D}_1$ and $\hat{D}_3$ or to $\hat{D}_2$ and $\hat{D}_4$. Indeed, on
these loci we have
\begin{equation}
	\begin{split}
		\Xi^{-1} \bigl( V_x \, \dd\psi - \widetilde{V}_x \, \dd\phi \bigr) \Bigr|_{x=\XAB} &= \pm \sij \, \frac{\Delta_x'(\XAB) \det(\mathbf{J})}{2\Xi a^2} \, \dd\psi_\pm = \frac{|\Delta_x'(\XAB) \det(\mathbf{J})|}{2|\Xi| a^2} \, \dd\psi_\pm \,, \\
		\Xi^{-1} \bigl( V_y \, \dd\psi - \widetilde{V}_y \, \dd\phi \bigr) \Bigr|_{y=\YAB} &= \pm \sik \, \frac{\Delta_y'(\YAB) \det(\mathbf{K})}{2\Xi a^2} \, \dd\chi_\pm = \frac{|\Delta_y'(\YAB) \det(\mathbf{K})|}{2|\Xi| a^2} \, \dd\chi_\pm \,,
	\end{split}
\end{equation}
where we placed the absolute values because the conditions for a well-defined metric imply $\sign(x^2-y^2) = \pm\sign[\Delta_x'(\XAB)] = \mp\sign[\Delta_y'(\YAB)]$.

\subsection{Complex structure and divisors}
\label{subsec:divisors}

A valuable insight into the orbifold structure of~$\Morb_4$ can be gained studying the four loci~$\hat{D}_a$ in~\eqref{divisors}, viewing them as \emph{divisors} from the point of view of the complex geometry associated with~$\Morb_4$.
Let us first show that indeed there exists an integrable complex structure compatible with our metrics~\eqref{equal_metric-M4}. This can be done
defining the holomorphic $(2,0)$-form
\begin{equation}
	\Omega = (\hat{e}^1 + \ii\,\hat{e}^2) \wedge (\hat{e}^3 + \ii\,\hat{e}^4) \equiv  \Omega^{(12)} \wedge \Omega^{(34)} \,,
\end{equation}
which, by construction, is compatible with the metrics~\eqref{equal_metric-M4}. The fact that the associated complex structure is integrable follows from the relation
\begin{equation}
	\dd\Omega = \ii P_\rho \wedge \Omega \,,
\end{equation}
where $P_\rho$  is the real Ricci form potential, which reads
\begin{equation}
	P_\rho = \frac{H \, \partial_y \bigl( H^{-2} \Delta_y \bigr)}{2\Xi (x^2 - y^2)} \bigl( V_x \, \dd\psi - \widetilde{V}_x \, \dd\phi \bigr)
	- \frac{H \, \partial_x \bigl( H^{-2} \Delta_x \bigr)}{2\Xi (x^2 - y^2)} \bigl( V_y \, \dd\psi - \widetilde{V}_y \, \dd\phi \bigr) \,,
	\label{pirro}
\end{equation}
thus establishing that $\Morb_4$ are \emph{complex orbifolds}. Although in principle we could introduce complex coordinates, this is technically involved, and ultimately not necessary. We will then continue with the natural coordinates in which the metric~\eqref{equal_metric-M4} was originally presented, where the $(x,y)$ coordinates play the role of ``moment maps'', despite the lack of a compatible symplectic structure.
The Ricci form $\rho = \dd P_\rho$ defines the first Chern class of the tangent bundle of~$\Morb_4$ as $\rho = 2\pi c_1(T\Morb_4)$, and will be used in the analysis later.

It is then clear that the four (real) codimension two loci $\hat{D}_a$ are  divisors in $\Morb_4$ and we will now describe these in more detail. We first zoom in on~$\hat{D}_1$, determined by the condition $x=\XA$. To this end, we define the coordinate $\zeta$ such that $x = \XA + \lambda \zeta$ and take the $\lambda\to0$ limit. Reaching the point $x=\XA$ the whole space~$\Morb_4$ degenerates to a two-dimensional orbifold, whose metric, expressed in terms of the coordinates $\psi_\pm^{(1)}$, is
\begin{equation} \label{div_metric1}
	\dd s_{(1)}^2 = \frac{y^2 - \XA^2}{\Delta_y} \, \dd y^2 + \frac{\Delta_x'(\XA)^2 \det(\mathbf{J})^2}{4\Xi^2 a^4 H(\XA,y)^2} \frac{\Delta_y}{y^2 - \XA^2} \, (\dd\psi_+^{(1)})^2 \,.
\end{equation}
$\hat{D}_1$ is a complex orbifold of (complex) dimension one, therefore it is natural to define the holomorphic $(1,0)$-form
\begin{equation}
	\Omega_{(1)} = \sqrt{\frac{y^2 - \XA^2}{\Delta_y}} \, \dd y + \ii \, \frac{|\Delta_x'(\XA) \det(\mathbf{J})|}{2|\Xi| a^2 H(\XA,y)} \sqrt{\frac{\Delta_y}{y^2 - \XA^2}} \, \dd\psi_+^{(1)} \,.
\end{equation}
Notice that, defined $\Omega^{(12)} \equiv \hat{e}^1+\ii\,\hat{e}^2$, with $\hat{e}^1$ and $\hat{e}^2$ given in~\eqref{M4-frame}, $\Omega_{(1)}$ is simply $\Omega^{(12)}$ restricted to $\hat{D}_1$.
It can be checked that $\dd\Omega_{(1)} = \ii P_{(1)} \wedge \Omega_{(1)}$, where, by definition, $P_{(1)}$ is the Ricci form potential, whose expression is given by
\begin{equation}
	P_{(1)} = \frac{|\Delta_x'(\XA) \det(\mathbf{J})|}{4|\Xi| a^2} \biggl[ \frac{\Delta_y'}{(\XA^2 - y^2) H(\XA,y)} - \frac{\Delta_y \, \partial_y \bigl((\XA^2 - y^2) H(\XA,y)^2 \bigr)}{(\XA^2 - y^2)^2 H(\XA,y)^3} \biggr] \dd\psi_+^{(1)} \,.
\end{equation}
This potential allows us to determine the Ricci form of~$\hat{D}_1$, namely $\rho_{(1)} = \dd P_{(1)}$, and, in turn, the first Chern class of its tangent bundle $c_1(T\hat{D}_1) = \rho_{(1)}/2\pi$.
Denoting, in general, with $\hat{\chi}_{(a)}$ the orbifold Euler characteristic of~$\hat{D}_a$, this can be determined by the integral
\begin{equation}
	\hat{\chi}_{(a)} = \frac{1}{2\pi} \int_{\hat{D}_a} \rho_{(a)} = \int_{\hat{D}_a} c_1(T\hat{D}_a) \,.
\end{equation}
In the case at hand we have
\begin{equation} \label{div_euler1}
	\hat{\chi}_{(1)} = \frac{\Delta\psi_+^{(1)}}{2\pi} \frac{|\Delta_x'(\XA) \det(\mathbf{J})|}{4 |\Xi| a^2}
	\biggl[ \frac{\Delta_y'(\YB)}{(\XA^2 - \YB^2) H(\XA,\YB)} + \frac{-\Delta_y'(\YA)}{(\XA^2 - \YA^2) H(\XA,\YA)} \biggr] \,.
\end{equation}

Endowed with metric~\eqref{div_metric1}, $\hat{D}_1$ is a compact surface parameterized by the periodic azimuthal coordinate $\psi_+^{(1)}$ and the compact ``polar'' coordinate~$y$, with $y\in[\YA,\YB]$. As $y$ approaches one of the endpoints of this interval, the line element becomes
\begin{equation}
	\dd s_{(1)}^2 \underset{\YBA}{\simeq} \dd\varrho^2 + \varrho^2 \, \frac{\Delta_x'(\XA)^2 \Delta_y'(\YBA)^2 \det(\mathbf{J})^2}{16\Xi^2 a^4 (\XA^2 - \YBA^2)^2 H(\XA,\YBA)^2} \, (\dd\psi_+^{(1)})^2 \,,
\end{equation}
where we defined $\varrho^2 = |y-\YBA|$.
In order to have a smooth orbifold metric on~$\hat{D}_1$ we must impose the following conditions at the north ($\YA$) and south ($\YB$) poles, respectively,
\begin{equation} \label{quant1_poles}
	\begin{aligned}
		\frac{|\Delta_x'(\XA) \det(\mathbf{J})|}{4|\Xi| a^2} \frac{\Delta_y'(\YA)}{(\XA^2 - \YA^2) H(\XA,\YA)} \, \Delta\psi_+^{(1)} &= -\frac{2\pi}{\mathtt{m}_-^{(1)}} \,,  \qquad\quad  && \mathtt{m}_-^{(1)} \in \NN \,, \\
		\frac{|\Delta_x'(\XA) \det(\mathbf{J})|}{4|\Xi| a^2} \frac{\Delta_y'(\YB)}{(\XA^2 - \YB^2) H(\XA,\YB)} \, \Delta\psi_+^{(1)} &= \frac{2\pi}{\mathtt{m}_+^{(1)}} \,,  \qquad\quad  && \mathtt{m}_+^{(1)} \in \NN \,,
	\end{aligned}
\end{equation}
where the minus in the first relation is due to the fact that $\Delta_y'(\YA)/(\XA^2 - \YA^2) < 0$.
These constraints ensure that the divisor in question has the orbifold structure of a spindle, specifically $\hat{D}_1 = \mathbb{WCP}^1_{[\mathtt{m}_-^{(1)},\mathtt{m}_+^{(1)}]}$,%
\footnote{Here, $\mathtt{m}_+^{(1)}$ and $\mathtt{m}_-^{(1)}$ must be natural numbers, but they need not be coprime. In order to be more precise we should write $\hat{D}_1 = \mathbb{WCP}^1_{[\bar{\mathtt{m}}_-^{(1)},\bar{\mathtt{m}}_+^{(1)}]}/\ZZ_{\mathtt{m}_0^{(1)}}$, where $\mathtt{m}_0^{(1)} = \gcd(\mathtt{m}_-^{(1)},\mathtt{m}_+^{(1)})$ and $\bar{\mathtt{m}}_\pm^{(1)} = \mathtt{m}_\pm^{(1)}/\mathtt{m}_0^{(1)}$.}
and in terms of the~$\mathtt{m}_\pm^{(1)}$ its Euler characteristic~\eqref{div_euler1} can be written as
\begin{equation}
	\hat{\chi}_{(1)} = \frac{1}{\mathtt{m}_-^{(1)}} + \frac{1}{\mathtt{m}_+^{(1)}} \,.
\end{equation}

A similar  analysis can be carried out for the remaining divisors.
Introducing the one-form $\Omega^{(34)} \equiv \hat{e}^3+\ii\,\hat{e}^4$, we define the holomorphic $(1,0)$-forms $\Omega_{(2)}=\Omega^{(34)}|_{\hat{D}_2}$, $\Omega_{(3)}=\Omega^{(12)}|_{\hat{D}_3}$ and $\Omega_{(4)}=\Omega^{(34)}|_{\hat{D}_4}$. Then, we compute the corresponding Ricci forms $\rho_{(a)}$ and the Euler characteristic of each divisor
\begin{align}
	\hat{\chi}_{(2)} &= \frac{\Delta\chi_+^{(2)}}{2\pi} \frac{|\Delta_y'(\YA) \det(\mathbf{K})|}{4 |\Xi| a^2}
	\biggl[ \frac{-\Delta_x'(\XB)}{(\XB^2 - \YA^2) H(\XB,\YA)} + \frac{\Delta_x'(\XA)}{(\XA^2 - \YA^2) H(\XA,\YA)} \biggr] \,, \\
	\hat{\chi}_{(3)} &= \frac{\Delta\psi_-^{(3)}}{2\pi} \frac{|\Delta_x'(\XB) \det(\mathbf{J})|}{4 |\Xi| a^2}
	\biggl[ \frac{\Delta_y'(\YB)}{(\XB^2 - \YB^2) H(\XB,\YB)} + \frac{-\Delta_y'(\YA)}{(\XB^2 - \YA^2) H(\XB,\YA)} \biggr] \,, \\
	\hat{\chi}_{(4)} &= \frac{\Delta\chi_-^{(4)}}{2\pi} \frac{|\Delta_y'(\YB) \det(\mathbf{K})|}{4 |\Xi| a^2}
	\biggl[ \frac{-\Delta_x'(\XB)}{(\XB^2 - \YB^2) H(\XB,\YB)} + \frac{\Delta_x'(\XA)}{(\XA^2 - \YB^2) H(\XA,\YB)} \biggr] \,.
\end{align}
Restricting to each single divisor and zooming in on the associated ``poles'', it is possible to impose suitable quantization conditions in order to give  to the divisor the structure of a spindle. Specifically, we will have
\begin{equation} \label{div_euler1-int}
	\hat{\chi}_{(a)} = \frac{1}{\mathtt{m}^{(a)}_-} + \frac{1}{\mathtt{m}^{(a)}_+} \,,
\end{equation}
where $\mathtt{m}^{(a)}_\pm$ are eight integer parameters. We will see shortly that these parameters will be entirely determined in terms of the toric data.

\subsection{Toric data}
\label{subsect:toric-data}

As we discussed in section~\ref{sec:toric}, given a toric orbifold~$\Morb_4$, its  toric data are  encoded in a fan with labels, \ie\ a set of ordered non-primitive vectors~$\vec{v}_a$ with
integer-valued components. These data, however, are basis-dependent and therefore not unique. For quadrilaterals a set of ``gauge invariant'' toric data is given by the $\hat d_{a,b}$ and the labels $m_a$. The
 vectors ~$\vec{v}_a$ can be extracted from the degenerate Killing vectors~$\xi_a$ associated with the metric on~$\Morb_4$ once a basis of an effective two-torus action $\{E_1,E_2\}$ is fixed, namely
 \begin{equation}
	\xi_a = \vec{v}_a \cdot (E_1,E_2) \,.
\end{equation}
We will postpone determining such a basis $\{E_1,E_2\}$ and instead work in a ``reference'' basis $\{e_1,e_2\}$, in which we will be able to determine all the parameters of the solution in terms of
the gauge invariant toric data. Specifically, we introduce the $2\pi$-periodic coordinates $\nu_1 = \frac{2\pi}{\Delta\psi}\psi$, $\nu_2 = \frac{2\pi}{\Delta\phi}\phi$ and consider the basis $\{e_1,e_2\}=\{\partial_{\nu_1},\partial_{\nu_2}\}$. From the Killing vectors \eqref{killing-j} and~\eqref{killing-k} we can derive the vectors
\begin{equation} \label{vec-n}
	\begin{aligned}
		\vec{V}_1 &= \Bigl( \frac{2\pi}{\Delta\psi} \, J^{(\psi)}_-, \frac{2\pi}{\Delta \phi} \, J^{(\phi)}_- \Bigr) \,,  \qquad  &
		\vec{V}_2 &= \Bigl( \frac{2\pi}{\Delta\psi} \, K^{(\psi)}_-, \frac{2\pi}{\Delta \phi} \, K^{(\phi)}_- \Bigr) \,, \\
		\vec{V}_3 &= \Bigl( \frac{2\pi}{\Delta\psi} \, J^{(\psi)}_+, \frac{2\pi}{\Delta \phi} \, J^{(\phi)}_+ \Bigr) \,,  \qquad  &
		\vec{V}_4 &= \Bigl( \frac{2\pi}{\Delta\psi} \, K^{(\psi)}_+, \frac{2\pi}{\Delta \phi} \, K^{(\phi)}_+ \Bigr) \,.
	\end{aligned}
\end{equation}
Since we will not be able to determine the periodicities $\Delta\psi$ and  $\Delta\phi$ separately, but only in the combination $\Delta\psi\,\Delta\phi$, we actually  do not know the explicit form of the vectors in~\eqref{vec-n}, thus, in particular, we do not know whether they  belong to $\ZZ^2$. We earlier referred to these vectors as providing a ``fake'' fan.
Nevertheless, the proper basis will be related to $\{e_1,e_2\}$   through an $SL(2,\RR)$ rotation $S$, acting  on the vectors and the basis
as $v_{a} ^{I}=S_{IJ} V_{a}^{J}$ and $E_I = S_{JI}^{-1} e_J$, respectively. Since $\det(\vec{V}_a,\vec{V}_b)$ are  invariant under these transformations
 we can still use the vectors~\eqref{vec-n} to compute the matrix $D_{ab}$ and trust the results obtained.
There is only one \textit{caveat}: the vectors~\eqref{vec-n} must be dual to a convex polytope and, in our conventions, must be ordered counter-clockwise, which imply $\det(\vec{V}_a,\vec{V}_{a+1})>0$ for any~$a$. If this condition is not met, the vectors can be ordered in the correct way by means of a reflection about a line in the $\ZZ^2$ plane, which can be realized, \eg, swapping the two components of each vector.
This transformation accounts in exchanging $\partial_{\nu_1}$ and $\partial_{\nu_2}$ and the two bases $\{\partial_{\nu_2},\partial_{\nu_1}\}$ and $\{E_1,E_2\}$ will now be related through a matrix with determinant equal to~$-1$. As a consequence, the intersection matrices computed starting from the two sets of vectors extracted from the two aforementioned bases will have opposite sign.
In order to keep track of this fact we introduce the sign $\sid=\pm$, telling whether the vectors are ordered as in~\eqref{vec-n} ($+$) or with the components swapped ($-$). $\sid$ will multiply every determinant computed from the vectors~\eqref{vec-n} and its value will be fixed shortly.

The orbifold information of~$\Morb_4$ are encoded in the determinants of the vectors $\vec{V}_a$, defined before as $d_{a,b} = \det(\vec{V}_a, \vec{V}_b)$. Indeed, $d_{a,a+1}$ gives the order of the quotient singularity at the intersection of the adjacent divisors~$D_a$ and~$D_{a+1}$, namely $\CC^2/\ZZ_{d_{a,a+1}}$.
Using the set of vectors~\eqref{vec-n} we compute
\begin{align} \label{quant_div}
	\begin{split}
		d_{1,2} &= -\sid \frac{2\pi}{\Delta\psi} \frac{2\pi}{\Delta\phi} \frac{4\Xi a^2 (\XA^2 - \YA^2) H(\XA,\YA)}{\Delta_x'(\XA) \Delta_y'(\YA)} \,, \\
		d_{2,3} &= \sid \frac{2\pi}{\Delta\psi} \frac{2\pi}{\Delta\phi} \frac{4\Xi a^2 (\XB^2 - \YA^2) H(\XB,\YA)}{\Delta_x'(\XB) \Delta_y'(\YA)} \,, \\
		d_{3,4} &= -\sid \frac{2\pi}{\Delta\psi} \frac{2\pi}{\Delta\phi} \frac{4\Xi a^2 (\XB^2 - \YB^2) H(\XB,\YB)}{\Delta_x'(\XB) \Delta_y'(\YB)} \,, \\
		d_{4,1} &= \sid \frac{2\pi}{\Delta\psi} \frac{2\pi}{\Delta\phi} \frac{4\Xi a^2 (\XA^2 - \YB^2) H(\XA,\YB)}{\Delta_x'(\XA) \Delta_y'(\YB)} \,,
	\end{split}
\end{align}
where we made use of the identity
\begin{equation}
	J_{\sigma_1}^{(\psi)} K_{\sigma_2}^{(\phi)} - K_{\sigma_2}^{(\psi)} J_{\sigma_1}^{(\phi)} =
	-\frac{4\Xi a^2 (x_{\sigma_1}^2 - y_{\sigma_2}^2) H(x_{\sigma_1}, y_{\sigma_2})}{\Delta_x'(x_{\sigma_1}) \Delta_y'(y_{\sigma_2})} \,,  \qquad  \sigma_1, \sigma_2 = \pm \,.
\end{equation}
Imposing the condition $\det(\vec{\vv}_a,\vec{\vv}_{a+1})>0$ we can fix the value of~$\sid$, obtaining\footnote{In order to do so we must use the relations between the signs of $\Delta_x'(\XBA)$ and $\Delta_y'(\YBA)$.}
\begin{equation} \label{kD}
	\sid = \sign\bigl[ \Xi (x^2 - y^2) \bigr] \,.
\end{equation}
Recalling the definition of~$\sij$ and~$\sik$ in~\eqref{cond_signs}, we then have
\begin{equation} \label{kD-kJK}
	\sid \sij = \sign\bigl[ \det(\mathbf{J}) \bigr] \,,  \qquad
	\sid \sik = -\sign\bigl[ \det(\mathbf{K}) \bigr] \,.
\end{equation}
Following the notation of~\cite{Faedo:2022rqx}, we also define
\begin{equation} \label{def_t}
	t_J \equiv d_{1,3} = -\sid \, \frac{2\pi}{\Delta\psi} \frac{2\pi}{\Delta\phi} \det(\mathbf{J}) \,,  \qquad
	t_K \equiv d_{2,4} = -\sid \, \frac{2\pi}{\Delta\psi} \frac{2\pi}{\Delta\phi} \det(\mathbf{K}) \,.
\end{equation}
The four equations~\eqref{quant_div} together with the two equations~\eqref{def_t} provide the ``quantization conditions'' for the parameters of the solution, in terms of the gauge invariant toric data
and we postpone their analysis  to subsection~\ref{subsec:quantization}.
Note that the six $d_{a,b}$ defined above automatically satisfy the relation~\eqref{drelation}.

The intersection matrix describing a given set of toric divisors was defined in~\eqref{inter-num}. In our construction, with a bit of computation the diagonal terms can be cast in the form
\begin{align} \label{D11v}
	D_{11} &= -\sid \frac{\Delta\psi}{2\pi} \frac{\Delta\phi}{2\pi} \frac{\Delta_x'(\XA)^2}{4\Xi a^2} \biggl[ \frac{1}{(\XA^2 - \YB^2) H(\XA,\YB)} - \frac{1}{(\XA^2 - \YA^2) H(\XA,\YA)} \biggr] \,, \\
	D_{22} &= -\sid \frac{\Delta\psi}{2\pi} \frac{\Delta\phi}{2\pi} \frac{\Delta_y'(\YA)^2}{4\Xi a^2} \biggl[ \frac{1}{(\XB^2 - \YA^2) H(\XB,\YA)} - \frac{1}{(\XA^2 - \YA^2) H(\XA,\YA)} \biggr] \,, \\
	D_{33} &= \sid \frac{\Delta\psi}{2\pi} \frac{\Delta\phi}{2\pi} \frac{\Delta_x'(\XB)^2}{4\Xi a^2} \biggl[ \frac{1}{(\XB^2 - \YB^2) H(\XB,\YB)} - \frac{1}{(\XB^2 - \YA^2) H(\XB,\YA)} \biggr] \,, \\
	D_{44} &= \sid \frac{\Delta\psi}{2\pi} \frac{\Delta\phi}{2\pi} \frac{\Delta_y'(\YB)^2}{4\Xi a^2} \biggl[ \frac{1}{(\XB^2 - \YB^2) H(\XB,\YB)} - \frac{1}{(\XA^2 - \YB^2) H(\XA,\YB)} \biggr] \,,
\end{align}
whereas the off-diagonal terms are simply ($D_{ab}=D_{ba}$)
\begin{equation} \label{Dabv}
	D_{12} = \frac{1}{d_{1,2}} \,,  \qquad  D_{23} = \frac{1}{d_{2,3}} \,,  \qquad
	D_{34} = \frac{1}{d_{3,4}} \,,  \qquad  D_{41} = \frac{1}{d_{4,1}} \,,
\end{equation}
with $d_{a,a+1}$ given in~\eqref{quant_div}.

The toric orbifold~$\Morb_4$ is complex, therefore the intersection matrix can also be computed from its complex structure. In particular, the diagonal terms are given by
\begin{equation}
	D_{aa} = \frac{1}{m_a} \int_{\hat{D}_a} \bigl[ c_1(T\Morb_4) - c_1(T\hat{D}_a) \bigr] \,,
\end{equation}
where, following the original formula~\eqref{chern-D-1}, the integration should be performed over the ramification divisors, related to the branch divisors~\eqref{divisors} as $D_a = \hat{D}_a\times \mathrm{pt}/\ZZ_{m_a}$, from which it follows that $\int_{D_a}\!\Lambda = \frac{1}{m_a} \int_{\hat{D}_a}\!\Lambda$ for any two-form $\Lambda$. However, from the loci~\eqref{divisors} we are able to extract only quantities referred to the branch divisors and, as a consequence, we have to perform the integration over $\hat{D}_a$.
Here, we also made use of the property $c_1(TD_a) = c_1(T\hat{D}_a)$, valid for any toric divisor. This relation follows from the fact that the first Chern class of a tangent bundle may be computed from the curvature of the underlying base space (a local quantity) and is, therefore, insensitive to the action of the cyclic group~$\ZZ_{m_a}$.
While $c_1(T\hat{D}_a)$ can be obtained from the Ricci form~$\rho_{(a)}$ of~$\hat{D}_a$, as explained in subsection~\ref{subsec:divisors}, $c_1(T\Morb_4)$ is obtained from $c_1(T\Morb_4) =\rho/2\pi$, where $P_\rho$ is given in~\eqref{pirro}. Taking, as an example, the divisor~$\hat{D}_1$, an explicit computation gives
\begin{equation}
	\begin{split}
		\rho\bigr|_{\hat{D}_1} - \rho_{(1)} &= \sij \frac{\Delta_x'(\XA)}{\Delta_x'(\XB)} \, \partial_y \biggl[ \frac{(\XB^2 - y^2) H(\XB,y)}{(\XA^2 - y^2) H(\XA,y)} \biggr] \dd y \wedge \dd\psi_+^{(1)} \\
		& - \frac{|\Delta_x'(\XA) \det(\mathbf{J})|}{2|\Xi| a^2} \, \partial_y \biggl[ \frac{y \, \Delta_y}{(\XA^2 - y^2)^2 H(\XA,y)} \biggr] \dd y \wedge \dd\psi_+^{(1)} \,,
	\end{split}
\end{equation}
hence, performing the integration,
\begin{align} \label{D11rho}
	D_{11} &= \sij \frac{1}{m_1} \frac{\Delta\psi_+^{(1)}}{2\pi} \frac{\Delta_x'(\XA)}{\Delta_x'(\XB)} \biggl[ \frac{(\XB^2 - \YB^2) H(\XB,\YB)}{(\XA^2 - \YB^2) H(\XA,\YB)} - \frac{(\XB^2 - \YA^2) H(\XB,\YA)}{(\XA^2 - \YA^2) H(\XA,\YA)} \biggr] \\
	&= -\sij \frac{1}{m_1} \frac{\Delta\psi_+^{(1)}}{2\pi} \frac{\Delta_x'(\XA)^2 \det(\mathbf{J})}{4\Xi a^2} \biggl[ \frac{1}{(\XA^2 - \YB^2) H(\XA,\YB)} - \frac{1}{(\XA^2 - \YA^2) H(\XA,\YA)} \biggr] , \nonumber
\end{align}
where, in the last line, we used the identity
\begin{equation}
	\begin{split}
		& \frac{(\XBA^2 - \YB^2) H(\XBA,\YB)}{(\XAB^2 - \YB^2) H(\XAB,\YB)} - \frac{(\XBA^2 - \YA^2) H(\XBA,\YA)}{(\XAB^2 - \YA^2) H(\XAB,\YA)} = \\
		& \qquad\qquad = \mp \frac{\det(\mathbf{J})}{4\Xi a^2} \biggl[ \frac{\Delta_x'(\XA) \Delta_x'(\XB)}{(\XAB^2 - \YB^2) H(\XAB,\YB)} - \frac{\Delta_x'(\XA) \Delta_x'(\XB)}{(\XAB^2 - \YA^2) H(\XAB,\YA)} \biggr] \,.
	\end{split}
\end{equation}
We can now compare the expressions for~$D_{11}$ presented in~\eqref{D11v} and~\eqref{D11rho}, thus obtaining
\begin{equation} \label{quant1_label0}
	\sid \frac{\Delta\psi}{2\pi} \frac{\Delta\phi}{2\pi} = \sij \frac{1}{m_1} \frac{\Delta\psi_+^{(1)}}{2\pi} \det(\mathbf{J}) \,.
\end{equation}
By means of~\eqref{periods} this relation becomes (see also~\eqref{kD-kJK})
\begin{equation} \label{quant1_label}
	\sid \frac{\Delta\psi_+^{(1)}}{2\pi} \frac{\Delta\psi_-^{(1)}}{2\pi} \bigl|\det(\mathbf{J})\bigr| = \sij \frac{1}{m_1} \frac{\Delta\psi_+^{(1)}}{2\pi} \det(\mathbf{J})  \qquad  \implies  \qquad
	\frac{\Delta\psi_-^{(1)}}{2\pi} = \frac{1}{m_1} \,,
\end{equation}
Similarly, the inspection of the components~$D_{22}$, $D_{33}$ and~$D_{44}$ gives, respectively,
\begin{equation} \label{m234}
	\frac{\Delta\chi_-^{(2)}}{2\pi} = \frac{1}{m_2} \,,  \qquad\quad
	\frac{\Delta\psi_+^{(3)}}{2\pi} = \frac{1}{m_3} \,,  \qquad\quad
	\frac{\Delta\chi_+^{(4)}}{2\pi} = \frac{1}{m_4} \,.
\end{equation}
We note that these formulas correlate the label~$m_a$ associated with a given divisor~$\hat{D}_a$ with the periodicity of the null coordinate generated precisely by $\xi_a$ on the same divisor. As we
will see below, they will be  crucial for determining the fluxes in terms of the toric data, and ultimately for implementing the extremization procedure of section~\ref{sec:extremization}.

Before computing the off-diagonal terms of the intersection matrix, we combine equations~\eqref{quant1_poles} and~\eqref{quant1_label0}. Comparing the results with~\eqref{quant_div} we find the useful relations	$d_{1,2} = m_1 \mathtt{m}_-^{(1)}$ and $d_{4,1} = m_1 \mathtt{m}_+^{(1)}$. Equivalently, after using~\eqref{dvslabels}, we have
\begin{equation} \label{dhat-vs-m}
	\mathtt{m}_-^{(1)} = m_2 \, \hat{d}_{1,2} \,,  \qquad  \mathtt{m}_+^{(1)} = m_4 \, \hat{d}_{4,1} \,.
\end{equation}
Similar formulas can be proven, relating the order of the orbifold singularities of the remaining divisors to the labels $m_a$ and the positive integers $\hat{d}_{a,a+1}$, namely
\begin{equation}
	\begin{aligned}
    \mathtt{m}_-^{(2)} &= m_1 \, \hat{d}_{1,2} \,,  \qquad  & \mathtt{m}_+^{(2)} &= m_3 \, \hat{d}_{2,3} \,, \\
    \mathtt{m}_-^{(3)} &= m_2 \, \hat{d}_{2,3} \,,  \qquad  & \mathtt{m}_+^{(3)} &= m_4 \, \hat{d}_{3,4} \,, \\
		\mathtt{m}_-^{(4)} &= m_1 \, \hat{d}_{4,1} \,,  \qquad  & \mathtt{m}_+^{(4)} &= m_3 \, \hat{d}_{3,4} \,.
	\end{aligned}
\end{equation}

Going back to the intersection matrix, we can rewrite relation~\eqref{chern-D-2} as
\begin{equation} \label{chern-D-2b}
	D_{a\,a-1} + D_{a\,a+1} = \frac{1}{m_a} \int_{\hat{D}_a} c_1(T\hat{D}_a) = \frac{1}{m_a} \, \hat{\chi}_{(a)} \,.
\end{equation}
It is now straightforward to compare the expressions obtained from the toric geometry and from the complex structure. Indeed, if we take as an example $a=1$, the left-hand side can be read from~\eqref{Dabv}, while $\hat{\chi}_{(1)}$ is given in~\eqref{div_euler1-int}, and the equality is satisfied by means of~\eqref{dhat-vs-m}. In the same way it is possible to prove the consistency of~\eqref{chern-D-2b} for the other divisors.

Let us now return to the issue of finding a basis for an effective torus action. With the information that we have obtained so far, it is possible to show that the following  $SL(2,\RR)$ matrix
\begin{equation} \label{SL-matrx}
	S=\begin{pmatrix}
		\frac{2\pi}{\Delta\phi} (a_+ K^{(\phi)}_+ - a_- K^{(\phi)}_-)  &  -\frac{2\pi}{\Delta\psi} (a_+ K^{(\psi)}_+ - a_- K^{(\psi)}_-) \\
	 	-\frac{2\pi}{\Delta\phi} \frac{J^{(\phi)}_-}{m_1}  &  \frac{2\pi}{\Delta\psi} \frac{J^{(\psi)}_-}{m_1}
	\end{pmatrix} \,,
\end{equation}
acting on the ``fake'' vectors~\eqref{vec-n} transforms them in the  $\ZZ^2$-valued set (recall that $d_{1,a}=m_1 m_a \, \hat{d}_{1,a}$ from~\eqref{dvslabels})
\begin{equation} \label{new-vec-n}
	\begin{aligned}
		\vec{v}_1 &=(m_1,0) \,,  \qquad  &
		\vec{v}_2 &= (a_+ d_{2,4},d_{1,2}/m_1) \,, \\
		\vec{v}_3 &= (a_- d_{2,3}+a_+ d_{3,4}, d_{1,3}/m_1) \,,  \qquad  &
		\vec{v}_4 &= (a_- d_{2,4},-d_{4,1}/m_1) \, ,
	\end{aligned}
\end{equation}
where $a_{\pm}\in\ZZ$ are integers such that\footnote{Here, for simplicity we assumed that gcd$( \mathtt{m}_+^{(1)}, \mathtt{m}^{(1)}_-)=1$ and that $\sid=1$.}
\begin{equation}
a_-  \mathtt{m}_-^{(1)} + a_+  \mathtt{m}_+^{(1)} =-1\, ,
\end{equation}
 which always exist by Bézout's lemma. The resulting basis reads
 \begin{equation}
	\begin{split}
		E_1 &= \frac{1}{m_1} \bigl( J^{(\psi)}_- \partial_\psi + J^{(\phi)}_- \partial_\phi \bigr) \, , \\
		E_2 &= \bigl( a_+ K^{(\psi)}_+ - a_- K^{(\psi)}_- \bigr) \, \partial_{\psi} + \bigl( a_+ K^{(\phi)}_+ - a_- K^{(\phi)}_- \bigr) \, \partial_{\phi} \, ,
	\end{split}
\end{equation}
and the fact that the $\vec{v}_a$ are  $\ZZ^2$-valued indicates that this is a basis for an effectively acting torus. Notice that, remarkably,  although the basis $\{e_1,e_2\}$ and the vectors in~\eqref{vec-n} depend on the periodicities $\Delta \psi$ and $\Delta \phi$ separately (that we do not know),  the new basis $\{E_1,E_2\}$ and the new vectors  $\vec{v}_a$ will be explicitly determined in terms of the gauge invariant toric data, once we complete the analysis of the quantization conditions, to which we now turn. Lastly, since $d_{a,b}=m_a m_b \, \hat{d}_{a,b}$, it is straightforward to observe that a factor $m_a$ can be collected from each vector $\vec{v}_a$, confirming that the $m_a$ serve as labels for the divisors $\hat{D}_a$.

\subsection{Quantization conditions}
\label{subsec:quantization}

The analysis of the toric data
 of~$\Morb_4$ performed in the previous subsection shows clearly that this space is a well-defined orbifold only if the $d_{a,b}$ computed explicitly in~\eqref{quant_div} and~\eqref{def_t} are integers.
In this subsection we will show how to solve these ``quantization'' conditions by writing all the parameters of the solution, namely $a$, $\Nx$, $\Ny$ and $\Delta\psi\,\Delta\phi$,%
\footnote{In the supersymmetric case, the one we shall focus on, the charge $\delta$ is fixed in terms of~$a$ by the supersymmetry condition $ags^2 = -\kk$.}
in terms of the six integers $d_{a,b}$.

Instead of tackling directly the aforementioned quantization conditions, we follow an alternative path and begin studying the total magnetic fluxes across the divisors. They are defined as
\begin{equation} \label{qh_def}
	\hat{\flq}_i^a \equiv \frac{g_c}{2\pi} \int_{\hat{D}_a} F_i \,,
\end{equation}
where $g_c$ is the gauge coupling constant and the field strengths~$F_i$ can be conveniently written as
\begin{equation}
	F_i = \frac{c \tilde{c}}{\Xi \, s H^2} \bigl[ \partial_x H \, \dd x \wedge \bigl( V_y \, \dd\psi - \widetilde{V}_y \, \dd\phi \bigr)
	+ \partial_y H \, \dd y \wedge \bigl( V_x \, \dd\psi - \widetilde{V}_x \, \dd\phi \bigr) \bigr] \,.
\end{equation}
We start considering, as an example, the first divisor~$\hat{D}_1$, whose associated magnetic fluxes are
\begin{equation}
	\hat{\flq}_1^1 = \hat{\flq}_2^1 = -\frac{\Delta\psi_+^{(1)}}{2\pi} \, \frac{g_c c \tilde{c}}{s} \frac{|\Delta_x'(\XA) \det(\mathbf{J})|}{2|\Xi| a^2} \biggl[ \frac{1}{H(\XA,\YB)} - \frac{1}{H(\XA,\YA)} \biggr] \,.
\end{equation}
We now introduce the magnetic fluxes across the ramification divisors~$D_a$
\begin{equation} \label{magnetic_def}
	\flq_i^a \equiv \frac{\hat{\flq}_i^a}{m_a} = \frac{g_c}{2\pi} \int_{D_a} F_i \,,
\end{equation}
where, we recall, $m_a\int_{D_a}\!\Lambda = \int_{\hat{D}_a}\!\Lambda$ for any two-form $\Lambda$.
By means of~\eqref{quant1_label} and~\eqref{periods}, the $R$-symmetry flux $\flq_R^a =	\flq_1^a + \flq_2^a$ across~$D_1$ reads
\begin{equation} \label{qR1}
	\flq_R^1 = -\frac{\Delta\psi}{2\pi} \frac{\Delta\phi}{2\pi} \, \frac{g_c c \tilde{c}}{s} \frac{|\Delta_x'(\XA)|}{|\Xi| a^2} \biggl[ \frac{1}{H(\XA,\YB)} - \frac{1}{H(\XA,\YA)} \biggr] \,.
\end{equation}
When supersymmetry is imposed ($ags^2 = -\kk$), one can take advantage of the relations presented at the end of subsection~\ref{subsec:equal-charges} and, employing~\eqref{root_H}, the expression of $\flq_R^1$ boils down to\footnote{Here, we took $\delta>0$; for negative values the sign is flipped and similarly happens for the other fluxes.}
\begin{equation}
	\flq_R^1 = \sid \, \frac{\Delta\psi}{2\pi} \frac{\Delta\phi}{2\pi} \, \frac{g\,g_c \Delta_x'(\XA)}{\Xi a^2} \bigl( \tau^{(\YB)} \YB - \tau^{(\YA)} \YA \bigr) \,.
\end{equation}
Exploiting the simplifications implied by supersymmetry, it can be shown that
\begin{equation} \label{quant_qR1}
	\flq_R^1 = -\frac{\tau^{(\YB)}}{d_{4,1}} - \frac{\tau^{(\YA)}}{d_{1,2}} - \tau^{(\XA)} \, \frac{t_K}{d_{4,1} d_{1,2}} \,,
\end{equation}
where, crucially, we employed relation~\eqref{quant1_label}, obtained using the adjunction formula.
This relation implies that, once the quantization conditions~\eqref{quant_div} and~\eqref{def_t} are satisfied, the $R$-symmetry flux $\flq_R^1$ is automatically quantized and its expression is given in terms of integers.

In a similar way we can compute the total magnetic fluxes across the remaining divisors
\begin{equation} \label{qR234}
	\begin{split}
		\flq_R^2 &= -\frac{\Delta\psi}{2\pi} \frac{\Delta\phi}{2\pi} \, \frac{g_c c \tilde{c}}{s} \frac{|\Delta_y'(\YA)|}{|\Xi| a^2} \biggl[ \frac{1}{H(\XB,\YA)} - \frac{1}{H(\XA,\YA)} \biggr] \,, \\
		\flq_R^3 &= -\frac{\Delta\psi}{2\pi} \frac{\Delta\phi}{2\pi} \, \frac{g_c c \tilde{c}}{s} \frac{|\Delta_x'(\XB)|}{|\Xi| a^2} \biggl[ \frac{1}{H(\XB,\YB)} - \frac{1}{H(\XB,\YA)} \biggr] \,, \\
		\flq_R^4 &= -\frac{\Delta\psi}{2\pi} \frac{\Delta\phi}{2\pi} \, \frac{g_c c \tilde{c}}{s} \frac{|\Delta_y'(\YB)|}{|\Xi| a^2} \biggl[ \frac{1}{H(\XB,\YB)} - \frac{1}{H(\XA,\YB)} \biggr] \,.
	\end{split}
\end{equation}
Imposing supersymmetry, the $R$-symmetry fluxes take the form
\begin{equation}
	\begin{split}
		\flq_R^2 &= -\sid \, \frac{\Delta\psi}{2\pi} \frac{\Delta\phi}{2\pi} \, \frac{g\,g_c \Delta_y'(\YA)}{\Xi a^2} \bigl( \tau^{(\XB)} \XB - \tau^{(\XA)} \XA \bigr) \,, \\
		\flq_R^3 &= -\sid \, \frac{\Delta\psi}{2\pi} \frac{\Delta\phi}{2\pi} \, \frac{g\,g_c \Delta_x'(\XB)}{\Xi a^2} \bigl( \tau^{(\YB)} \YB - \tau^{(\YA)} \YA \bigr) \,, \\
		\flq_R^4 &= \sid \, \frac{\Delta\psi}{2\pi} \frac{\Delta\phi}{2\pi} \, \frac{g\,g_c \Delta_y'(\YB)}{\Xi a^2} \bigl( \tau^{(\XB)} \XB - \tau^{(\XA)} \XA \bigr) \,,
	\end{split}
\end{equation}
and the following quantization formulas can be proven
\begin{equation} \label{quant_qR234}
	\begin{split}
		\flq_R^2 &= -\frac{\tau^{(\XB)}}{d_{2,3}} - \frac{\tau^{(\XA)}}{d_{1,2}} + \tau^{(\YA)} \, \frac{t_J}{d_{1,2} d_{2,3}} \,, \\
		\flq_R^3 &= -\frac{\tau^{(\YB)}}{d_{3,4}} - \frac{\tau^{(\YA)}}{d_{2,3}} + \tau^{(\XB)} \, \frac{t_K}{d_{2,3} d_{3,4}} \,, \\
		\flq_R^4 &= -\frac{\tau^{(\XB)}}{d_{3,4}} - \frac{\tau^{(\XA)}}{d_{4,1}} - \tau^{(\YB)} \, \frac{t_J}{d_{3,4} d_{4,1}} \,.
	\end{split}
\end{equation}
Remarkably, the above conditions take exactly the form
\begin{equation}
	\flq_R^a = \sum_b D_{ab} \, \sigma^b \,,
\end{equation}
with the $\sigma^a$ being signs, \ie\ $\sigma^a = \pm1$, as conjectured in \cite{Faedo:2022rqx}. These solve the topological constraints $\sum_a \flq_R^a \vec{V}_a = 0$, with the  $\sigma^a$ specifying the type of supersymmetry-preserving  twist.
In addition, we note the following useful relations
\begin{equation} \label{qR-ratios}
	\frac{\flq_R^1}{\flq_R^3} = -\frac{\Delta_x'(\XA)}{\Delta_x'(\XB)} \,,  \qquad
	\frac{\flq_R^2}{\flq_R^4} = -\frac{\Delta_y'(\YA)}{\Delta_y'(\YB)} \,.
\end{equation}
Notice that the $R$-symmetry flux $\hat{\flq}_R^1= m_1 \flq_R^1$ across~$\hat{D}_1$, descending from~\eqref{quant_qR1} and~\eqref{dhat-vs-m}, takes the form
\begin{equation}
	\hat{\flq}_R^1 = -\frac{\tau^{(\YB)}}{\mathtt{m}_+^{(1)}} - \frac{\tau^{(\YA)}}{\mathtt{m}_-^{(1)}} - \tau^{(\XA)} \, \frac{t_K}{\mathtt{m}_+^{(1)} \mathtt{m}_-^{(1)} m_1} \,.
\end{equation}
This expression is consistent with the expected quantization condition for the total magnetic flux through $\hat{D}_1 = \mathbb{WCP}^1_{[\mathtt{m}_-^{(1)},\mathtt{m}_+^{(1)}]}$. In particular, the first two terms can be identified with twist or anti-twist~\cite{Ferrero:2021etw} through  $\hat{D}_1$, while the factor $m_1$ in the denominator of  the last term is due to the $\ZZ_{m_1}$ singularity in the normal direction to~$\hat{D}_1$~\cite{Cheung:2022ilc}.
Similar computations for the other fluxes $\hat{\flq}_R^a$ show that these can be correctly expressed as
\begin{equation}
	\begin{split}
 		\hat{\flq}_R^2 &= -\frac{\tau^{(\XB)}}{\mathtt{m}_+^{(2)}} - \frac{\tau^{(\XA)}}{\mathtt{m}_-^{(2)}} + \tau^{(\YA)} \, \frac{t_J}{\mathtt{m}_+^{(2)} \mathtt{m}_-^{(2)} m_2} \,, \\
 		\hat{\flq}_R^3 &= -\frac{\tau^{(\YB)}}{\mathtt{m}_+^{(3)}} - \frac{\tau^{(\YA)}}{\mathtt{m}_-^{(3)}} + \tau^{(\XB)} \, \frac{t_K}{\mathtt{m}_+^{(3)} \mathtt{m}_-^{(3)} m_3} \,, \\
 		\hat{\flq}_R^4 &= -\frac{\tau^{(\XB)}}{\mathtt{m}_+^{(4)}} - \frac{\tau^{(\XA)}}{\mathtt{m}_-^{(4)}} - \tau^{(\YB)} \, \frac{t_J}{\mathtt{m}_+^{(4)} \mathtt{m}_-^{(4)} m_4} \,.
	\end{split}
\end{equation}

Coming back to the quantization conditions for the metric parameters, the strategy for their resolution consists in expressing $a$, $\Nx$, $\Ny$ and the product $\Delta\psi\,\Delta\phi$ in terms of the roots~$\XBA$ and~$\YBA$, and, subsequently, in writing the latter in terms of the integers~$d_{a,b}$ and of the quantized fluxes~$\flq_R^a$.
To avoid clumsy notation, until the end of this subsection we shall consider $0<y<x$. As explained in subsection~\ref{subsec:existence}, in this case the roots are organized such that $\Delta_x^-(\XB)=0=\Delta_x^-(\XA)$ and $\Delta_y^-(\YB)=0=\Delta_y^+(\YA)$, which amounts in taking $\tau^{(\XB)}=\tau^{(\XA)}=\tau^{(\YB)}=-1$ and $\tau^{(\YA)}=+1$ where needed.
With this choice, all the $R$-symmetry fluxes are negative, a fact that can be easily seen trading the~$\sid$ for the absolute values.
Combining appropriately the pair of equations $\Delta_x^-(\XBA)=0$, from~\eqref{Delta+-} we obtain
\begin{equation} \label{from-x}
	\begin{split}
		a &= -\kk \, \frac{\XB^{D-4} (1 - g \XB) - \XA^{D-4} (1 - g \XA)}{\XB^{D-5} (1 - g \XB) - \XA^{D-5} (1 - g \XA)} \,, \\
		\Nx &= (-\kk \, a) \, \frac{\XB^{D-5} \XA^{D-5} (\XB - \XA) (1 - g \XB) (1 - g \XA)}{2 [\XB^{D-5} (1 - g \XB) - \XA^{D-5} (1 - g \XA)]} \,,
	\end{split}
\end{equation}
whereas applying the same method to $\Delta_y^\pm$ we are able to isolate $a$ and $\Ny$ in terms of~$\YBA$
\begin{equation} \label{from-y}
	\begin{split}
		a &= -\kk \, \frac{\YB^{D-4} (1 - g \YB) + \YA^{D-4} (1 + g \YA)}{\YB^{D-5} (1 - g \YB) - \YA^{D-5} (1 + g \YA)} \,, \\
		\Ny &= (-\kk \, a) \, \frac{\YB^{D-5} \YA^{D-5} (\YB + \YA) (1 - g \YB) (1 + g \YA)}{2 [\YB^{D-5} (1 - g \YB) - \YA^{D-5} (1 + g \YA)]} \,.
	\end{split}
\end{equation}
Consider, now, conditions~\eqref{quant_div}. Taking the ratios first/fourth and second/third and imposing supersymmetry, we obtain the two equations
\begin{equation}
	\frac{d_{1,2} \, \flq_R^2}{d_{4,1} \, \flq_R^4} = \frac{\XA + \YA}{\XA - \YB} \,,  \qquad
	\frac{d_{2,3} \, \flq_R^2}{d_{3,4} \, \flq_R^4} = \frac{\XB + \YA}{\XB - \YB} \,,
\end{equation}
which are solved by
\begin{equation} \label{roots-x}
	\XA = \frac{d_{1,2} \flq_R^2 \, \YB + d_{4,1} \flq_R^4 \, \YA}{d_{1,2} \flq_R^2 - d_{4,1} \flq_R^4} \,,  \qquad
	\XB = \frac{d_{2,3} \flq_R^2 \, \YB + d_{3,4} \flq_R^4 \, \YA}{d_{2,3} \flq_R^2 - d_{3,4} \flq_R^4} \,.
\end{equation}
All the other possible ratios of~\eqref{quant_div} and~\eqref{def_t} are identically satisfied by~\eqref{roots-x} making use of~\eqref{quant_qR1}, \eqref{quant_qR234} and~\eqref{qR-ratios}.
Now that we have the roots~$\XBA$ as functions of~$\YBA$, we can express all the physical parameters in terms of the latter.
In order to proceed, we write down the expressions of $t_J$ and $t_K$ in the supersymmetric case, which simplify to
\begin{equation}
	\begin{split}
		t_J &= \sid \frac{2\pi}{\Delta\psi} \frac{2\pi}{\Delta\phi} \frac{4\Xi a^2 (1 - \kk\,ag)}{g} \frac{\XB - \XA}{\Delta_x'(\XB) \Delta_x'(\XA)} \,, \\
		t_K &= \sid \frac{2\pi}{\Delta\psi} \frac{2\pi}{\Delta\phi} \frac{4\Xi a^2 (1 - \kk\,ag)}{g} \frac{\YB + \YA}{\Delta_y'(\YB) \Delta_y'(\YA)} \,.
	\end{split}
\end{equation}
Notice that, when $0<y<x$, both $t_J$ and $t_K$ are negative.
Having the expressions of $t_J$ and $t_K$, we can immediately derive the product $\Delta\psi\,\Delta\phi$ in terms of the roots~$\YBA$ and of the other integers characterizing the system
\begin{equation} \label{Dpsi*Dphi}
	\frac{\Delta\psi}{2\pi} \frac{\Delta\phi}{2\pi} = -\sid \, \frac{a^2 (1 + \kk\,ag)}{4g\,g_c^2 (\YB + \YA)^3} \frac{(t_J \flq_R^1 \flq_R^3)^2}{t_K \flq_R^2 \flq_R^4} \,.
\end{equation}
The next step is to determine the roots~$\YBA$ in terms of the integers $d_{a,b}$. Combining in an appropriate way the fluxes~$\flq_R^a$ with $t_J$, we obtain
\begin{equation} \label{Delta'}
	\begin{split}
		\Delta_y'(\YA) &= -\frac{4 \flq_R^2}{t_J \flq_R^1 \flq_R^3} \, g_c (1 - \kk\,ag) (\YB + \YA)^2 \,, \\
		\Delta_y'(\YB) &= \frac{4 \flq_R^4}{t_J \flq_R^1 \flq_R^3} \, g_c (1 - \kk\,ag) (\YB + \YA)^2 \,,
	\end{split}
\end{equation}
which is a pair of equations containing only $\YBA$, once $a$ is substituted as in~\eqref{from-y}.

In order to solve this system of equations we need to specify the number of spacetime dimensions. Starting with $D=6$, we substitute in~\eqref{Delta'} the expression of~$\Delta_y'$ given in~\eqref{root_Delta'} and parameterize the roots of $\Delta_y$ as $\YBA = w (1 \pm \x)$. When choosing the integers that describe~$\Morb_4$ we shall need to impose $w>0$ and $0<\x<1$ in order to have $0<\YA<\YB$. Solving the resulting equations for~$w$ and~$\x$ we obtain
\begin{equation} \label{6d-w}
	w = -\frac{3(\mathfrak{Q}_+ + \mathfrak{Q}_-) \x - \x^3 \pm \sqrt{3 - 6(\mathfrak{Q}_+ + \mathfrak{Q}_-) - 3 (1 - 3(\mathfrak{Q}_+ + \mathfrak{Q}_-)^2) \x^2 + \x^4}}{g [3 + \x^4 - 6(\mathfrak{Q}_+ + \mathfrak{Q}_-) (1 + \x^2)]} \,,
\end{equation}
where we defined $\mathfrak{Q}_+ = \frac{\flq_R^4}{t_J \flq_R^1 \flq_R^3}$ and $\mathfrak{Q}_- = \frac{\flq_R^2}{t_J \flq_R^1 \flq_R^3}$ and $\x$ is solution of the cubic equation
\begin{equation} \label{6d-x}
	\x^3 - 3(\mathfrak{Q}_+ - \mathfrak{Q}_-) \x^2 - 3 [1 - 2(\mathfrak{Q}_+ + \mathfrak{Q}_-)] \x + 3(\mathfrak{Q}_+ - \mathfrak{Q}_-) = 0 \,.
\end{equation}
The two signs in~\eqref{6d-w} generate the two sets of physical parameters connected by the inversion symmetry~\eqref{inv_sym} and, in particular, the plus yields the configuration with $\Xi<0$. This last statement can be proven noting that $\Xi = (1-\kk\,ag)(1+\kk\,ag)$ and rearranging the second term as
\begin{equation} \label{6d-kag}
	1 + \kk\,ag = \mp \frac{2g w \sqrt{3 - 6(\mathfrak{Q}_+ + \mathfrak{Q}_-) - 3 (1 - 3(\mathfrak{Q}_+ + \mathfrak{Q}_-)^2) \x^2 + \x^4}}{1 - \x^2} \,.
\end{equation}
The first factor of~$\Xi$ is positive due to supersymmetry, while the second one contributes to the sign and tells whether $\Xi$ is positive or negative.
The last thing to do is to equate the expression for $a$ obtained by means of~\eqref{from-x} and~\eqref{from-y}. This gives the following Diophantine equation for the integers~$d_{a,b}$
\begin{equation} \label{diofa_6d}
	t_K \flq_R^2 \flq_R^4 - 3 (\flq_R^1 - \flq_R^3) = 0 \,,
\end{equation}
whose analysis is postponed to appendix~\ref{app:diofa}.
The origin of this constraint can be traced back in the specific structure of our solution. Indeed, the quantization conditions involve the four parameters~$a$, $\Nx$, $\Ny$ and $\Delta\psi\,\Delta\phi$, but, as we discussed,
the integers $d_{a,b}$ characterizing the system are five. This means that there must be a redundancy among them and equation~\eqref{diofa_6d} precisely resolves it.
Considering a system with different charges~$\delta_i$ would increase by one the number of parameters, but would also bring an additional ``quantum number'' into play, thus not changing the balance.
It is possible that the constraint~\eqref{diofa_6d} would be eliminated if a more general solution
could be constructed, but we will not pursue this here. Similar unnatural Diophantine constraints have been observed in previous  solutions associated with orbifolds
\cite{Cheung:2022ilc,Couzens:2022lvg,Faedo:2022rqx}.

We now focus on $D=7$ and solve system~\eqref{Delta'} following the same method adopted in the six-dimensional case. Parameterizing again the two roots as $\YBA = w (1 \pm \x)$, we obtain
\begin{align}
	\label{7d-w}
	w &= -\frac{4(\mathfrak{Q}_+ + \mathfrak{Q}_-) \x \pm \sqrt2 \sqrt{[1 - 2(\mathfrak{Q}_+ + \mathfrak{Q}_-)] [1 + (1 - 4(\mathfrak{Q}_+ + \mathfrak{Q}_-)) \x^2]}}{2g [1 + \x^2 - 2(\mathfrak{Q}_+ + \mathfrak{Q}_-) (1 + 3\x^2)]} \,, \\
	\label{7d-x}
	\x &= \frac{\mathfrak{Q}_+ - \mathfrak{Q}_-}{1 - 3(\mathfrak{Q}_+ + \mathfrak{Q}_-)} \,,
\end{align}
where $\mathfrak{Q}_\pm$ are defined as in the previous case.
As before, the two signs in~\eqref{7d-w} correspond to the two configurations related by~\eqref{inv_sym} and the plus sign gives $\Xi<0$. This can be seen considering the analogous expressions of~\eqref{6d-kag}, but in $D=7$,
\begin{equation}
	1 + \kk\,ag = \mp \, 2\sqrt2 \, g w \sqrt{[1 - 2(\mathfrak{Q}_+ + \mathfrak{Q}_-)] [1 + (1 - 4(\mathfrak{Q}_+ + \mathfrak{Q}_-)) \x^2]} \,.
\end{equation}
Imposing the consistency condition that comes equating~\eqref{from-x} with~\eqref{from-y} we obtain the following Diophantine equation
\begin{equation} \label{diofa_7d}
	\begin{split}
		& 2 \flq_R^1 \flq_R^3 t_J \bigl[ (\flq_R^1)^3 d_{1,2} d_{4,1} (d_{1,2} - d_{4,1}) - (\flq_R^3)^3 d_{2,3} d_{3,4} (d_{2,3} - d_{3,4}) \bigr] \\
		& \qquad + \flq_R^1 \flq_R^3 t_J \bigl[8 \bigl( (\flq_R^1)^2 d_{1,2} d_{4,1} - (\flq_R^3)^2 d_{2,3} d_{3,4} \bigr) - \bigl( (\flq_R^1)^4 d_{1,2}^2 d_{4,1}^2 - (\flq_R^3)^4 d_{2,3}^2 d_{3,4}^2 \bigr) \bigr] \\
		& \qquad + 4 (\flq_R^1)^3 d_{1,2} d_{4,1} \bigl( \flq_R^2 d_{4,1} - \flq_R^4 d_{1,2} \bigr) - 4 (\flq_R^3)^3 d_{2,3} d_{3,4} \bigl( \flq_R^2 d_{3,4} - \flq_R^4 d_{2,3} \bigr) \\
		& \qquad + 3 (\flq_R^2 + \flq_R^4) \bigl[ (\flq_R^1)^4 d_{1,2}^2 d_{4,1}^2 - (\flq_R^3)^4 d_{2,3}^2 d_{3,4}^2 \bigr] = 0 \,.
	\end{split}
\end{equation}
Although this is surprisingly much more complicated than the analogous constraint~\eqref{diofa_6d} in $D=6$, similar comments apply.  We refer to appendix~\ref{app:diofa} for its resolution.

We close this subsection with a brief recap of the procedure that, starting from the set of integers characterizing a given solution, allows to reconstruct all its parameters.
The first step is to choose four positive integers $d_{a,a+1}$ and two integers $d_{1,3}=t_J$ and $d_{2,4}=t_K$ that satisfy constraint~\eqref{drelation} and the Diophantine equation~\eqref{diofa_6d}, if $D=6$, or~\eqref{diofa_7d}, for $D=7$.
In six dimensions we compute the auxiliary quantities $\x$ and $w$ using~\eqref{6d-x} and~\eqref{6d-w}, while for the seven-dimensional solution the corresponding relations are~\eqref{7d-x} and~\eqref{7d-w}. In both cases, $\mathfrak{Q}_+ = \frac{\flq_R^4}{t_J \flq_R^1 \flq_R^3}$, $\mathfrak{Q}_- = \frac{\flq_R^2}{t_J \flq_R^1 \flq_R^3}$ and the charges $\flq_R^a$ are given by~\eqref{quant_qR1} and~\eqref{quant_qR234}, where, for the system in question, $\tau^{(\XB)}=\tau^{(\XA)}=\tau^{(\YB)}=-1$ and $\tau^{(\YA)}=+1$.
The endpoints of the range of the coordinate~$y$ read $\YBA = w (1 \pm \x)$, while $\XBA$ can be obtained from~\eqref{roots-x}.
Lastly, the parameters $a$, $\Nx$ and $\Ny$ follow from~\eqref{from-x} and~\eqref{from-y}, while the product $\Delta\psi\,\Delta\phi$ in given by~\eqref{Dpsi*Dphi}, where the sign~$\sid$ is fixed so to have a positive result.

\section{Off-shell free energies and extremization}
\label{sec:extremization}

In this section we present the off-shell free energy proposed in~\cite{Faedo:2022rqx}, whose construction is based on the idea of gluing ``gravitational blocks'', introduced in~\cite{Hosseini:2019iad}. This recipe extends the results of~\cite{Faedo:2021nub} to four-dimensional toric orbifolds
with an arbitrary number of fixed points.
For $D=7$, the field theory counterpart of this construction was given  in~\cite{Martelli:2023oqk},  integrating on $\Morb_4$ the anomaly polynomial of the six-dimensional SCFTs associated with M5-branes. In $D=6$ a solid proof is still
lacking, nevertheless the gravitational block conjectured in~\cite{Faedo:2022rqx} was retrieved in~\cite{Colombo:2023fhu} through the integration of the equivariant volume of the associated geometry, with the addition of higher times. Below we will be primarily interested in four-dimensional toric orbifolds, with $\Morb_4$ represented by the solutions in~\ref{subsec:equal-charges}. We speculate that these $D=6,7$ solutions are holographically dual to $d=1,2$ theories, obtained compactifying on $\Morb_4$ the five-dimensional $\mathcal{N}=1$ SCFTs dual to \cite{Brandhuber:1999np} and  the six-dimensional $\mathcal{N}=(0,2)$ dual to a stack of M5-branes.
Supporting evidence is provided by the extremization of our off-shell free energy, which reproduces exactly the entropy/central charge of the (putative) black hole/black string with $\AdS_{D-4}\times\Morb_4$ near-horizon.

\subsection{Recap of the recipe}

We now summarize the key ingredients necessary to construct the off-shell free energy, whose extremization should reproduce the entropy and the central charge of the dual gravity theories in $D=6$ and $D=7$, respectively. For further details, see~\cite{Faedo:2022rqx}. Recall that an opposite convention for the $D_a$ and $\hat D_a$ divisors, as well as for the other related quantities, is employed here. In particular, the vectors $\vec{\vv}_a$ need not be primitive and the information carried by the labels is now encoded in the vectors themselves.

On the field theory side, the systems under exam result from the twisted compactification of five- and six-dimensional SCFTs on a four-dimensional toric orbifold~$\Morb_4$, completely characterized by the counter-clockwise ordered vectors~$\vec{\vv}_a \in \ZZ^2$.
The twisting is realized coupling the SCFTs to two background gauge fields~$A_i$, with field strengths $F_i = \dd A_i$.
As a first step, we define the ``physical fluxes'' as%
\footnote{Here and in the following we shall rename the background gauge fields as $g_c A_i \mapsto A_i$. The nomenclature ``physical fluxes'' should not be confused with the fact that the natural fluxes in supergravity are computed on the branch divisors, see~\eqref{qh_def}.}
\begin{equation} \label{q_def}
	\flq_i^a \equiv \frac{1}{2\pi} \int_{D_a} F_i \,,
\end{equation}
where $a=1,\ldots,\fpN$ runs over the number of facets.
The one-forms $-A_i$ can also be viewed as the connections on two line orbibundles~$E_i$. Since the set of $c_1(L_{a})$ form a basis for
 $H^2(\Morb_4,\QQ)$, we can decompose the first Chern class of~$E_i$ as
\begin{equation} \label{p_def}
	c_1(E_i)  = - \frac{\dd A_i}{2\pi} = -\sum_a \flp_i^a c_1(L_a) \,,
\end{equation}
where $\flp_i^a \in \QQ$. As a consequence
\begin{equation} \label{physfluxes}
	\flq_i^a = \sum_b D_{ab} \, \flp_i^b \,,  \qquad  \sum_a \flq_i^a \vec{\vv}_a = 0 \,,
\end{equation}
where $D_{ab}$ is the intersection matrix defined in~\eqref{inter-num}.
The latter equation, which follows from~\eqref{homo-rel}, is a constraint on the physical fluxes and implies that, for fixed~$i$, only $\fpN-2$ of the $\flq_i^a$ are independent. On the other hand, $H^2(\Morb_4,\QQ)$ has dimension $\fpN-2$, thus, for fixed~$i$, only $\fpN-2$ of the $\flp_i^a$ are linearly independent---see the discussion around~\eqref{cohomo-rel}. Therefore, the former system of equations can be inverted to obtain $\flp_i^a$ in terms of~$\flq_i^a$ once the redundant equations are eliminated.
A similar condition applies also to the $R$-symmetry fluxes~$\flq_R^a$, specifically $\sum_a \flq_R^a \vec{\vv}_a = 0$, and a possible choice to solve it is
\begin{equation} \label{physfluxes-relation}
	\flq_R^a = \sum_b D_{ab} \, \sigma^b \,,
\end{equation}
where, we conjectured, $\sigma^a$ are $n$ arbitrary signs, \ie\ $\sigma^a = \pm1$. This assumption is supported by several examples where the spindles are explicitly involved~\cite{Ferrero:2021etw,Faedo:2022rqx} in both $D=6,7$, and the solutions presented in~\ref{subsec:equal-charges} are no exception, as we will see later.

We also conjecture that the entropy/central charge corresponding, respectively, to a general class of $d=5,6$ SCFTs compactified on~$\Morb_4$ and with twists parameterized by $\sigma^a$, can be computed by the constrained extremization of the following  \emph{off-shell free energy}
\begin{equation} \label{F-extr}
	\Ispindle(\varphi_i, \epsilon_i; \flq_i^a) = k_d \sum_a \frac{\eta_d^{\,a}}{d_{a,a+1}} \frac{\mathcal{F}_d(\Phi_i^a)}{\epsilon_1^a \epsilon_2^a} \, ,
\end{equation}
where the sum is over the number of fixed points, that for planar polytopes coincides with the number of facets.
Here, $\mathcal{F}_d$ are the usual gravitational blocks  (\cf\ table~2 of~\cite{Faedo:2021nub})
\begin{equation}
	\mathcal{F}_5(\Delta_i) = -\frac{4\sqrt2 \pi \, N^{5/2}}{15 \sqrt{8 - N_f}} (\Delta_1 \Delta_2)^{3/2} \,,  \qquad  \mathcal{F}_6(\Delta_i) = -\frac{9 N^3}{256} (\Delta_1 \Delta_2)^2 \,,
\end{equation}
the variables $\Phi_i ^a$ are defined as
\begin{equation} \label{def_Phi}
	\Phi_i^a = \varphi_i - \flp_i^a \epsilon_1^a - \flp_i^{a+1} \epsilon_2^a \,,
\end{equation}
and the auxiliary quantities $\epsilon_{1,2}^a$ are constructed from the fugacities $\vec{\epsilon}=(\epsilon_1,\epsilon_2)$ associated with the $U(1)^2$ rotational symmetry of $\Morb_4$ as
\begin{equation} \label{def_epsilon}
	\epsilon^a_1 = -\frac{\det(\vec{\vv}_{a+1}, \vec{\epsilon})}{d_{a,a+1}} \,,  \qquad
	\epsilon^a_2 = \frac{\det(\vec{\vv}_a, \vec{\epsilon})}{d_{a,a+1}} \,.
\end{equation}
We recall that $d_{a,b} = \det(\vec{\vv}_a, \vec{\vv}_{b})$, whereas $k_d$ are numerical constants which assume, \textit{a posteriori}, the values
\begin{equation}
	k_5 = -1 \,,  \qquad  k_6 = \frac{64}{9} \,.
\end{equation}
Also the values of $\eta^{\,a}_d$ is one of the ingredients of the prescription. Specifically, for $d=5$ we assume they are related to the type of twist as $\eta^{\,a}_5=\sigma^{a}\sigma^{a+1}$, while for $d=6$ we take all of them with the same value, which is, ultimately, related to the chirality of the SCFT in question.
In particular, $\eta^{\,a}_6=\kk$, with $\kk=\pm1$ so that the extremization of the off-shell free energy reproduces the central charge of the $d=2$ SCFTs with $\mathcal{N}=(0,2)$ or $\mathcal{N}=(2,0)$, respectively, extracted from the anomaly polynomial of the $d=6$ theory.
This $\kk$ is also related to the sign in the equation $\hat{\nabla}_{\hat{\mu}}\vartheta = \frac{\kk}{2}\beta_{\hat{\mu}}\vartheta$ satisfied by the Killing spinor along
$\AdS_3$.

The variables of the off-shell free energy, namely $\varphi_i$ and $\epsilon_i$, are subject to the constraint
\begin{equation} \label{constraint}
	\varphi_1 + \varphi_2 - \det(\vec{W}, \vec{\epsilon}) = 2 \,,
\end{equation}
where the auxiliary two-dimensional vector $\vec{W}$ is determined imposing
\begin{equation} \label{relation-p-W}
	\flp_1^a + \flp_2^a =\sigma^a + \det(\vec{W}, \vec{\vv}_a) \,.
\end{equation}
The vector $\vec{W}$ parameterizes a  ``gauge symmetry'' featured by our construction and can be set to any desired value by means of suitable transformation~\cite{Faedo:2022rqx}.
The present construction enjoys an additional symmetry. Once the fluxes~$\flq_i^a$ are fixed, if an $SL(2,\ZZ)$ transformation is applied to $\vec{\vv}_a$, $\vec{\epsilon}$ and $\vec{W}$, the off-shell free energy and the constraint can be shown to preserve their form, thus leaving the extremization problem unaltered~\cite{Faedo:2022rqx}.

The last step of the procedure is the constrained extremization of $\Ispindle(\varphi_i,\epsilon_i)$, which can be performed defining the function
\begin{equation}\label{entropy_funct}
	\mathcal{S}(\varphi_i, \epsilon_i, \Lambda; \flq_i^a) = \Ispindle(\varphi_i, \epsilon_i; \flq_i^a) + \Lambda \bigl( \varphi_1 + \varphi_2 - \det(\vec{W}, \vec{\epsilon}) - 2 \bigr)
\end{equation}
and extremizing it with respect to~$\varphi_i$, $\epsilon_i$ and the Lagrangian multiplier~$\Lambda$.
As a consequence of Euler's theorem, $\Ispindle(\varphi_i^*,\epsilon_i^*)=-\frac{2}{h}\Lambda^*$, with $h=1$ for $d=5$, $h=2$ for $d=6$~\cite{Faedo:2022rqx}.

\subsection{Application to the new AdS$_{D-4}\times\Morb_4$ solutions}

We now apply  the gravitational blocks prescription to the $\AdS_{D-4}\times\Morb_4$ backgrounds with equal charges presented in subsection~\ref{subsec:equal-charges}.
The first ingredient needed is the toric data describing the toric orbifold~$\Morb_4$. As we mentioned, the recipe to construct the off-shell free energy is formally invariant under $SL(2,\RR)$ transformations of the vectors~$\vec{\vv}_a$ comprising the toric data, although this group is broken to $SL(2,\ZZ)$ when integer-valued vectors are considered. Thanks to this property, we use vectors~\eqref{vec-n} as toric data, assuming that a set of proper (integer-valued) toric data exists and is related to~\eqref{vec-n} by an~$SL(2,\RR)$ transformation---see the discussion below~\eqref{vec-n}.
The physical fluxes defined in~\eqref{q_def} coincide with the magnetic fluxes~\eqref{magnetic_def}, whose explicit forms are given in~\eqref{qR1} and~\eqref{qR234}.
The vector of twists can be deduced comparing~\eqref{physfluxes-relation} with the quantized expressions for the fluxes~\eqref{quant_qR1} and~\eqref{quant_qR234}. When $\delta$ is positive we obtain
\begin{equation} \label{twist-par}
	\sigma^1 = -\tau^{(\XA)} \,,  \qquad  \sigma^2 = -\tau^{(\YA)} \,,  \qquad
	\sigma^3 = -\tau^{(\XB)} \,,  \qquad  \sigma^4 = -\tau^{(\YB)} \,,
\end{equation}
which becomes $\sigma^a = [\sign(x^2-y^2),-\sign(x^2-y^2),+,+]$ if we restrict to $x,y>0$.
The results of subsection~\ref{subsec:existence} imply that three out of the four components of $\sigma^a$ have always the same sign, while the fourth one is the opposite. This property agrees with the vector of twists found for the $\AdS \times \spindle \ltimes \spindle$ solutions studied in~\cite{Faedo:2022rqx}.
Taking advantage of the gauge symmetry enjoyed by our prescription, one can always pick a gauge in which the auxiliary vector $\vec{W}$ vanishes. When the charges are equal, as in the case of interest, this choice is particularly useful as~\eqref{relation-p-W} implies the simple relation
\begin{equation}
	\flp_i^a = \frac{\sigma^a}{2} \,.
\end{equation}

An important feature characterizing all the systems with equal charges is that the extremizing values $\varphi_i^*$ are equal, namely $\varphi_1^* = \varphi_2^*$. Indeed, the off-shell free energy~\eqref{F-extr} is symmetric under the exchange $\varphi_1 \leftrightarrow \varphi_2$ and, in particular, $\Phi_1^a -\Phi_2^a = \varphi_1 - \varphi_2$, which does not depend on the index $a$. Then, the necessary condition $(\partial_{\varphi_1}-\partial_{\varphi_2})\mathcal{S}=0$ forces the values at the extremum to be equal.
Additionally, in the gauge we are considering, the constraint~\eqref{constraint} with $\vec{W}=0$ fixes them to $\varphi_1^* = \varphi_2^* = 1$.
Extremizing the off-shell free energy with respect to the remaining variables~$\epsilon_{1,2}$ we obtain the following critical values for the fugacities%
\footnote{These results hold when $\sid>0$. In the opposite case, we need to exchange the two components of the vectors in~\eqref{vec-n} and, therefore, to exchange $\epsilon_1^*$ and $\epsilon_2^*$ accordingly.}
\begin{alignat}{3}
	& d = 5:  \qquad\quad  & \epsilon_1^* &= \frac{2}{g} \, \frac{2\pi}{\Delta\psi} \,,  \qquad  & \epsilon_2^* &= -\kk \, 2a \, \frac{2\pi}{\Delta\phi} \,, \\
	& d = 6:  \qquad\quad  & \epsilon_1^* &= \frac{1}{g} \, \frac{2\pi}{\Delta\psi} \,,  \qquad  & \epsilon_2^* &= -\kk \, a \, \frac{2\pi}{\Delta\phi} \,.
\end{alignat}
These expressions have the same form of the results of~\cite{Faedo:2022rqx} once we perform the proper $SL(2,\RR)$ transformations connecting our basis of the torus action $\{\partial_{\nu_1},\partial_{\nu_2}\}$ with the ones adopted in~\cite{Faedo:2022rqx}. Plugging these critical values in~\eqref{F-extr}, we get the off-shell free energy at its extremum%
\footnote{In order to take into account a possible exchange in the components of $\vec{v}_a$, which  would affect $d_{a,a+1}$ in the denominator of~\eqref{F-extr}, we multiplied the off-shell free energy by $\sid$.}
\begin{align}
	\Ispindle^*_{d=5}& = \frac{9\sqrt2 \pi N^{5/2}}{5 \sqrt{8 - N_f}} \, \frac{g^3 (1 - \kk\,ag)}{|\Xi| a^2} \frac{\Delta\psi}{2\pi} \frac{\Delta\phi}{2\pi} \bigl| (x_+^2 +\tau^{(x_-)} x_-^2) \Delta y - \Delta x (y_+^2 - \tau^{(x_-)} y_-^2) \bigr| \,,
	\\
	\Ispindle^*_{d=6}& = \kk \, 4N^3 \, \frac{g^4 (1 - \kk\,ag)}{|\Xi| a^3} \frac{\Delta\psi}{2\pi} \frac{\Delta\phi}{2\pi} \bigl| (x_+^3 + \tau^{(x_-)} x_-^3) (y_+^2 - y_-^2) -  (x_+^2 - x_-^2)(y_+^3 - \tau^{(x_-)} y_-^3) \bigr| \, .
\end{align}
To compare these formulas with their gravitational counterparts, we must recall that the roots $\XBA$ and $\YBA$ are related through the constraint obtained equating $a$ in~\eqref{from-x} and in~\eqref{from-y}.
For what concerns $\Ispindle^*_{d=5}$, its structure can be shown to agree with the gravitational entropy~\eqref{entropy_branes}, whereas $\Ispindle^*_{d=6}$ matches the gravitational central charge~\eqref{central-charge_branes}, but with a different sign.
We believe that the origin of the opposite sign can be traced back to the chirality of the dual two-dimensional SCFTs,
but since we have not solved the Killing spinor equations of $D=7$ supergravity, we have no control over this information.

\section{Discussion}
\label{sec:discussion}

In this paper we constructed new families of supersymmetric solutions in $D=10$ and $D=11$ supergravities, corresponding to D4 and M5-branes wrapping four-dimensional orbifolds $\Morb_4$, respectively. These generalize to arbitrary toric orbifolds with four fixed points the examples presented in \cite{Cheung:2022ilc,Couzens:2022lvg,Faedo:2022rqx}, where the corresponding orbifolds were the total space of a spindle fibred over another spindle.
After extracting the necessary data, we fed these into the off-shell free energies conjectured in our previous paper
\cite{Faedo:2022rqx} (and then proved in~\cite{BenettiGenolini:2023yfe,Colombo:2023fhu}), obtaining a perfect match between the gravitational observables calculated directly from the solutions and those derived  from the corresponding extremization problems. This is further evidence that the extremal problems proposed in \cite{Faedo:2022rqx}   encapsulate the necessary and sufficient conditions for the existence of the supergravity solutions. Furthermore, by construction, the gravitational central charge associated with the AdS$_3$ solutions agrees with the central charge of the dual two-dimensional SCFTs, obtained by integrating the M5-brane anomaly polynomial   on the orbifolds \cite{Martelli:2023oqk}. It would be nice to derive in field theory the entropy associated with the AdS$_2$ solutions, by computing the supersymmetric partition function of the dual five-dimensional theories compactified on $\Morb_4\times S^1$ and then taking its large $N$ limit.

Obvious extensions of the present paper include  looking for solutions involving  four-dimensional orbifolds $\Morb_4$ with one or more of the following features.
Firstly, one can strive
 to tame the solution with  unequal charges.   It is also natural
to wonder whether there exist solutions where the Diophantine constraints need not be imposed, as these appear to be just an artefact of the ansatz employed.
Although explicit solutions corresponding to toric orbifolds with more than four fixed points are likely to be inaccessible, we expect that it should
be relatively simple to construct solutions describing branes wrapped on  orbifolds with three fixed points, namely weighted projective spaces $\mathbb{WCP}^2_{[n_1,n_2,n_3]}$.
It is possible that such solutions may be obtained in the special case of equal NUT parameters ($\Nx=\Ny$), that we have not analysed in this paper.%
\footnote{In this scenario a singularity occurs when $x=y$ and,  in order to investigate this case, it may be useful to employ a change of coordinates analogous to that used in \cite{Martelli:2007pv} and \cite{Martelli:2013aqa} in different contexts.}
Finally, for completeness, it would be desirable to check explicitly the supersymmetry conditions for the $D=7$ solutions. We leave these unglamorous tasks as possible projects for  eager students.

The results of this paper add  evidence that the discovery of supergravity solutions displaying orbifold singularities  has led to a vaster arena for exploring  quantum gravity via holography.
We are convinced that it is worth pursuing these explorations in different directions.
On the one hand, it should be rewarding to construct further explicit examples of orbifold solutions, with AdS factors or otherwise. For example, using the results in
 \cite{Colombo:2023fhu} one can anticipate the existence of numerous solutions in type~II or M-theory and it would be nice to construct some of these explicitly.
On the other hand, it is of uttermost importance to address the issue of proving existence of the solutions, for the
choices of topological/equivariant data determined by the associated  extremal problems. Among the several
 extremal problems discussed in \cite{Martelli:2023oqk,Colombo:2023fhu}, the GK geometry setup \cite{Gauntlett:2007ts,Couzens:2018wnk}
  is perhaps the simplest one. We are confident
 that the ideas of equivariant localization put forward in
 \cite{BenettiGenolini:2023kxp,Martelli:2023oqk,BenettiGenolini:2023ndb,Colombo:2023fhu} will play a key role in tackling  these  problems, and more generally in unveiling novel aspects  of supergravity.

 Finally, holography implies that the solutions discussed in this paper, as well as all the previous ones in the recent surge of orbifold supergravity
 solutions,  must have a field theory counterpart in terms of localized partition functions
 on related rigidly supersymmetric backgrounds. Based on the findings of  \cite{Inglese:2023wky,Inglese:2023tyc}, we expect that this will be a fruitful direction of work in the near future.

\section*{Acknowledgments}

We thank  M.~Abreu, C.~Casagrande, L.~Cassia, E.~Colombo, E.~Martinengo, A.~Pittelli and A.~Zaffaroni for useful discussions and comments.
AF and DM are partially supported  by the INFN.

\appendix

\section{The AdS${} \times \spindle_1 \ltimes \spindle_2$ limits}
\label{app:spindle-limit}

The $D$-dimensional system presented in section~\ref{subsec:sugra-solutions} includes, as particular limits, the $\AdS_{D-4} \times \spindle_1 \ltimes \spindle_2$ solutions studied in~\cite{Faedo:2022rqx,Couzens:2022lvg}, for $D=6$, and in~\cite{Cheung:2022ilc}, when $D=7$. As we shall see, these backgrounds can be retrieved only if the (putative) supersymmetry condition $a g s_1 s_2 = \pm1$ is imposed, with the two charges~$\delta_i$ having the same sign.

\subsection{The AdS$_2 \times \spindle_1 \ltimes \spindle_2$ limit}
\label{subsect:6d-spindle-limit}

We shall now focus on $D=6$ and see how the $\AdS_2 \times \spindle_1 \ltimes \spindle_2$ backgrounds are recovered.
The first step is to perform the following substitutions
\begin{equation}
	y \mapsto \epsilon^2 y \,,  \quad  \phi \mapsto \epsilon^2 \phi \,,  \quad  a \mapsto \epsilon^2 a \,,  \quad  \Ny \mapsto \epsilon^6 \Ny \,,  \quad  \Nx \mapsto \epsilon^2 \Nx \,,  \quad  s_i \mapsto \frac{s_i}{\epsilon} \,,
\end{equation}
followed by the $\epsilon\to0^+$ limit. The underlying idea is to take $y\to0$, but keeping the ratio $y/a$ finite, preserving the supersymmetry condition and preventing many of the functions to become trivial.

A thorough analysis shows that the desired configurations can be obtained only when the two charges $\delta_i$ have the same sign. For the sake of simplicity, we shall restrict to $\delta_1,\delta_2 > 0$ and, in this case, introduce the new coordinates
\begin{equation} \label{limit-6d_coord}
	\tilde{y} = x \,,  \qquad  \tilde{x} = \frac{2y}{a} \,,  \qquad  \tilde{z} = \psi + \frac{\kk\,\phi}{2ag} \,,  \qquad  \tilde{\psi} = \frac{\kk\,\phi}{a} \,,
\end{equation}
with $\kk = \pm1$.
Imposing the supersymmetry condition $a g s_1 s_2 = -\kk$, the metric and the scalar fields reduce to the ones in (2.40) of~\cite{Faedo:2022rqx}, when $m=2g_c/3$ and with $q_i$ and $\mathtt{a}$ defined as
\begin{equation}
	q_i = -2\Nx s_i^2 \,,  \qquad  \mathtt{a} = \frac{8\Ny}{a^3} \,.
\end{equation}
In order to reconstruct the gauge potentials of~\cite{Faedo:2022rqx}, we first need to perform the gauge transformation $A_i \mapsto A_i - \dd\psi$ and, then, the correct expression follows. Computing the limit of the two-form $B$ we obtain
\begin{equation}
	B = \kk \, \frac{\mathtt{a} \tilde{y}}{2g} \, \vol{\AdS_2} \,.
\end{equation}
When $\kk=-1$, the origin of the opposite sign with respect to~\cite{Faedo:2022rqx} can be traced in a different orientation of the spacetime induced by the change of coordinates~\eqref{limit-6d_coord}.

Notice that the global analysis of the $\AdS_2 \times \spindle_1 \ltimes \spindle_2$ solutions implies that $q_1 q_2 < 0$ \cite{Faedo:2021nub}, whereas with our identifications $q_1$ and $q_2$ must have the same sign, unless an analytic continuation of these parameters is made.

\subsection{The AdS$_3 \times \spindle_1 \ltimes \spindle_2$ limit}

We now move to the seven-dimensional system and show how our solutions reduce to the $\AdS_3 \times \spindle_1 \ltimes \spindle_2$ backgrounds of~\cite{Cheung:2022ilc}, setting here $g=1/2$, \ie\ $g_c=1$, and imposing the supersymmetry condition.
We begin computing the limit $y\to0$, which, in this case, is parameterized as
\begin{equation}
	y \mapsto \epsilon^2 y \,,  \quad  \phi \mapsto \epsilon^2 \phi \,,  \quad  a \mapsto \epsilon^2 a \,,  \quad  \Ny \mapsto \epsilon^8 \Ny \,,  \quad  \Nx \mapsto \epsilon^2 \Nx \,,  \quad  s_i \mapsto \frac{s_i}{\epsilon} \,,
\end{equation}
with $\epsilon\to0^+$.
Once again, we restrict to positive charges~$\delta_i$ and impose $a g s_1 s_2 = -\kk$.
Introducing the new coordinates
\begin{equation} \label{limit-7d_coord}
	\tilde{y} = x^2 \,,  \qquad  \tilde{x} = \frac{9y^2}{4a^2} \,,  \qquad  \tilde{\phi} = \psi + \frac{\kk\,\phi}{3ag} \,,  \qquad  \tilde{\psi} = \frac{2\kk\,\phi}{a} \,,
\end{equation}
the metric, gauge potentials\footnote{Also in this case we had to perform the gauge transformation $A_i \mapsto A_i - \dd\psi$.} and scalar fields assume the expressions given in~(3.7) and~(3.8) of~\cite{Cheung:2022ilc}, where $q_i$ and $\mathtt{a}$ are defined as\footnote{Here, we relabelled the constant~$a$ in~\cite{Cheung:2022ilc} as~$\mathtt{a}$ in order to avoid confusion.}
\begin{equation}
	q_i = -2\Nx s_i^2 \,,  \qquad  \mathtt{a} = \frac{27\Ny}{2a^4} \,.
\end{equation}
After the limit and the change of coordinates, the three-form $B$ becomes
\begin{equation}
	B = \kk \, \frac{4 \mathtt{a} \tilde{y}}{27g} \, \vol{\AdS_3} + \frac{\mathtt{a}}{12\tilde{x}} \, \dd\tilde{\phi} \wedge \dd\tilde{\psi} \wedge \dd\tilde{y} \,.
\end{equation}
In order to compare this result with equation~(3.8) of~\cite{Cheung:2022ilc}, we need to dualize $B$ to the new three-form
\begin{equation}
	S_{(3)} \equiv (X_1 X_2)^2 \star\!H = -\frac{8\mathtt{a} \tilde{y}}{27} \, \vol{\AdS_3} + \kk \, \frac{\mathtt{a} \tilde{y} Q}{6g \tilde{x}^2 h_1 h_2} \, \dd\tilde{x} \wedge \dd\tilde{\psi} \wedge \dd\tilde{\phi} \,,
\end{equation}
where the function $Q(\tilde{y})$ is defined in equation~(2.5) of~\cite{Cheung:2022ilc}.
This expression agrees with the one in~\cite{Cheung:2022ilc} for $g=1/2$. Also here, when $\kk=-1$ the different sign is due to the opposite orientation of the spacetime induced by the change of coordinates~\eqref{limit-7d_coord}.

\section{Uplift of the solutions}
\label{app:uplifts}

In this appendix we uplift the six- and seven-dimensional NUT--$\AdS$ backgrounds presented in section~\ref{subsec:sugra-solutions} to solutions of massive type~IIA and 11d supergravity, respectively.  All the details of the truncation ans\"atze are discussed in~\cite{Faedo:2022rqx,Couzens:2022lvg} and in~\cite{Cheung:2022ilc}; here, we report only the relevant ingredients required for the computation of the entropy and the central charge of the two systems.

The higher-dimensional theories we consider are described by the equations of motion that descend from the massive type~IIA action, written in the string frame,%
\footnote{Here we use the shortcut $B_{(2)}^n$ to denote the wedge product of $B_{(2)}$ with itself $n$ times, divided by~$n!$.}
\begin{align}
	S_\mathrm{mIIA} = \frac{1}{16\pi G_{(10)}} \biggl\{ &\int \dd^{10}x \sqrt{-g} \Bigl[ \ee^{-2\Phi} \bigl( R + 4|\dd\Phi|^2 - \frac12|H_{(3)}|^2 \bigr) -\frac12\bigl( F_{(0)}^2 + |F_{(2)}|^2 + |F_{(4)}|^2 \bigl) \Bigr] \nonumber \\
	- \frac12 &\int \bigl( B_{(2)} \wedge \dd C_{(3)} \wedge \dd C_{(3)} + 2F_{(0)} B_{(2)}^3 \wedge \dd C_{(3)} + 6F_{(0)}^2 B_{(2)}^5 \bigr) \biggr\} \,,
\end{align}
and the 11d supergravity action
\begin{equation}
	S_\mathrm{11d} = \frac{1}{16\pi G_{(11)}} \biggl\{ \int \dd^{11}x \sqrt{-g} \Bigl[ R -\frac12|F_{(4)}|^2 \Bigr] + \frac16 \int \bigl( F_{(4)} \wedge F_{(4)} \wedge C_{(3)} \bigr) \bigg\} \,.
\end{equation}
In these conventions, the ten-dimensional field strengths are
\begin{equation}
	H_{(3)} = \dd B_{(2)} \,,  \quad  F_{(2)} = \dd C_{(1)} + F_{(0)} B_{(2)} \,,  \quad  F_{(4)} = \dd C_{(3)} - H_{(3)} \wedge C_{(1)} + \frac12 F_{(0)} B_{(2)} \wedge B_{(2)} \,,
\end{equation}
while the eleven-dimensional four-form flux is simply $F_{(4)} = \dd C_{(3)} $.

Many of the expressions we present in the rest of this appendix are similar in both massive type~IIA and 11d supergravity, therefore we shall write them in a unified fashion in order to avoid repetitive formulas.
The uplifted metrics, written in the string frame for massive type~IIA, read
\begin{align}\label{uplifted_metrics}
	\begin{split}
		\dd s_{D+4}^2 &= \lambda^2 \mu_0^{(D-7)/3} \Delta_H^{1/(D-4)} \bigl\{  (H_1 H_2)^{-1/(D-2)}\dd s_D^2 \\
		& + g_{c}^{-2} \Delta_H^{-1}\bigl[ \dd\mu_0^2 + H_1 \bigl(\dd\mu_1^2 + \mu_1^2 \sigma_1^2\bigr) + H_2 \bigl(\dd\mu_2^2 + \mu_2^2 \sigma_2^2\bigr) \bigr] \bigr\} \,,
	\end{split}
\end{align}
where $\dd s^2_D$ is the line element of $M_D$, the quadrilateral solutions of section~\ref{subsec:sugra-solutions}, the one-forms $\sigma_i \equiv \dd \phi_i - g_c A_i$ are built up from the $D$-dimensional gauge fields~\eqref{nut_gauge} and $\lambda$ is a strictly positive constant which realizes the scaling symmetry of the supergravity theories (see section 3 of~\cite{Faedo:2022rqx} and 3.2 of~\cite{Cheung:2022ilc}). While in 11d supergravity its introduction is not necessary, in massive type~IIA the additional parameter $\lambda$ plays a crucial role for the correct quantization of the ten-dimensional solutions.
The coordinates $\mu_a$, with $a=0,1,2$, satisfy the constraint $\sum \mu_a ^2=1$ and can be conveniently parameterized as
\begin{equation}
	\mu_0 = \sin\xi \,,  \qquad  \mu_1 = \cos\xi \sin\eta \,,  \qquad  \mu_2 = \cos\xi \cos\eta \,.
\end{equation}
The angular coordinates~$\phi_1$, $\phi_2$ have canonical $2\pi$ periodicities and $\eta \in [0,\pi/2]$.
The presence of the overall $\mu_0$ factor in the ten-dimensional metric implies $\xi \in (0,\pi/2]$ and the line element~\eqref{uplifted_metrics} describes, at each point of $M_6$, a four-dimensional hemisphere, that we denote as~$\hemi$. On the contrary, in eleven dimensions $\mu_0$ is not present, therefore $\xi\in[-\pi/2,\pi/2]$ and the dimensional reduction is performed on a (squashed) four-sphere, that we continue to denote as~$\hemi$.
For convenience, we defined the function
\begin{equation}
	\Delta_H = H_1 H_2 \, \mu_0^2 + H_2 \, \mu_1^2 + H_1 \, \mu_2^2 \,.
\end{equation}
The four-form flux, common to both higher-dimensional supergravities, can be written as
\begin{align} \label{uplifted_F4}
	F_{(4)} =& -\frac{\lambda^{(2D-11)}}{g_c^3} \frac{H_1 H_2\,U_H}{\Delta_H^2} \frac{\mu_1 \mu_2}{\mu_0^{(D-4)/3}} \, \dd\mu_1 \wedge \dd\mu_2 \wedge \sigma_1 \wedge \sigma_2 \\
	& - 2 (D-5) \lambda^{(2D-11)} \mu_0^{(10-D)/3} \frac{g s_1 s_2}{\Xi a^2}\frac{\Ny x^{D-3} - \Nx y^{D-3}}{(x y)^{D-4}} \, \dd y \wedge \dd \phi \wedge \dd x \wedge \dd \psi + \ldots \,, \nonumber
\end{align}
where we defined the function
\begin{equation}
	U_H = 2 \bigl[ (1 - H_1)(1 - H_2) \mu_0^2 - 1 \bigr] - \frac{10-D}{3} \Delta_H \,.
\end{equation}
The dots in~\eqref{uplifted_F4} represent terms of $F_{(4)}$ omitted because they do not contribute to the flux quantization through $\hemi$ or $\Morb_4$. Finally, the ten-dimensional dilaton and Romans mass $F_{(0)}$ read
\begin{equation}
	\ee^\Phi = \lambda^2 \mu_0^{-5/6} \Delta_H^{1/4} \,,  \qquad
	F_{(0)} = \frac{m}{\lambda^3} = \frac{2 g_c}{3 \lambda^3} \,.
\end{equation}

In order to compute the entropy and the central charge of the uplifted solutions, we first need to write the higher-dimensional metric in the form
\begin{equation}
	\dd s_{D+4}^2 = \ee^{2A} \bigl( \dd s_{\AdS_{D-4}}^2 + \dd s_{M_8}^2 \bigr) \,,
\end{equation}
where $\dd s^2_{M_8}$ is the line element of the internal fibred space $\hemi \hookrightarrow M_8 \to \Morb_4$.
The entropy and the central charge can then be read from the $(D-4)$-dimensional effective Newton constant $G_{(D-4)}$ as\footnote{We are using conventions in which $G_{(11)}=(2\pi\ell_{s}) G_{(10)}$, that is the same to say that the eleven-dimensional Planck length is equal to $\ell_s$. }
\begin{equation} \label{entropy_and_central_def}
	S = \frac{1}{4G_{(2)}} = \frac{8\pi^2}{(2\pi\ell_s)^8} \! \int \ee^{8A-2\Phi} \, \vol{M_8} \,,  \qquad
	c = \frac{3}{2G_{(3)}} = \frac{48\pi^2}{(2\pi\ell_s)^9} \! \int \ee^{9A} \, \vol{M_8} \,.
\end{equation}
Plugging the relevant quantities, we obtain
\begin{align}
	\label{entropy}
	S &= \frac{1}{(2\pi\ell_s)^8} \frac{48\pi^6 \lambda^4}{5|\Xi| a^2 g_c^4} \frac{\Delta\psi}{2\pi} \frac{\Delta\phi}{2\pi} \bigl| (\XB^3 - \XA^3) (\YB - \YA) - (\XB - \XA) (\YB^3 - \YA^3) \bigr| \,, \\
	\label{central-charge}
	c &= \frac{1}{(2\pi\ell_s)^9} \frac{64\pi^6 \lambda^9}{|\Xi a^3| g_c^4} \frac{\Delta\psi}{2\pi} \frac{\Delta\phi}{2\pi} \bigl| (\XB^4 - \XA^4) (\YB^2 - \YA^2) - (\XB^2 - \XA^2) (\YB^4 - \YA^4) \bigr| \,.
\end{align}
To compare these results with the quantities found on the field theory side, we proceed to quantize the fluxes. In particular, the four-form flux yields
\begin{equation} \label{quantization_F4}
	\frac{1}{(2\pi\ell_s)^3} \int_\hemi F_{(4)} = N \in \NN \,,
	\qquad  \frac{1}{(2\pi\ell_s)^3} \int_{\Morb_4} F_{(4)} = K \in \NN \,,
\end{equation}
where $N$ can be interpreted as the number of D4 and M5-branes wrapped over $\Morb_4$, in ten and eleven dimensions respectively. In massive type~IIA supergravity one has to impose the additional condition
\begin{equation} \label{quantization_F0}
	(2\pi\ell_s) F_{(0)} = n_0 \in \NN \,,
\end{equation}
closely related to the presence of D8-branes. Indeed, our background has a boundary at $\xi=0$, where $\mu_0 =0$ and the warp factor is singular, which corresponds to the location of an O8-plane and $N_f =8-n_0 $ coincident D8-branes. Computing the integral
\begin{equation}
	\int_{\hemi} \frac{H_1 H_2 U_H}{\Delta_H^2} \frac{\mu_1 \mu_2}{\mu_0^{(D-4)/3}} \, \dd\mu_1 \wedge \dd\mu_2 \wedge \dd\phi_1 \wedge \dd\phi_2 = -(D-3)(D-5) \pi^2 \,,
\end{equation}
condition~\eqref{quantization_F4}, together with~\eqref{quantization_F0} in the case of massive type~IIA, can be solved to give $g_c$ and $\lambda$ in terms of $N$ (and $n_0$)
\begin{alignat}{3}
	& D = 6 :  \qquad\quad  & g_c^8 &= \frac{1}{(2\pi\ell_s)^8} \frac{18\pi^6}{N^3 n_0} \,,  \qquad  & \lambda^8 &= \frac{8\pi^2}{9N n_0^3} \,, \\
	& D = 7 :  \qquad\quad  & \frac{g_c^3}{\lambda^3} &= \frac{1}{(2\pi\ell_s)^3} \frac{8\pi^2}{N} \,. &&
\end{alignat}
These expressions are the same as for other known six- and seven-dimensional solutions (\cf~\cite{Faedo:2021nub,Faedo:2022rqx} and~\cite{Ferrero:2021wvk,Cheung:2022ilc}, respectively, and references therein).
In terms of the number of branes, the entropy~\eqref{entropy} and the central charge~\eqref{central-charge} take the form
\begin{align}
	\label{entropy_branes}
	S &= \frac{9\sqrt2 \pi N^{5/2}}{5\sqrt{8-N_f}} \frac{g^4}{|\Xi| a^2} \frac{\Delta\psi}{2\pi} \frac{\Delta\phi}{2\pi} \bigl| (\XB^3 - \XA^3) (\YB - \YA) - (\XB - \XA) (\YB^3 - \YA^3) \bigr| \,, \\
	\label{central-charge_branes}
	c &= 4N^3 \frac{g^5}{|\Xi a^3|} \frac{\Delta\psi}{2\pi} \frac{\Delta\phi}{2\pi} \bigl| (\XB^4 - \XA^4) (\YB^2 - \YA^2) - (\XB^2 - \XA^2) (\YB^4 - \YA^4) \bigr| \,.
\end{align}

As last step in the quantization of the fluxes, we consider the second constraint coming from~\eqref{quantization_F4}. The integration is performed along a representative of $\Morb_4$, which we take to be at $\xi=\pi/2$, namely the pole of the hemisphere $\hemi$, in $D=6$, or at one of the poles of the sphere $\hemi$, in $D=7$. In terms of the physical parameters, the total flux across $\Morb_4$ reads
\begin{equation}
	K = N \frac{4g_c^3 g s_1 s_2}{(D-3)(D-5) \Xi a^2} \frac{\Delta \psi}{2\pi} \frac{\Delta\phi}{2\pi} \bigl[ \Ny (\YB^{5-D} - \YA^{5-D}) (\XB^2 - \XA^2) - \Nx (\XB^{5-D} - \XA^{5-D}) (\YB^2 - \YA^2) \bigr] .
\end{equation}
Although the expression seems rather involved, making use of the results of subsection~\ref{subsec:quantization} it is possible to write the total flux $K$ in the following remarkable form
\begin{equation} \label{quant_K}
	K = \sid \, N \, \frac{d_{1,2} d_{2,3} (\flq_R^2)^2 - d_{3,4} d_{4,1} (\flq_R^4)^2}{4 (D-5) d_{1,3}}
	= \sid \, N \, \frac{d_{2,3} d_{3,4} (\flq_R^3)^2 - d_{4,1} d_{1,2} (\flq_R^1)^2}{4 (D-5) d_{2,4}} \,.
\end{equation}
When the physical parameters are properly quantized, all the elements appearing in \eqref{quant_K} are rational, thus $N$ can be tuned appropriately in order to make $K$ integer.
It is noteworthy that the Diophantine constraints play a crucial role in writing $K$ in terms of rational quantities only, and, in particular, in making the square roots in $\XBA$ and $\YBA$---see \eqref{6d-w} and \eqref{7d-w}---disappear from the final expression.

\section{Supersymmetry of the $D=6$ solutions with equal charges}
\label{app:6d_KSE}

In this appendix we check explicitly that the equal charges solution of section~\ref{subsec:equal-charges} is supersymmetric, \ie\ we construct a Killing spinor that satisfies the following Killing spinor equations~\cite{Faedo:2022rqx}
\begingroup
\allowdisplaybreaks
\begin{align}
	\begin{split} \label{KSE_grav}
		\mathcal{D}_\mu \epsilon^A + \frac18 \bigl[ g_c (X_1 + X_2) + m X_0 \bigr] \Gamma_\mu \epsilon^A & \\
		+ \frac{1}{32} \bigl[ m (X_1 X_2)^{-1/2} B_{\nu\lambda} \Gamma_7 \delta^A_B + \ii \bigl( X_1^{-1} F_1 + X_2^{-1} F_2 \bigr)_{\nu\lambda} (\sigma^3)^A_{\,\ B} \bigr] \bigl(\Gamma_\mu^{\ \nu\lambda} - 6\delta_\mu^\nu \, \Gamma^\lambda \bigr) \epsilon^B & \\
		- \frac{1}{96} (X_1 X_2) H_{\nu\lambda\rho} \Gamma_7 \bigl(\Gamma_\mu^{\ \nu\lambda\rho} - 3\delta_\mu^\nu \, \Gamma^{\lambda\rho} \bigr) \epsilon^A &= 0 \,,
	\end{split} \\[0.5em]
	\begin{split} \label{KSE_dila}
		\frac14 \bigl( X_1^{-1} \partial_\mu X_1 + X_2^{-1} \partial_\mu X_2 \bigr) \Gamma^\mu \epsilon^A - \frac18 \bigl[ g_c (X_1 + X_2) - 3m X_0 \bigr] \epsilon^A & \\
		+ \frac{1}{32} \bigl[ m (X_1 X_2)^{-1/2} B_{\mu\nu} \Gamma_7 \delta^A_B - \ii \bigl( X_1^{-1} F_1 + X_2^{-1} F_2 \bigr)_{\mu\nu} (\sigma^3)^A_{\,\ B} \bigr] \Gamma^{\mu\nu} \epsilon^B & \\
		+ \frac{1}{96} (X_1 X_2) H_{\mu\nu\lambda} \Gamma_7 \Gamma^{\mu\nu\lambda} \epsilon^A &= 0 \,,
	\end{split} \\[0.5em]
	\begin{split} \label{KSE_gauge}
		\frac12 \bigl( X_1^{-1} \partial_\mu X_1 - X_2^{-1} \partial_\mu X_2 \bigr) \Gamma^\mu (\sigma^3)^A_{\,\ B} \epsilon^B - g_c (X_1 - X_2) (\sigma^3)^A_{\,\ B} \epsilon^B & \\
		- \frac{\ii}{4} \bigl( X_1^{-1} F_1 - X_2^{-1} F_2 \bigr)_{\mu\nu} \Gamma^{\mu\nu} \epsilon^A &= 0 \,,
	\end{split}
\end{align}
\endgroup
where
\begin{equation} \label{cov-D}
	\mathcal{D}_\mu \epsilon^A \equiv \partial_\mu \epsilon^A + \frac14 \, \omega_\mu^{\ ab} \Gamma_{ab} \epsilon^A - \frac{\ii}{2} \, g_c (A_1 + A_2)_\mu (\sigma^3)^A_{\,\ B} \epsilon^B \,.
\end{equation}
These follow from setting to zero the supersymmetry variations of the fermionic fields of the theory with three vector multiplets \cite{DAuria:2000afl} that do not vanish automatically in the sub-truncation that we are considering. Here $(\sigma^3)^A_{\,\ B}$ is the usual third Pauli matrix, $\{\Gamma_a,\Gamma_b\}=2\eta_{ab}$ and $\Gamma_7 \equiv \Gamma^0 \Gamma^1 \Gamma^2 \Gamma^3 \Gamma^4 \Gamma^5$. The $SU(2)$ indices $A,B$ are raised and lowered as $\epsilon^A=\varepsilon^{AB}\epsilon_B$ and $\epsilon_A=\epsilon^B\varepsilon_{BA}$, where $\varepsilon_{AB}=-\varepsilon_{BA}$ and its inverse matrix $\varepsilon^{AB}$ is defined such that $\varepsilon^{AB}\varepsilon_{AC}=\delta^B_C$. The supersymmetry parameter $\epsilon^A$ is an eight-component symplectic-Majorana spinor, hence it satisfies the condition
\begin{equation} \label{symp-Maj}
	\varepsilon^{AB} \epsilon_B^* = \mathcal{B}_6 \epsilon_A \,,
\end{equation}
where $\mathcal{B}_6$ is related to the six-dimensional charge conjugation matrix $\mathcal{C}_6$ by $\mathcal{B}_6=-\ii\,\mathcal{C}_6\Gamma^0$.

When the charges are equal, equation~\eqref{KSE_gauge} automatically vanishes. To specialize the other equations to the solution of section~\ref{subsec:equal-charges}  we employ the following six-dimensional orthonormal frame
\begin{equation}
	e^{\hat{a}} = H^{1/4} \, \frac{x y}{a} \, \hat{e}^{\hat{a}} \,, \qquad  e^{\hat{\imath}+1}=H^{1/4} \, \hat{e}^{\hat{\imath}}\,,
\end{equation}
where $\hat{e}^{\hat{a}}$, $\hat{a}=0,1$, is the zweibein on $\AdS_2$, $\hat{e}^{\hat{\imath}}$, $\hat{\imath}=1,\ldots,4$, is the vierbein on $\Morb_4$ introduced in~\eqref{M4-frame} and the coordinates are denoted as $x^{\hat{\mu}}$ and $x^{\hat{\alpha}}$, respectively.
Then, equation~\eqref{KSE_dila} reduces to the constraint
\begin{equation} \label{rel0}
	\sqrt{\frac{\Delta_y}{y^2 - x^2}} \, \Gamma^3 \epsilon^A + \sqrt{\frac{\Delta_x}{x^2 - y^2}} \, \Gamma^5 \epsilon^A + \ii \, \frac{c \tilde{c}}{s} \, (\sigma^3)^A_{\,\ B} \epsilon^B - g H \bigl( y \, \Gamma^{23} + x \, \Gamma^{45} \bigr) \epsilon^A = 0 \,,
\end{equation}
where, we recall, $g=m=2g_c/3$.
Writing this algebraic equation schematically as $\mathcal{M}\,\epsilon = 0$, we need to impose $\det(\mathcal{M})=0$ in order to have non-trivial solutions, \ie\ $\epsilon\neq0$. This necessary condition is satisfied for any value of $x$ and $y$ if and only if $(a g s^2)^2=1$, which agrees with \cite{Chow:2011fh}.%
\footnote{Notice that a factor $c_2 ^2 \tilde{c}_1^2$ is missing in (4.8) of~\cite{Chow:2011fh}, which is asymmetric under $1\leftrightarrow2$. }
From now on, we shall impose $ags^2 = -\kk$, with $\kk=\pm 1$.
Multiplying~\eqref{rel0} by its complex conjugate we obtain
\begin{equation} \label{rel1}
	\frac{1}{y} \sqrt{\frac{\Delta_y}{y^2 - x^2}} \, \Gamma^2 \epsilon^A + \frac{1}{x} \sqrt{\frac{\Delta_x}{x^2 - y^2}} \, \Gamma^4 \epsilon^A + g H \epsilon^A + \biggl[ g \bigl( y \, \partial_x H + x \, \partial_y H \bigr) - \frac{\kk\,a}{x y } \biggr] \Gamma^{2345} \epsilon^A = 0 \,,
\end{equation}
and, thanks to this equation, the $\AdS_2$ components of~\eqref{KSE_grav} can be written in a simple fashion as
\begin{equation} \label{KSE_ads}
	\partial_{\hat{\mu}} \epsilon^A + \frac14 \, \hat{\omega}_{\hat{\mu}}^{\ \hat{a}\hat{b}}\, \Gamma_{\hat{a}\hat{b}}\epsilon^A + \frac{\kk}{2} \, \hat{e}^{\hat{c}}_{\hat{\mu}} \,\Gamma_{\hat{c}}^{\ 2345} \epsilon^A = 0 \,,
\end{equation}
where $\hat{\omega}_{\hat{\mu}}^{\ \hat{a}\hat{b}}$ is the spin connection on $\AdS_2$.
The components along the coordinates $y$ and $x$ read
\begin{align}
	\begin{split} \label{KSE_y}
		\partial_y \epsilon^A + \frac38 H^{-1} \partial_y H \, \epsilon^A + \frac12 \sqrt{\frac{y^2 - x^2}{\Delta_y}} \, \Gamma^2 \Biggl[ g \, \partial_y (y H) - g x \, \partial_y H \, \Gamma^{2345} & \\
		+ \sqrt{\frac{\Delta_x}{x^2 - y^2}} \, \frac{1}{x^2 - y^2} \bigl( y \, \Gamma^{235} + x \, \Gamma^4 \bigr) \Biggr] \epsilon^A &= 0 \,,
	\end{split} \\
	\begin{split} \label{KSE_x}
		\partial_x \epsilon^A + \frac38 H^{-1} \partial_x H \, \epsilon^A + \frac12 \sqrt{\frac{x^2 - y^2}{\Delta_x}} \, \Gamma^4 \Biggl[ g \, \partial_x (x H) - g y \, \partial_x H \, \Gamma^{2345} & \\
		- \sqrt{\frac{\Delta_y}{y^2 - x^2}} \, \frac{1}{x^2 - y^2} \bigl( y \, \Gamma^2 + x \, \Gamma^{345} \bigr) \Biggr] \epsilon^A &= 0 \,,
	\end{split}
\end{align}
whilst the remaining components along the angular directions $\phi$ and $\psi$ are, respectively,
\small
\begin{align} \label{KSE_phi}
	\partial_\phi \epsilon^A &= \frac{3\ii g}{2} \biggl[\alpha+\frac{2sc\tilde{c}(\Nx y\widetilde{V}_y - \Ny x \widetilde{V}_x)}{\Xi \, x y (x^2 - y^2) H}\biggr] (\sigma^3)^A_{\,\ B} \epsilon^B
	+ \frac{1}{2\Xi \, H}  \Biggl\{ \sqrt{\frac{-\Delta_x \, \Delta_y}{(x^2 - y^2)^2}} \, \frac{H}{a^2} \bigl( y \, \Gamma^{25} - x \, \Gamma^{34} \bigr) \nonumber \\
	& + g \widetilde{V}_x \sqrt{\frac{\Delta_y}{y^2 - x^2}} \bigl[ \partial_y (y H) \, \Gamma^3 + x \, \partial_y H \, \Gamma^{245} \bigr]
	+ g \widetilde{V}_y \sqrt{\frac{\Delta_x}{x^2 - y^2}} \bigl[ \partial_x (x H) \, \Gamma^5 + y \, \partial_x H \, \Gamma^{234} \bigr] \nonumber \\
	& - \biggl[ \frac{x (\Delta_y \widetilde{V}_x - \Delta_x \widetilde{V}_y)}{(x^2 - y^2)^2} + \frac{\Delta_x' \widetilde{V}_y}{2 (x^2 - y^2)} \biggr] \Gamma^{45}
	+ \biggl[ \frac{y (\Delta_y \widetilde{V}_x - \Delta_x \widetilde{V}_y)}{(x^2 - y^2)^2} + \frac{\Delta_y' \widetilde{V}_x}{2 (x^2 - y^2)} \biggr] \Gamma^{23} \Biggr\} \, \epsilon^A \,,
\end{align}
\begin{align} \label{KSE_psi}
	\partial_\psi \epsilon^A &=	\frac{3\ii g}{2} \biggl[\beta-\frac{2sc\tilde{c}(\Nx y{V}_y - \Ny x {V}_x)}{\Xi \, x y (x^2 - y^2) H}\biggr] (\sigma^3)^A_{\,\ B} \epsilon^B
	- \frac{1}{2\Xi \, H} \Biggl\{ \sqrt{\frac{-\Delta_x \, \Delta_y}{(x^2 - y^2)^2}} \, g^2 H \bigl( y \, \Gamma^{25} - x \, \Gamma^{34} \bigr) \nonumber \\
	& + g V_x \sqrt{\frac{\Delta_y}{y^2 - x^2}} \bigl[ \partial_y (y H) \, \Gamma^3 + x \, \partial_y H \, \Gamma^{245} \bigr]
	+ g V_y \sqrt{\frac{\Delta_x}{x^2 - y^2}} \bigl[ \partial_x (x H) \, \Gamma^5 + y \, \partial_x H \, \Gamma^{234} \bigr] \nonumber \\
	& - \biggl[ \frac{x (\Delta_y V_x - \Delta_x V_y)}{(x^2 - y^2)^2} + \frac{\Delta_x' V_y}{2 (x^2 - y^2)} \biggr] \Gamma^{45}
	+ \biggl[ \frac{y (\Delta_y V_x - \Delta_x V_y)}{(x^2 - y^2)^2} + \frac{\Delta_y' V_x}{2 (x^2 - y^2)} \biggr] \Gamma^{23} \Biggr\} \, \epsilon^A \,.
\end{align}
\normalsize
Here, we performed the gauge transformation $A_i \mapsto A_i + \alpha\,\dd\phi + \beta\,\dd\psi$, with $\alpha$ and $\beta$ real constants. As we shall see, the latter can be tuned appropriately to make the Killing spinor independent of the angular coordinates.

With all the equations made explicit, we can now proceed to decompose the six-dimensional gamma matrices. In particular, we adopt the following decomposition
\begin{equation} \label{gamma-deco}
	\Gamma^{\hat{a}} = \beta^{\hat{a}} \otimes \gamma_* \,,  \qquad\quad  \Gamma^{\hat{\imath}+1} = I_2 \otimes \gamma^{\hat{\imath}} \,,
\end{equation}
where $\beta^{\hat{a}}$ are the (Lorentzian) gamma matrices in $D=2$ and $\gamma^{\hat{\imath}}$ are the (Euclidean) gamma matrices in $D=4$. The related chiral matrices are $\beta_* = -\beta^0 \beta^1$ and $\gamma_* = -\gamma^1 \gamma^2 \gamma^3 \gamma^4$, respectively.
As a consequence, the six-dimensional matrices $\mathcal{B}_6$ and $\Gamma_*$ decompose as\footnote{$\mathcal{B}_4$ is related to the four-dimensional charge conjugation matrix $\mathcal{C}_4$ by $\mathcal{B}_4 = (\mathcal{C}_4)^t$.}
\begin{equation}
	\mathcal{B}_6 = \mathcal{B}_2 \otimes (\mathcal{B}_4 \gamma_*) \,,  \qquad\quad  \Gamma_* = \beta_* \otimes \gamma_* \,.
\end{equation}
The ansatz for the six-dimensional Killing spinor is
\begin{equation} \label{killing-deco}
	\epsilon^A = \vartheta \otimes \zeta^A \,,
\end{equation}
where $\vartheta=\vartheta(x^{\hat{\mu}})$ is a Majorana spinor on~$\AdS_2$, hence $\vartheta^* = \mathcal{B}_2\vartheta$, and $\zeta^A=\zeta^A(x^{\hat{\alpha}})$ are two four-component spinors defined on~$\Morb_4$.
In this way, equation~\eqref{KSE_ads} reduces to
\begin{equation}
	\biggl( \hat{\nabla}_{\hat{\mu}} \vartheta - \frac{\kk}{2} \, \beta_{\hat{\mu}} \vartheta \biggr) \otimes \zeta^A = 0  \qquad  \implies  \qquad
	\hat{\nabla}_{\hat{\mu}} \vartheta = \frac{\kk}{2} \, \beta_{\hat{\mu}} \vartheta \,,
\end{equation}
which implies that $\vartheta$ must be a Killing spinor on~$\AdS_2$.

Employing the decompositions~\eqref{gamma-deco} and~\eqref{killing-deco}, equation~\eqref{rel0} becomes
\begin{equation} \label{KSE4_constr}
	\sqrt{\frac{\Delta_y}{y^2 - x^2}} \, \gamma^2 \zeta^A + \sqrt{\frac{\Delta_x}{x^2 - y^2}} \, \gamma^4 \zeta^A + \ii \, \frac{c \tilde{c}}{s} \, (\sigma^3)^A_{\,\ B} \zeta^B - g H \bigl( y \, \gamma^{12} + x \, \gamma^{34} \bigr) \zeta^A = 0 \,.
\end{equation}
Equations~\eqref{KSE_y} and~\eqref{KSE_x} boil down to
\begin{align}
	\begin{split} \label{KSE4_y}
		\partial_y \bigl(H^{3/8} \zeta^A \bigr) + \frac12 \sqrt{\frac{y^2 - x^2}{\Delta_y}} \, \gamma^1 \Biggl[ g \, \partial_y (y H) + g x \, \partial_y H \, \gamma_* & \\
		+ \sqrt{\frac{\Delta_x}{x^2 - y^2}} \, \frac{1}{x^2 - y^2} \bigl( y \, \gamma^{124} + x \, \gamma^3 \bigr) \Biggr] \bigl(H^{3/8} \zeta^A \bigr) &= 0 \,,
	\end{split} \\
	\begin{split} \label{KSE4_x}
		\partial_x \bigl(H^{3/8} \zeta^A \bigr) + \frac12 \sqrt{\frac{x^2 - y^2}{\Delta_x}} \, \gamma^3 \Biggl[ g \, \partial_x (x H) + g y \, \partial_x H \, \gamma_* & \\
		- \sqrt{\frac{\Delta_y}{y^2 - x^2}} \, \frac{1}{x^2 - y^2} \bigl( y \, \gamma^1 + x \, \gamma^{234} \bigr) \Biggr] \bigl(H^{3/8} \zeta^A \bigr) &= 0 \,,
	\end{split}
\end{align}
while~\eqref{KSE_phi} and~\eqref{KSE_psi} give
\small
\begin{align} \label{KSE4_phi}
	\partial_\phi \zeta^A &= \frac{3\ii g}{2} \biggl[\alpha + \frac{2sc\tilde{c}(\Nx y\widetilde{V}_y - \Ny x \widetilde{V}_x)}{\Xi \, x y (x^2 - y^2) H}\biggr] (\sigma^3)^A_{\,\ B} \zeta^B
		+ \frac{1}{2\Xi \, H}  \Biggl\{ \sqrt{\frac{-\Delta_x \, \Delta_y}{(x^2 - y^2)^2}}
	\, \frac{H}{a^2} \bigl( y \, \gamma^{14} - x \, \gamma^{23} \bigr) \nonumber \\
	& + g \widetilde{V}_x \sqrt{\frac{\Delta_y}{y^2 - x^2}} \bigl[ \partial_y (y H) \, \gamma^2 + x \, \partial_y H \, \gamma^{134} \bigr]
	+ g \widetilde{V}_y \sqrt{\frac{\Delta_x}{x^2 - y^2}} \bigl[ \partial_x (x H) \, \gamma^4 + y \, \partial_x H \, \gamma^{123} \bigr] \nonumber \\
	& - \biggl[ \frac{x (\Delta_y \widetilde{V}_x - \Delta_x \widetilde{V}_y)}{(x^2 - y^2)^2} + \frac{\Delta_x' \widetilde{V}_y}{2 (x^2 - y^2)} \biggr] \gamma^{34}
	+ \biggl[ \frac{y (\Delta_y \widetilde{V}_x - \Delta_x \widetilde{V}_y)}{(x^2 - y^2)^2} + \frac{\Delta_y' \widetilde{V}_x}{2 (x^2 - y^2)} \biggr] \gamma^{12} \Biggr\} \zeta^A \,,
\end{align}
\begin{align} \label{KSE4_psi}
	\partial_\psi \zeta^A &= \frac{3\ii g}{2} \biggl[\beta - \frac{2sc\tilde{c}(\Nx y{V}_y - \Ny x {V}_x)}{\Xi \, x y (x^2 - y^2) H}\biggr] (\sigma^3)^A_{\,\ B} \zeta^B
	- \frac{1}{2\Xi \, H} \Biggl\{ \sqrt{\frac{-\Delta_x \, \Delta_y}{(x^2 - y^2)^2}} \, g^2 H \bigl( y \, \gamma^{14} - x \, \gamma^{23} \bigr) \nonumber \\
	& + g V_x \sqrt{\frac{\Delta_y}{y^2 - x^2}} \bigl[ \partial_y (y H) \, \gamma^2 + x \, \partial_y H \, \gamma^{134} \bigr]
	+ g V_y \sqrt{\frac{\Delta_x}{x^2 - y^2}} \bigl[ \partial_x (x H) \, \gamma^4 + y \, \partial_x H \, \gamma^{123} \bigr] \nonumber \\
	& - \biggl[ \frac{x (\Delta_y V_x - \Delta_x V_y)}{(x^2 - y^2)^2} + \frac{\Delta_x' V_y}{2 (x^2 - y^2)} \biggr] \gamma^{34}
	+ \biggl[ \frac{y (\Delta_y V_x - \Delta_x V_y)}{(x^2 - y^2)^2} + \frac{\Delta_y' V_x}{2 (x^2 - y^2)} \biggr] \gamma^{12} \Biggr\} \zeta^A \,.
\end{align}
\normalsize

From now on, in order to solve the Killing spinor equations~\eqref{KSE4_constr}--\eqref{KSE4_psi} we shall employ an explicit representation for the four-dimensional gamma matrices $\gamma^{\hat{\imath}}$, namely
\begin{equation} \label{gamma-rep}
	\gamma^1 = \sigma^1 \otimes \sigma^3 \,,  \qquad  \gamma^2 = \sigma^2 \otimes \sigma^3 \,,  \qquad
	\gamma^3 = I_2 \otimes \sigma^1 \,,  \qquad  \gamma^4 = I_2 \otimes \sigma^2 \,.
\end{equation}
In this representation we have
\begin{equation}
	\mathcal{B}_4 = -\sigma^1 \otimes \sigma^2 \,,  \qquad\quad  \gamma_* = \sigma^3 \otimes \sigma^3 \,.
\end{equation}
Since the symplectic-Majorana condition \eqref{symp-Maj} relates the two Killing spinors $\epsilon^A$ as
\begin{equation} \label{symp-Maj4}
	\epsilon^2 = (\mathcal{B}_6 \epsilon^1)^* = \vartheta \otimes (\mathcal{B}_4 \gamma_* \zeta^1)^* \,,
\end{equation}
we can focus uniquely on the spinor $\zeta^1$. Writing it as
\begin{equation} \label{zeta1}
	\zeta^1 = H^{-3/8} \bigl( \chi_1, \chi_2, \chi_3, \chi_4 \bigr) \,,
\end{equation}
with $\chi_i=\chi_i(y,\phi,x,\psi)$ complex functions of the coordinates on~$\Morb_4$, equation~\eqref{KSE4_constr} is solved by\footnote{Here and in what follows we assume $\delta>0$. When $\delta<0$ it is sufficient to exchange $\Delta_\bullet^+ \leftrightarrow \Delta_\bullet^-$.}
\begin{equation} \label{chis}
	\chi_1 = -\sqrt{\frac{\Delta_y}{y^2 - x^2}} \, \frac{y - x}{\Delta_y^-} \, \chi_3 \,,  \quad
	\chi_2 = -\sqrt{-\frac{\Delta_y}{\Delta_x}} \, \frac{\Delta_x^-}{\Delta_y^-} \, \chi_3 \,,  \quad
	\chi_4 = -\sqrt{\frac{x^2 - y^2}{\Delta_x}} \, \frac{\Delta_x^-}{x + y} \, \chi_3 \,,
\end{equation}
where the functions $\Delta^{\pm}_{\bullet}$ have been introduced in~\eqref{Delta+-} and satisfy $\Delta_\bullet = \Delta_\bullet^+ \, \Delta_\bullet^-$. In terms of the unique unknown function $\chi_3$, equations~\eqref{KSE4_y} and~\eqref{KSE4_x} read
\begin{equation}
	\partial_y \chi_3 = \frac12 \, \partial_y \biggl[ \log\biggl| \frac{\Delta_y^-}{y - x} \biggr| \biggr] \chi_3 \,, \qquad
	\partial_x \chi_3 = \frac12 \, \partial_x \biggl[ \log\biggl| \frac{\Delta_x^-}{x- y} \biggr| \biggr] \chi_3  \,,
\end{equation}
while~\eqref{KSE4_phi} and~\eqref{KSE4_psi} boil down respectively to
\begin{equation}
	\partial_\phi \chi_3 = \frac{\ii g}{2} \Bigl(3\alpha  - \frac{\kk}{ag}\Bigr) \, \chi_3 \,, 	 \qquad
	\partial_\psi \chi_3 = \frac{\ii g}{2} (3\beta + 1) \, \chi_3 \,.
\end{equation}
Tuning the gauge parameters to
\begin{equation}
	\alpha = \frac{\kk}{3ag} = -\frac{s^2}{3} \,,  \qquad  \beta = -\frac13 \,,
\end{equation}
the system can be easily solved, finding an angular-independent Killing spinor with components given by
\begin{equation}
	\begin{aligned}
		\chi_1 &= -\xi \, \biggl( \frac{\Delta_x^+ \Delta_y^+}{y + x} \biggr)^{1/2} \,,  \qquad
		& \chi_2 &= \xi \, \biggl( \frac{\Delta_x^- \Delta_y^+}{x - y} \biggr)^{1/2} \,, \\
		\chi_3 &= \xi \, \biggl( \frac{\Delta_x^+ \Delta_y^-}{y - x} \biggr)^{1/2} \,,  \qquad
		& \chi_4 &= -\xi \, \sign(x-y) \biggl( -\frac{\Delta_x^- \Delta_y^-}{x+y} \biggr)^{1/2} \,,
	\end{aligned}
\end{equation}
with $\xi$ a complex constant. From the symplectic-Majorana condition~\eqref{symp-Maj4}, we obtain
\begin{equation}
	\zeta^2 = (\mathcal{B}_4 \gamma_* \zeta^1)^* = \ii \, H^{-3/8} \bigl( -\chi_4^*, -\chi_3^*, \chi_2^*, \chi_1^* \bigr) \,,
\end{equation}
which, as it can be shown explicitly, satisfies all the related Killing spinor equations.

We can now count the amount of supersymmetry preserved by our $\AdS_2\times\Morb_4$ background. Since $\vartheta$ is a two-dimensional Majorana spinor, it has two real independent degrees of freedom. Similarly, $\zeta^A$ is completely determined by the complex constant $\xi$, accounting for two real degrees of freedom. Therefore, there are in total four real independent Killing spinors and thus, being sixteen the number of supersymmetries of the six-dimensional $\mathcal{N}=(1,1)$ theory, our solution is $1/4$-BPS.

As a consistency check, we perform the $y\to0$ limit explained in appendix~\ref{subsect:6d-spindle-limit} that, at least formally, reproduces the $\AdS_2 \times \spindle_1 \ltimes \spindle_2$ background of~\cite{Faedo:2022rqx}, with $q_1=q_2$. After rescaling the overall complex constant as $\xi \mapsto \xi/a$, the result is\footnote{We must take $\kk=+1$ in order for $\vartheta$ to be the correct $\AdS_2$ Killing spinor of~\cite{Faedo:2022rqx}.}
\begin{equation}
	\begin{aligned}
		(\zeta^1)_1 &= -\frac{\xi}{2} \, x^{-1/2} y^{1/8} h_i^{-3/8} Q_1^{1/2} f_1^{1/2} \,,  \qquad
		& (\zeta^1)_2 &= \frac{\xi}{2} \, x^{-1/2} y^{1/8} h_i^{-3/8} Q_1^{1/2} f_2^{1/2} \,, \\
		(\zeta^1)_3 &= \frac{\xi}{2} \, x^{-1/2} y^{1/8} h_i^{-3/8} Q_2^{1/2} f_1^{1/2} \,,  \qquad
		& (\zeta^1)_4 &= -\frac{\xi}{2} \, x^{-1/2} y^{1/8} h_i^{-3/8} Q_2^{1/2} f_2^{1/2} \,,
	\end{aligned}
\end{equation}
in agreement with the combination of~(2.19) and~(2.49) of~\cite{Faedo:2022rqx} with $\xi=-2\xi_\eta \xi_\chi$.

\section{Some solutions to the Diophantine equations}
\label{app:diofa}

In this appendix we deal with the Diophantine equations~\eqref{diofa_6d} and~\eqref{diofa_7d}, presenting some combinations of integers~$d_{a,b}$ that satisfy them. In both cases, we do not attempt to give an analytic resolution, since the expression of the constraints is involved.
Rather, we apply a brute-force method, considering all the possible combinations of $d_{1,2}$, $d_{2,3}$, $d_{3,4}$, $d_{4,1}$ and $|d_{2,4}|$ in the range 1--200, with $d_{1,3}$ given by~\eqref{drelation}, and checking which of them solve the Diophantine equation.
Among all the possible configurations, we examine only the ones yielding negative $R$-symmetry fluxes~$\flq_R^a$ and $d_{1,3},d_{2,4} < 0$, conditions required for a correct signature of the metric when restricting to $0<y<x$.
Additionally, before testing the constraint, a preliminary check is performed in order to select only the configurations giving $w>0$ and $0<\x<1$, necessary to have $0<\YA<\YB$.
Lastly, due to the structure of relation~\eqref{drelation} and of the Diophantine equations, given a specific solution, we can generate an infinite family of solutions multiplying the original configuration by any natural number. These have the same values of $a$, $\Nx$ and~$\Ny$, while the product $\Delta\psi\,\Delta\phi$ gets divided by the natural number in question.

Focussing on $D=6$, the first solutions to the Diophantine equation~\eqref{diofa_6d} are collected in table~\ref{tab:quant_6d}. Many more exist for higher values of~$d_{a,b}$.
\begin{table}[ht]
	\centering
	\begin{tabular}[t]{ | C{1.8em} | C{1.8em} | C{1.8em} | C{1.8em} | C{1.8em} | C{1.8em} | }
		\hline
		$d_{1,2}$ & $d_{2,3}$ & $d_{3,4}$ & $d_{4,1}$ & $d_{1,3}$ & $d_{2,4}$ \\
		\hline\hline
		$6$  & $9$  & $18$ & $4$  & $-24$ & $-3$ \\
		$8$  & $10$ & $17$ & $8$  & $-28$ & $-2$ \\
		$8$  & $12$ & $22$ & $5$  & $-29$ & $-4$ \\
		$10$ & $15$ & $26$ & $6$  & $-34$ & $-5$ \\
		$10$ & $20$ & $27$ & $9$  & $-45$ & $-2$ \\
		$12$ & $15$ & $23$ & $11$ & $-37$ & $-3$ \\
		$12$ & $16$ & $30$ & $11$ & $-46$ & $-4$ \\
		\hline
	\end{tabular}
	\qquad
	\begin{tabular}[t]{ | C{1.8em} | C{1.8em} | C{1.8em} | C{1.8em} | C{1.8em} | C{1.8em} | }
		\hline
		$d_{1,2}$ & $d_{2,3}$ & $d_{3,4}$ & $d_{4,1}$ & $d_{1,3}$ & $d_{2,4}$ \\
		\hline\hline
		$12$ & $18$ & $30$ & $7$  & $-39$ & $-6$  \\
		$13$ & $16$ & $24$ & $14$ & $-44$ & $-2$  \\
		$13$ & $25$ & $42$ & $7$  & $-53$ & $-7$  \\
		$14$ & $19$ & $24$ & $15$ & $-51$ & $-1$  \\
		$14$ & $21$ & $34$ & $8$  & $-44$ & $-7$  \\
		$14$ & $21$ & $40$ & $9$  & $-53$ & $-7$  \\
		$14$ & $24$ & $28$ & $14$ & $-56$ & $-1$  \\
		\hline
	\end{tabular}
	\caption{Examples of physical solutions to the Diophantine equation in $D=6$.}
	\label{tab:quant_6d}
\end{table}

In the seven-dimensional case we found only one configuration in the range considered, namely $d_{1,2}=40$, $d_{2,3}=56$, $d_{3,4}=100$, $d_{4,1}=44$, $d_{2,4}=-9$. Then, we had to multiply it by 3 in order to have an integer value of $t_J$ as well, hence
\begin{equation}
	\begin{aligned}
		d_{1,2} &= 120 \,,  \qquad  & d_{2,3} &= 168 \,,   & \qquad  d_{3,4} &= 300 \,, \\
		d_{4,1} &= 132 \,,  \qquad  & d_{1,3} &= -512 \,,  & \qquad  d_{2,4} &= -27 \,.
	\end{aligned}
\end{equation}

\section{Proof of the adjunction formula for toric orbifolds}
\label{app:adjunction}

In the mathematical literature the adjunction formula is well-known in the case of toric varieties, \ie\ within the context of algebraic geometry.
However, to the best of our knowledge, a proof for complex toric orbifolds in arbitrary dimension is lacking. Below we fill this small gap and present a proof  of the following theorem.
\begin{theorem}
\label{thm:adjunction}
	Let us consider a toric K\"ahler orbifold $\Morb$, a ramification divisor $D_a$ and the line bundle $L_a$ associated with it through Poincaré duality. The following \emph{adjunction formula} holds
	\begin{equation}
		c_1(T\Morb) \bigr|_{D_a} = c_1(TD_a) + c_1(L_a) \bigr|_{D_a} \,.
	\end{equation}
\end{theorem}
The proof of this formula will be carried out within the context of symplectic or, alternatively, complex geometry, using the formalism adopted by Guillemin and Abreu (see, \eg, \cite{Guillemin:1994kae,Abreu:2000xhs,Abreu:2001to}).
On the one hand, the symplectic coordinates will play a fundamental role in the computation of the various blocks entering the adjunction formula. On the other hand, the complex coordinates will give a shortcut to compute the Ricci form of a divisor starting from some characteristic quantities of the total space.\\

Let us consider the toric symplectic orbifold $(\Morb_{2m},\omega)$ described in subsection~\ref{subsec:toric-intro}, which  can be completely characterized by its labelled polytope $\mathcal{P}$~\eqref{polytope} or, alternatively, by its fan~$\vec{v}_a \in \ZZ^m$, with $a=1,\ldots,\fan$.
In the reminder of this appendix we shall focus on K\"ahler orbifolds, on which any $\mathbb{T}^m$-invariant metric in symplectic coordinates $(y_I,\phi_I)$, $I=1,\ldots,m$, takes the form~\cite{Abreu:2000xhs}
\begin{equation} \label{toric_metric}
	\dd s^2 = G_{IJ}(y) \, \dd y_I \dd y_J + G^{IJ}(y) \, \dd\phi_I \dd\phi_J \,,
\end{equation}
where $G_{IJ}$ is determined by a symplectic potential $G(y)$ as
\begin{equation}
	G_{IJ} = \frac{\partial^2 G}{\partial y_I \partial y_J}
\end{equation}
and $G^{IJ} = (G^{-1})_{IJ}$ is the inverse matrix.
The canonical metric is given by~\cite{Guillemin:1994kae}
\begin{equation}
	G(y) = \frac12 \sum_a l_a \log l_a  \qquad  \implies  \qquad
	G_{IJ} = \frac12 \sum_a \frac{v_a^I v_a^J}{l_a} \,,
\end{equation}
with $l_a(y) = y_I v_a^I - \lambda_a$.
It can be proven that~\cite{Abreu:2000xhs}
\begin{equation} \label{toric_determinant}
	\det(G_{IJ}) = f(y) \prod_a l_a(y)^{-1} \,,
\end{equation}
with $f$ some smooth and strictly positive function on $\mathcal{P}$.

\subsection{Adapted coordinates}
\label{sec:coord-on-div}

The strategy to prove the adjunction formula will be to pick a divisor, fix a system of coordinates adapted to it and zoom in on the chosen divisor in order to explicitly compute the different quantities involved. Even though the coordinates to be chosen are different on each divisor, we shall not include the index of the corresponding divisor in order to have less cluttered expressions.
To proceed, we need the following lemma.
\begin{lemma} \label{lem:align}
	Given a generic vector $\vec{v} = (v^1,\ldots,v^m) \in \ZZ^m$, it is always possible to align it along a preferred axis by means of an $SL(m,\ZZ)$ transformation. In particular, we can always rotate it to $\vec{w} = (0,\ldots,0,k)$, with $k = \gcd(v^1,\ldots,v^m)$.
\end{lemma}

\begin{proof}
	Let us begin with a simpler model considering a two-dimensional vector $\vec{v}_{(2)} = (v^1,v^2)$, with $k = \gcd(v^1,v^2)$. This vector can be cast in the form $\vec{w}_{(2)} = (0,k)$ by means of the left-multiplication by the matrix
	\begin{equation}
		S_{(2)} = \begin{pmatrix}
			v^2/k & -v^1/k \\
			r_1   & r_2
		\end{pmatrix} \in SL(2,\ZZ) \,,
	\end{equation}
	where $r_1$ and $r_2$ are two integers satisfying $r_1 v^1 + r_2 v^2 = k$, which always exist due to Bézout's lemma.

	Going back to the original vector $\vec{v} \in \ZZ^m$, it is possible to set to zero its first component left-multiplying it by the matrix
	\begin{equation}
		U_1 = \begin{pmatrix}
			S_1 & 0 \\
			0   & I_{m-2}
		\end{pmatrix} \in SL(m,\ZZ) \,,
	\end{equation}
	where $S_1 \in SL(2,\ZZ)$ is chosen appropriately. Explicitly, we get $U_1 \vec{v} = (0, k_{12}, v^3, \ldots, v^m)$, with $k_{12} = \gcd(v^1,v^2)$. We can now multiply the resulting vector by
	\begin{equation}
		U_2 = \begin{pmatrix}
			1 & 0   & 0 \\
			0 & S_2 & 0 \\
			0 & 0   & I_{m-3}
		\end{pmatrix} \in SL(m,\ZZ) \,,
	\end{equation}
	and obtain a new vector with vanishing second component	$U_2 U_1 \vec{v} = (0, 0, k_{123}, v^4, \ldots, v^m)$, with $k_{123} = \gcd(k_{12},v^3) = \gcd(v^1,v^2,v^3)$. Iterating this procedure, \ie\ repeatedly multiplying $\vec{v}$ by suitable block diagonal matrices, it is possible to set to zero all the $v^i$, except for the last one, which will assume the value of the greatest common divisor of all the entries of $\vec{v}$.
\end{proof}

Consider a divisor $D_b$, determined by the vector $\vec{v}_b$, and, using the results of lemma~\ref{lem:align}, align this vector along the $m$-th component by means of a specific matrix $U \in SL(m,\ZZ)$
\begin{equation}
	\vec{w}_b \equiv U \vec{v}_b = (0, \ldots, 0, m_b) \,,
\end{equation}
where $m_b = \gcd(v_b^1,\ldots,v_b^n)$ and, geometrically, represents the label associated with the divisor $D_b$.
Moreover, we define $w_a^I \equiv U_{IJ} v_a^J$ for any vector $\vec{v}_a$ of the fan describing $\mathcal{P}$.
Defining the new coordinates $y'_I = U^{JI} y_J$, with $U^{IJ} = (U^{-1})_{IJ}$, the linear relations $l_a$ retain the same expression, now in terms of the new vectors $\vec{w}_a$,
\begin{equation}
	l_a = y'_I w_a^I - \lambda_a \,,
\end{equation}
but the condition defining the facet $\facet_b$ simplifies to
\begin{equation}
	l_b = y'_m m_b - \lambda_b = 0  \qquad  \implies  \qquad  y'_m = \frac{\lambda_b}{m_b} \,.
\end{equation}
Additionally, we introduce the new angular coordinates $\phi'_I = U_{IJ} \phi_J$, in terms of which metric~\eqref{toric_metric} takes the form
\begin{equation} \label{toric_metric-rot}
	\dd s^2 = \hat{G}_{IJ} \, \dd y'_I \dd y'_J + \hat{G}^{IJ} \, \dd\phi'_I \dd\phi'_J \,,
\end{equation}
where $\hat{G}_{IJ} = \bigl( U G U^T \bigr)_{IJ}$ and $\hat{G}^{IJ} = (\hat{G}^{-1})_{IJ}$.
The matrix $\hat{G}_{IJ}$ can be obtain from the same symplectic potential as before, but with $(y,\vec{v}_a)$ replaced by $(y',\vec{w}_a)$,
\begin{equation}
	\hat{G}_{IJ} = \frac{\partial^2 \hat{G}}{\partial y'_I \partial y'_J} \,,
\end{equation}
where $\hat{G}(y';\vec{w}_a) = G(y \mapsto y';\vec{v}_a \mapsto \vec{w}_a)$.

\subsubsection*{The metric: part I}

Having chosen a convenient system of coordinates, we can now restrict to the divisor $D_b$.
First of all, we focus on the $\dd y'$ part of metric~\eqref{toric_metric-rot} and expand its expression, splitting the index $I$ as $I = (i,m)$, with $i=1,\ldots,m-1$,%
\footnote{In the second line, and for the rest of this appendix, we adopt the Einstein notation and drop the explicit summation over the repeated indices $i$ and $j$. Moreover, we adopt the notation $\dd s^2|_{\dd y'}$ and $\dd s^2|_{\dd\phi'}$ to indicate when we look at the part of the metric containing only $\dd y'^2$ or $\dd\phi'^2$, respectively, ignoring the remaining terms. Notice that we do not fix the other coordinates, but simply not look at the associated pieces in the line element.}
\begin{align} \label{metric-y}
	\dd s^2\bigr|_{\dd y'} &= \sum_{i,j=1}^{m-1} \hat{G}_{ij} \, \dd y'_i \dd y'_j + 2\sum_{i=1}^{m-1} \hat{G}_{im} \, \dd y'_i \dd y'_m + \hat{G}_{mm} (\dd y'_m)^2 \\
	&= \frac12 \sum_{a \neq b} \frac{w_a^i w_a^j}{l_a} \, \dd y'_i \dd y'_j + \sum_{a \neq b} \frac{w_a^i w_a^m}{l_a} \, \dd y'_i \dd y'_m + \frac12 \Biggl( \sum_{a \neq b} \frac{(w_a^m)^2}{l_a} + \frac{m_b^2}{l_b} \Biggr) (\dd y'_m)^2 \nonumber \,,
\end{align}
where, in the last step, we used the fact that $w_b^I = 0$ for $I \neq m$. In order to zoom in on the divisor, we define the new coordinate $\zeta$ as
\begin{equation} \label{zoom-yn}
	y'_m = \frac{\lambda_b}{m_b} + \epsilon \, \zeta^2 \,,
\end{equation}
which gives
\begin{equation}
	\dd s^2\bigr|_{\dd y'} = \frac12 \sum_{a \neq b} \frac{w_a^i w_a^j}{l_a} \, \dd y'_i \dd y'_j + 2\epsilon\,\zeta \sum_{a \neq b} \frac{w_a^i w_a^m}{l_a} \, \dd y'_i \dd\zeta + 2\epsilon \Biggl( \epsilon\,\zeta^2 \sum_{a \neq b} \frac{(w_a^m)^2}{l_a} + m_b \Biggr) \dd\zeta^2 \,.
\end{equation}
Functions $l_a$, with $a \neq b$, are non-zero on $D_b$, therefore, taking the $\epsilon \to 0$ limit, we obtain
\begin{equation}
	\dd s_{(b)}^2\bigr|_{\dd y'} = \frac12 \sum_{a \neq b} \frac{w_a^i w_a^j}{k_a} \, \dd y'_i \dd y'_j = G'_{ij} \, \dd y'_i \dd y'_j \,,
\end{equation}
where we defined the metric $G'_{ij}$ and the functions $k_a$ as
\begin{equation}
	G'_{ij}(y') = \frac12 \sum_{a \neq b} \frac{w_a^i w_a^j}{k_a} \,,  \qquad\quad
	k_a(y') = y'_i w_a^i + \frac{\lambda_b w_a^m}{m_b} - \lambda_a \equiv y'_i w_a^i - \lambda_a' \,.
\end{equation}

The facet $\facet_b$ of the labelled polytope $\mathcal{P}$ is, in turn, a new polytope in $\RR^{m-1}$, which is delimited by the intersection of $\facet_b$ itself with its (say $N$) adjacent facets $\facet_{a_u}$, with $a_u \in \{a_1,\ldots,a_N\} \equiv A_b$. These intersections are $(m-2)$-dimensional facets of $\facet_b$ and will be denoted as $\mathcal{F}'_{b|a_u}$.
Since $\facet_b$ and $\facet_{a_u}$ are determined in $\RR^m$ by the linear equations $l_b(y)=0$ and $l_{a_u}(y)=0$, $\mathcal{F}'_{b|a_u}$ will be defined by their intersection which, once restricted to the $\RR^{m-1}$ space where $\facet_b$ lives, is exactly $k_{a_u}(y')=0$.
Additionally, since $\facet_b$ is part of the polytope $\mathcal{P}$, on the former we must have $l_{a_u}(y) \geq 0$, and non-negative must be also its restriction to the $\RR^{m-1}$ hyperplane where $\facet_b$ lies, \ie\ $k_{a_u}(y')$. As a consequence, the polytope $\facet_b$ can be defined as
\begin{equation} \label{poly-F}
	\facet_b = \bigcap_{a_u \in A_b} \{ y' \in \RR^{m-1} \; : \; k_{a_u}(y') \geq 0 \} \,,
\end{equation}
and can be completely characterized by the fan of $N$ vectors $w_{a_u}^i \in \ \ZZ^{m-1}$.

\begin{proposition}
	Given a labelled polytope $\mathcal{P}$, any facet $\facet$ is itself a labelled polytope.
\end{proposition}
\begin{proof}
	Considering a specific facet $\facet_b$, this polytope is \emph{convex} because defined as the finite intersection of half-spaces, see~\eqref{poly-F}.\\
	$\facet_b$ is \emph{simple}. Since $\mathcal{P}$ is simple, exactly $m$ facets meet at every vertex. Focussing on a vertex $p$ involving $\facet_b$, we have $l_b(p) = l_{a_1}(p) = \ldots = l_{a_{m-1}}(p) = 0$, with $a_1,\ldots,a_{m-1} \in A_b$, otherwise the intersection would lie outside of the polytope and would not be a vertex of~$\mathcal{P}$.
	If we now restrict the whole set of coordinates to the $\RR^{m-1}$ hyperplane where $\facet_b$ lives, $p$, which is also a vertex of the polytope $\facet_b$, lies at the intersection
	\begin{equation} \label{Fb-vertex}
		k_{a_1}(p) = \ldots = k_{a_{m-1}}(p) = 0 \,,
	\end{equation}
	thus showing that exactly $m-1$ facets of $\facet_b$ meet at $p'$, as required.
	Moreover, the vectors $\{w_{a_1}^i,\ldots,w_{a_{m-1}}^i\}$ corresponding to the linear relations in~\eqref{Fb-vertex} form a $\QQ$-basis of $\ZZ^{m-1}$. This is because they are $m-1$ integer-valued vectors, which are linearly independent:
	\begin{equation}
		0 \neq \det(v_{a_1}^I, \ldots, v_{a_{m-1}}^I, v_b^I) = \det(w_{a_1}^I, \ldots, w_{a_{m-1}}^I, w_b^I)
		= m_b \det(w_{a_1}^i, \ldots, w_{a_{m-1}}^i) \,,
	\end{equation}
	therefore none of the vectors $w_{a_u}^i$ can be proportional to a linear combination of the others.\\
	Lastly, the condition that $\facet_b$ be \emph{rational} is equivalent to the fact that the vectors $w_{a_u}^i$ have integer entries, which is the case.
\end{proof}

We now move the attention to the facets not adjacent to $\facet_b$. The equations $k_{a_u}(y')=0$, with $a_u \not\in A_b$, emerge as the intersection of $l_b(y)=0$ and $l_{a_u}(y)=0$, whose solutions lie outside of $\mathcal{P}$ due to the convexity of the latter. As a consequence, on $\facet_b$ we must have $l_{a_u}(y) \neq 0$ and, therefore, $k_{a_u}(y') > 0$ when $a_u \not\in A_b$.
The last statement implies that we can split $G'_{ij}$ into a ``canonical'' term containing all the poles and a smooth part
\begin{equation} \label{divisor-G-deco}
	G'_{ij} = \frac12 \sum_{a \in A_b} \frac{w_a^i w_a^j}{k_a} + \frac12 \sum_{a \not\in A_b} \frac{w_a^i w_a^j}{k_a} = \frac{\partial^2 G'}{\partial y'_i \partial y'_j} \,,
\end{equation}
where
\begin{equation}
	G'(y') = \underbrace{ \frac12 \sum_{a \in A_b} k_a \log k_a }_{G'^{\mathrm{can}}} + \underbrace{ \frac12 \sum_{a \not\in A_b} k_a \log k_a }_{H} \,,
\end{equation}
with $G'^{\mathrm{can}}(y')$ canonical metric on the toric K\"ahler orbifold $\facet_b$ and $H(y')$ smooth on $\facet_b$.

The labels $m_b$ and $m_a$, with $a \neq b$, enter in different ways inside the analysis of the metric on $D_b$. The first one, the label associated with~$D_b$, lies ``outside'' the divisor $D_b$ itself and, in particular, it shows up in the normal direction to $D_b$ (see, as an example, metric~\eqref{2spin_stack-metric}). As a consequence, the metric on $D_b$ is insensitive to~$m_b$.
On the contrary, the labels associated with the divisors~$D_{a_u}$ adjacent to $D_b$ build up the labels corresponding to the fan~$w_{a_u}^i$, \ie\ the labels of the divisors of (the original divisor) $D_b$.

\subsubsection*{The metric: part II}

The analysis of the angular part is more involved, since it requires the knowledge of the inverse matrix $\hat{G}^{IJ}$. If we decompose $\hat{G}_{IJ}$ as
\begin{equation}
	\hat{G}_{IJ} = \begin{pmatrix}
		\hat{G}_{ij} & \hat{G}_{im} \\
		\hat{G}_{mj} & \hat{G}_{mm}
	\end{pmatrix} \,,
\end{equation}
we obtain the following components for $\hat{G}^{IJ}$
\begin{equation}
	\hat{G}^{ij} = (\mathcal{G}^{-1})_{ij} + \gamma \, \Gamma_i \, \Gamma_j \,,  \qquad
	\hat{G}^{im} = -\gamma \, \Gamma_i \,, \qquad  \hat{G}^{mm} = \gamma \,,
\end{equation}
where we defined the submatrix $\mathcal{G}_{ij} \equiv \hat{G}_{ij}$ and the two functions
\begin{equation}
	\gamma(y') = \frac{\det(\hat{G}_{ij})}{\det(\hat{G}_{IJ})} \,,  \qquad\quad
	\Gamma_i(y') = (\mathcal{G}^{-1})_{ik} \hat{G}_{km} \,.
\end{equation}
By means of~\eqref{toric_determinant}, function $\gamma$ can be written as
\begin{equation}
	\gamma = \frac{\tilde{f} \, \prod_{a \neq b} l_a^{-1}}{f \, \prod_a l_a^{-1}} = \tilde{g} \, l_b \,,
\end{equation}
with $\tilde{g}$ smooth and strictly positive on $\mathcal{P}$, hence on $\facet_b$.
In the new rotated coordinates, the $\dd\phi'$ part of the metric reads
\begin{equation}
	\dd s^2\bigr|_{\dd\phi'} = \bigl( (\mathcal{G}^{-1})_{ij} + \gamma \, \Gamma_i \Gamma_j \bigr) \, \dd\phi'_i \dd\phi'_j - 2\gamma \, \Gamma_i \, \dd\phi'_i \dd\phi'_m + \gamma \, (\dd\phi'_m)^2 \,.
\end{equation}
Since $\mathcal{G}_{ij}$ and $\hat{G}_{in}$ are smooth everywhere apart from the facets $\facet_a$, with $a \neq b$ (see their definition in~\eqref{metric-y}), and likewise $(\mathcal{G}^{-1})_{ij}$, when we zoom in on $D_b$ the function $l_b$ goes to zero, $\gamma$ vanishes and $\Gamma_i$ remains finite, thus the metric becomes
\begin{equation}
	\dd s_{(b)}^2\bigr|_{\dd\phi'} = G'^{ij} \, \dd\phi'_i \dd\phi'_j \,,
\end{equation}
where we introduced the inverse matrix $G'^{ij} \equiv (G'^{-1})_{ij}$. Indeed, when $y'_m \to \lambda_b/m_b$, the matrix $\mathcal{G}_{ij} = \hat{G}_{ij}$ smoothly goes to $G'_{ij}$, therefore $(\mathcal{G}^{-1})_{ij}$ smoothly goes to $(G'^{-1})_{ij}$.

\subsection{Adjunction formula}

We have now set all the ingredients to prove the adjunction formula stated in theorem~\ref{thm:adjunction}.
Let us consider the divisor $D_b$, equipped with the metric
\begin{equation}
	\dd s_{(b)}^2 = G'_{ij} \, \dd y'_i \dd y'_j + G'^{ij} \, \dd\phi'_i \dd\phi'_j \,,
\end{equation}
where the matrix $G'_{ij}$ is given by~\eqref{divisor-G-deco} and $G'^{ij}$ is its inverse.
The Ricci potential of this toric K\"ahler orbifold is%
\footnote{This expression can be derived from the Ricci form, and the corresponding Ricci potential, written in complex coordinates.}
\begin{equation}
	P_{(b)} = \frac12 \, G'^{ij} \frac{\partial}{\partial y'_j} \bigl(\log \det G'\bigr) \dd\phi'_i \,,
\end{equation}
where we denote $\det G' = \det(G'_{ij})$, from which we derive
\begin{equation} \label{c1TD}
	c_1(TD_b) = \frac{1}{2\pi} \, \dd P_{(b)} = \frac{1}{4\pi} \, \partial'_k \bigl[ G'^{ij} \partial'_j \bigl( \log \det G' \bigr) \bigr] \dd y'_k \wedge \dd\phi'_i \,.
\end{equation}
Applying~\eqref{toric_determinant}, we have
\begin{equation} \label{det-G'}
	\det(G'_{ij}) = \mathring{g}(y') \prod_{a \in A_b} k_a(y')^{-1} \,,
\end{equation}
with $\mathring{g}$ some smooth and strictly positive function on $\facet_b$, hence
\begin{equation} \label{c1TD-exp}
	\begin{split}
		c_1(TD_b) &= -\frac{1}{4\pi} \, \partial'_k \biggl[ G'^{ij} \partial'_j \log\biggl(\mathring{g}^{-1} \prod_{a \in A_b} k_a \biggr) \biggr] \dd y'_k \wedge \dd\phi'_i \\
		&= -\frac{1}{4\pi} \sum_{a \in A_b} \partial'_k \bigl( G'^{ij} \partial'_j \log k_a \bigr) \dd y'_k \wedge \dd\phi'_i \,,
	\end{split}
\end{equation}
up to an exact form.

On the other hand, $c_1(L_a)$ in a system of coordinates adapted to $D_b$ reads~\cite{Martelli:2023oqk}
\begin{align} \label{c1La}
	c_1(L_a) &= -\frac{1}{4\pi} \, \dd \biggl( \frac{\hat{G}^{IJ} w_a^J}{l_a} \biggr) \wedge \dd\phi'_I \nonumber \\
	&= -\frac{1}{4\pi} \biggl[ \dd \biggl( \frac{\hat{G}^{ij} w_a^j}{l_a} + \frac{\hat{G}^{im} w_a^m}{l_a} \biggr) \wedge \dd\phi'_i
	+ \dd \biggl(\frac{\hat{G}^{mj} w_a^j}{l_a} + \frac{\hat{G}^{mm} w_a^m}{l_a} \biggr) \wedge \dd\phi'_m \biggr] \\
	&= -\frac{1}{4\pi} \biggl[ \dd \biggl( \frac{((\mathcal{G}^{-1})_{ij} + \gamma \, \Gamma_i \Gamma_j) w_a^j}{l_a} - \frac{\gamma \, \Gamma_i w_a^m}{l_a} \biggr) \wedge \dd\phi'_i
	- \dd \biggl(\frac{\gamma \, \Gamma_j w_a^j}{l_a} - \frac{\gamma \, w_a^m}{l_a} \biggr) \wedge \dd\phi'_m \biggr] \,. \nonumber
\end{align}
Given a function $h(y')$ smooth in a neighbourhood of $\facet_b$, we have
\begin{equation}
	\dd(\gamma \, h) = \dd(\tilde{g} \, l_b \, h) = l_b \, \dd(\tilde{g} h) + \tilde{g} h \, \dd l_b \,.
\end{equation}
In the limit $y'_m \to \lambda_b/m_b$ this quantity vanishes, because the product $\tilde{g} h$ is smooth, while $l_b \to 0$ and $\dd l_b = m_b \, \dd y'_m = 2\epsilon \, m_b \zeta \, \dd\zeta \to 0$.
Functions $1/l_a$ and $\Gamma_i/l_a$ are smooth everywhere apart from the facet $\facet_a$, with $a \neq b$, therefore, when restricting to the divisor $D_b$, all the terms in~\eqref{c1La} containing $\gamma$ disappear.
Additionally, since $(\mathcal{G}^{-1})_{ij}/l_a$ is smooth, its derivative with respect to $y'_m$ is finite, hence $\partial'_m[(\mathcal{G}^{-1})_{ij}/l_a] \dd y'_m \to 0$ when approaching the facet $\facet_b$. As a result,
\begin{equation}
	c_1(L_a)\bigr|_{D_b} = -\frac{1}{4\pi} \, \frac{\partial}{\partial y'_k} \biggl( \frac{G'^{ij} w_a^j}{k_a} \biggr) \dd y'_k \wedge \dd\phi'_i \,.
\end{equation}
Analysing in more detail this expression, we have two different results according to whether $D_a$, the divisor dual to the line bundle $L_a$, is adjacent to $D_b$ or not. Indeed, in the latter case $(G'^{ij} w_a^j)/k_a$ is smooth and $c_1(L_a)\bigr|_{D_b}$ vanishes in cohomology, therefore
\begin{equation} \label{c1La-final}
	c_1(L_a)\bigr|_{D_b} = \begin{cases}
		-\dfrac{1}{4\pi} \, \dfrac{\partial}{\partial y'_k} \biggl( \dfrac{G'^{ij} w_a^j}{k_a} \biggr) \dd y'_k \wedge \dd\phi'_i  \quad &  a \in A_b \,, \\
		0  \quad &  a \not\in A_b \,.
	\end{cases}
\end{equation}
Recalling that $w_a^j = \partial'_j k_a$, we can now compute the quantity
\begin{equation}
	\sum_{a \neq b} c_1(L_a)\bigr|_{D_b} = -\frac{1}{4\pi} \sum_{a \in A_b} \partial'_k \bigl( G'^{ij} \partial'_j \log k_a \bigr) \dd y'_k \wedge \dd\phi'_i \,,
\end{equation}
which perfectly agrees with~\eqref{c1TD-exp}. As a result, we have the adjunction formula
\begin{equation}
	c_1(TD_b) = \sum_{a \neq b} c_1(L_a)\bigr|_{D_b} = \bigl[ c_1(T\Morb_{2n}) - c_1(L_b) \bigr] \bigr|_{D_b} \,,
\end{equation}
where we used the following relation for the orbifold canonical line bundle
\begin{equation}
	K_{\Morb_{2n}}^\mathrm{orb} = -\sum_a D_a  \qquad  \implies  \qquad
	c_1(T\Morb_{2n}) = \sum_a c_1(L_a) \,.
\end{equation}

\subsection{Ricci form of toric divisors}
We close this appendix describing a method to compute the Ricci form of $D_b$ taking advantage of the underlying complex structure. In order to make this structure manifest, we introduce the holomorphic coordinates~\cite{Guillemin:1994kae} (see also~\cite{Martelli:2005tp})
\begin{equation}
	z_I = x_I + \ii \phi_I \,,  \qquad  x_I = \frac{\partial G}{\partial y_I} \,.
\end{equation}
The K\"ahler potential $F$ of the K\"ahler orbifold $\Morb_{2m}$ is obtained from the symplectic potential through the Legendre transformation
\begin{equation}
	F(x) = y_I x_I - G(y)  \qquad  \text{with}  \qquad  y_I = \frac{\partial F}{\partial x_I} \,.
\end{equation}
Therefore, metric~\eqref{toric_metric} on $\Morb_{2m}$ becomes
\begin{equation}
	\dd s^2 = F_{IJ}(x) \, (\dd x_I \dd x_J + \dd\phi_I \dd\phi_J) = F_{IJ}(x) \, \dd z_I \dd \bar{z}_J \,,
\end{equation}
where $F_{IJ}$ is determined by the K\"ahler potential $F(x)$ as
\begin{equation}
	F_{IJ} = \frac{\partial^2 F}{\partial x_I \partial x_J}  \qquad  \implies  \qquad
	F_{IJ} = G^{IJ} \,,
\end{equation}
which also implies, \cf~\eqref{toric_determinant},
\begin{equation}
	\det(F_{IJ}) = f(x)^{-1} \prod_a l_a \,,
\end{equation}
with $f$ some smooth and strictly positive function on $\mathcal{P}$.
The Ricci form is given by
\begin{equation}
	\rho = -\ii \, \partial\bar\partial \log \det F \,,
\end{equation}
where we denote $\det F = \det (F_{IJ})$, and the corresponding (real) Ricci potential reads
\begin{equation} \label{ricci-pot}
	P = \frac{\ii}{2} \bigl( \partial \log \det F - \bar\partial \log \det F \bigr) \,,
\end{equation}
up to an exact form.
The Ricci potential~\eqref{ricci-pot} can also be obtained from the canonical $(m,0)$-form
\begin{equation}
	\Omega = \ee^{\ii\alpha} (\det F)^{1/2} \, \dd z_1 \wedge \ldots \wedge \dd z_m \,,
\end{equation}
with $\alpha$ a holomorphic function, by means of the relation $\dd\Omega = \ii P \wedge \Omega$.

Applying the $SL(m,\ZZ)$ transformation of section~\ref{sec:coord-on-div}, the real coordinates $x_I$ and the complex coordinates $z_I$ rotate as $x'_I = U_{IJ} x_J$ and $z'_I = U_{IJ} z_J$, with $x'_I = \partial \hat{G}/\partial y'_I$.
Accordingly, the metric becomes
\begin{equation}
	\dd s^2 = \hat{F}_{IJ} \, (\dd x'_I \dd x'_J + \dd\phi'_I \dd\phi'_J) = \hat{F}_{IJ} \, \dd z'_I \dd \bar{z}'_J \,,
\end{equation}
where we defined the matrix $\hat{F}_{IJ} = \bigl( (U^{-1})^T F U^{-1} \bigr)_{IJ} = \hat{G}^{IJ}$.
Because $\det U = 1$, the canonical $(m,0)$-form retains the same functional expression
\begin{equation}
	\Omega = \ee^{\ii\alpha} (\det \hat{F})^{1/2} \, \dd z'_1 \wedge \ldots \wedge \dd z'_m \,.
\end{equation}
Zooming in on the divisor $D_b$, \ie\ performing the change of coordinates~\eqref{zoom-yn} and taking the $\epsilon\to0$ limit, we have
\begin{equation}
	\dd x'_i = \hat{G}_{ij} \dd y'_j + O(\epsilon) \,,  \qquad  \dd x'_m = \hat{G}_{mj} \dd y'_j + \frac{m_b}{\zeta} \, \dd\zeta + O(\epsilon) \,,
\end{equation}
therefore, since $\hat{F}_{im}$ and $\hat{F}_{mm}$ vanish in this limit, the total metric becomes
\begin{equation}
	\dd s_{(b)}^2 = F'_{ij} \, (\dd x'_i \dd x'_j + \dd\phi'_i \dd\phi'_j) = F'_{ij} \, \dd z'_i \dd \bar{z}'_j \,,
\end{equation}
where we introduced the matrix $F'_{ij} = G'^{ij}$.
The canonical $(m-1,0)$-form associated with~$D_b$ is
\begin{equation}
	\Omega_{(b)} = \ee^{\ii\alpha'} (\det F')^{1/2} \, \dd z'_1 \wedge \ldots \wedge \dd z'_{m-1} \,,
\end{equation}
with $\det F' = \det(F'_{ij})$ and $\alpha'$ a holomorphic function.
Let us consider the total space $\Morb_{2m}$, whose Killing vector degenerating on $D_b$ is $\xi = \partial_{\phi'_m}$. If we now define the vector $V = \hat{F}_{mm}^{-1/2} \xi$ of unit norm, we notice that $\Omega_{(b)} = \imath_{V} \Omega|_{D_b}$, up to an overall holomorphic function.

The result of this subsection is the following prescription to compute the Ricci form of a given divisor $D_b$ of a toric K\"ahler orbifold $\Morb_{2m}$.
Consider the Killing vector $\xi$ degenerating on $D_b$ and define the vector $V = \langle\xi,\xi\rangle^{-1/2}\,\xi$ of unit norm. Let $\Omega$ be the canonical $(m,0)$-form on $\Morb_{2n}$.
The canonical $(m-1,0)$-form on $D_b$ is given by the interior product $\Omega_{(b)} = \imath_{V} \Omega|_{D_b}$. Then, $\Omega_{(b)}$ allows us to compute the Ricci potential $P_{(b)}$ and, lastly, the Ricci form $\rho_{(b)}$.

\begin{remark}
	Although this prescription was proven using a set of complex coordinates, it can be extended to systems in which explicit complex coordinates are not known.
\end{remark}

\bibliographystyle{JHEP}
\bibliography{biblio-nut}

\end{document}